\documentclass[11pt]{article}
\usepackage{amsmath}
\usepackage{amssymb}  
\usepackage{amsthm}
\usepackage{etoolbox}  
\usepackage{comment}
\usepackage{algorithm}
\usepackage{algpseudocode}
\usepackage{subfig}
\usepackage{color}
\usepackage[english]{babel}
\usepackage{graphicx}
\usepackage{wrapfig,epsfig}
\usepackage{epstopdf}
\usepackage{url}
\usepackage{graphicx}
\usepackage{color} 
\usepackage{epstopdf}
\usepackage{scrextend}
\usepackage[T1]{fontenc}
\usepackage{bbm}

\usepackage{tikz}
\usepackage{hyperref}
\hypersetup{colorlinks=true,citecolor=red,linkcolor=blue}

\usepackage[margin=1in]{geometry}

\DeclareMathOperator{\rev}{reverse}

\DeclareMathOperator{\LCS}{\mathsf{lcs}}

\DeclareMathOperator{\ED}{\mathsf{ed}}

\DeclareMathOperator{\poly}{poly}
\usepackage{etoolbox}

\makeatletter

\newcommand{\define}[4][ignore]{%
  \ifstrequal{#1}{ignore}{}{
  \@namedef{thmtitle@#2}{#1}}%
  \@namedef{thm@#2}{#4}%
  \@namedef{thmtypen@#2}{lemma}%
  \newtheorem{thmtype@#2}[theorem]{#3}%
  \newtheorem*{thmtypealt@#2}{#3~\ref{#2}}%
}

\newcommand{\state}[1]{%
  \@namedef{curthm}{#1}
  \@ifundefined{thmtitle@#1}{
  \begin{thmtype@#1}
    }{
  \begin{thmtype@#1}[\@nameuse{thmtitle@#1}]
  }
    \label{#1}
    \@nameuse{thm@#1}
  \end{thmtype@#1}
  \@ifundefined{thmdone@#1}{
  \@namedef{thmdone@#1}{stated}%
  }{}
}

\newcommand{\restate}[1]{%
  \@namedef{curthm}{#1}
  \@ifundefined{thmtitle@#1}{
    \begin{thmtypealt@#1}
    }{
  \begin{thmtypealt@#1}[\@nameuse{thmtitle@#1}]
  }
    \@nameuse{thm@#1}
  \end{thmtypealt@#1}
  \@ifundefined{thmdone@#1}{
  \@namedef{thmdone@#1}{stated}%
  }{}
}

\newcommand{\wmax}{\mathsf{w}_{\mathsf{max}}}
\newcommand{\wgap}{\mathsf{w}_{\mathsf{gap}}}
\newcommand{\wmin}{\mathsf{w}_{\mathsf{min}}}
\newcommand{\wlayers}{\mathsf{w}_{\mathsf{layers}}}
\newcommand{\lcs}{\mathsf{lcs}}
\newcommand{\ooutput}[1]{\mathsf{O}^{#1}}
\newcommand{\normal}[1]{||#1||}
\newcommand{\graph}[1]{\mathsf{G}^{#1}}

\newcommand{\tgraph}[1]{\hat{\mathsf{G}}^{#1}}
\newcommand{\opt}{\mathsf{opt}}
\newcommand{\lcsdata}{\mathsf{lcs}\textsf{-}\mathsf{cmp}}
\newcommand{\nonmetric}{\textsc{NonMetricEstimation}}

\newcommand{\lis}{\mathsf{lis}}
\newcommand{\ssmall}{\mathsf{sm}}
\newcommand{\llarge}{\mathsf{lg}}
\newcommand{\sa}{\mathsf{sa}}
\newcommand{\cdi}{\mathsf{cdi}}
\newcommand{\ps}{\mathsf{ps}}
\newcommand{\quality}{\mathsf{q}}
\newcommand{\gdelta}{\mathsf{NG}_{\lambda}}
\newcommand{\gjustdelta}{\mathsf{NG}}
\newcommand{\fdelta}{\mathsf{NF}_{\lambda}}

\newcommand{\thmlabel}[1]{
  \@ifundefined{thmdone@\@nameuse{curthm}}{\label{#1}
    }{\tag*{\eqref{#1}}}
}
\makeatother
\newcommand{\wh}{\widehat}
\newcommand{\wt}{\widetilde}

\newcommand{\eps}{\epsilon}

\DeclareMathOperator*{\E}{{\bf {E}}}

\renewcommand{\varepsilon}{\epsilon}
\renewcommand{\tilde}{\wt}
\renewcommand{\hat}{\wh}

\newcommand{\tO}{\tilde{O}}

\graphicspath{{./figs/}}

\newtheorem{theorem}{Theorem}[section]
\newtheorem{lemma}[theorem]{Lemma}

\newtheorem{fact}[theorem]{Fact}
\newtheorem{claim}[theorem]{Claim}
\newtheorem{corollary}[theorem]{Corollary}

\newtheorem{property}[theorem]{Property}
\newtheorem{challenge}[theorem]{Challenge}

\theoremstyle{definition}
\newtheorem{definition}[theorem]{Definition}

\definecolor{xiaorui}{rgb}{0.4,0.5,0.7}
\definecolor{zhao}{rgb}{0.7,0.4,0.5}
\definecolor{aviad}{rgb}{0.5,0.7,0.4}

\title{Approximation Algorithms for LCS and LIS with Truly Improved Running Times\thanks{A preliminary version of this work is appeared in FOCS'19.}}
\author{
Aviad Rubinstein\thanks{Stanford University.
  \texttt{aviad@stanford.edu}}
\and 
Saeed Seddighin\thanks{Toyota Technological Institute at Chicago. 
  \texttt{saeedreza.seddighin@gmail.com} }
\and Zhao Song\thanks{ Adobe Research.
	\texttt{zsong@adobe.com}  }
\and Xiaorui Sun\thanks{University of Illinois. 
	\texttt{xiaorui@uic.edu} }
}

\begin{document}

\begin{titlepage}
\maketitle
\begin{abstract}
Longest common subsequence ($\mathsf{LCS}$) is a classic and central problem in combinatorial optimization. While $\mathsf{LCS}$ admits a quadratic time solution, recent evidence suggests that solving the problem may be impossible in truly subquadratic time. 
A special case of $\mathsf{LCS}$ wherein each character appears at most once in every string is equivalent to the longest increasing subsequence problem ($\mathsf{LIS}$) which can be solved in quasilinear time. In this work, we present  novel algorithms for approximating $\mathsf{LCS}$ in truly subquadratic time and $\mathsf{LIS}$ in truly sublinear time. 
Our approximation factors depend on the ratio of the optimal solution size over the input size. We denote this ratio by $\lambda$
and obtain the following results for $\mathsf{LCS}$ and $\mathsf{LIS}$ without any prior knowledge of $\lambda$.
\begin{itemize}
	\item A truly subquadratic time algorithm for $\mathsf{LCS}$ with approximation factor $\Omega(\lambda^3)$.
	\item A truly sublinear time algorithm for $\mathsf{LIS}$ with approximation factor $\Omega(\lambda^3)$.
\end{itemize}

Triangle inequality was recently used by Boroujeni, Ehsani, Ghodsi, HajiAghayi and Seddighin \cite{beghs18} and Charkraborty, Das, Goldenberg, Koucky and Saks \cite{cdgks18} to present new approximation algorithms for edit distance. Our techniques for $\mathsf{LCS}$ extend the notion of triangle inequality to non-metric settings.

\end{abstract}

\thispagestyle{empty}
\end{titlepage}

% !TeX root = ../../main.tex

\section{Introduction}

We consider three problems in combinatorial optimization: the longest common subsequence (\textsf{LCS}), the edit distance (\textsf{ED}), and the longest increasing subsequence (\textsf{LIS}).  The \textsf{LCS} of two strings $A$ and $B$ is simply their longest (not necessarily contiguous) common subsequence. 
The edit distance is defined as the minimum number of character deletions, insertions, and substitutions required to transform $A$ into $B$. For the purpose of our discussion, we consider a more restricted definition of edit distance where substitutions are not allowed%
\footnote{Alternatively, the cost of a substitution is doubled as it requires a deletion and an insertion.}. Longest increasing subsequence is equivalent to a special case of \textsf{LCS} where the input strings are permutations. All three problems are very fundamental and have been subject to a plethora of studies in the past few decades and specially in recent years ~\cite{lms98,bjkk04,bes06,ao09,ako10,ss10,bi15,abw15,ahww16,ab17,ar18,cglrr18,beghs18,cdgks18,hsss19}.

If the strings have length $n$, both \textsf{LCS} and \textsf{ED} can be solved in quadratic time ($O(n^2)$) with dynamic programming. These running times are slightly improved to $O(n^2/\log^2(n))$ by Masek and Paterson~\cite{mp80}, however, efforts to improve the running time to $O(n^{2-\Omega(1)})$ for either edit distance or \textsf{LCS} were all futile.

In recent years, our understanding of the source of complexity for these problems improved thanks to a sequence of fine-grained complexity developments~\cite{abw15, ahww16}. We now know that assuming the strong exponential time hypothesis (\textsf{SETH})~\cite{abw15}, or even weaker assumptions such as the orthogonal vectors conjecture (\textsf{\textsf{OVC}})~\cite{abw15} or \textsf{branching-program-SETH}~\cite{ahww16}, there are no truly sub-quadratic\footnote{By {\em truly sub-quadratic} we mean $O(n^{2-\epsilon})$, for any constant $\epsilon >0$} time algorithms for \textsf{LCS}. Similar results also hold for edit distance \cite{bi15}.

The classic approach to break the quadratic barrier for these problems is approximation algorithms. Note that for (multiplicative) approximations, \textsf{LCS} and edit distance are no longer equivalent (much like we have a $2$-approximation algorithm for Vertex Cover, but Independent Set is NP-hard to approximate within near-linear factors).

For edit distance, an $\tilde{O}(n+\Delta^2)$-time algorithm of~\cite{lms98} (where $\Delta$ is the true edit distance between the strings) implies a linear-time $\sqrt{n}$-approximation algorithm.
The approximation factor has been significantly improved in a series of works to $O(n^{3/7})$~\cite{bjkk04}, to $O(n^{0.34})$~\cite{bes06}, to $O(2^{\tilde{O}(\sqrt{\log n})})$~\cite{ao09}\footnote{We define $\widetilde{O}(f)$ to be $f\cdot \log^{O(1)}(f)$.}, and finally to polylogarithmic~\cite{ako10}. A recent work of Boroujeni \textit{et al.}~\cite{beghs18} obtains a  constant factor approximation quantum algorithm for edit distance that runs in truly subquadratic time. % 
Finally, the breakthrough of Chakraborty \textit{et al.}~\cite{cdgks18} gave a classic (randomized) constant factor approximation for edit distance in truly subquadratic time. A key component in both of the latest constant factor approximation algorithms is the application of triangle inequality (for edit distance between certain substrings of the input). 
A particular challenge in extending these ideas to \textsf{LCS} is that \textsf{LCS} is not a metric and in particular does not satisfy the triangle inequality.

Our understanding of the complexity of approximate solutions for \textsf{LCS} is embarrassingly limited. For general strings, there are several linear-time $1/\sqrt{n}$-approximation algorithms based on sampling techniques. % (recently improved to $O(n^{1/2-\epsilon})$~\cite{hsss19}. 
For alphabet size $|\Sigma|$, there is a trivial $1/|\Sigma|$-approximation algorithm that runs in linear time. %(for binary strings, this approximation ratio was also recently impro$1/2+\epsilon$~\cite{RS19-LCS}). 
Whether or not these approximation factors can be improved by keeping the running time linear is one of the central problems in fine-grained complexity. 
Very recently, both the general $1/\sqrt{n}$-approximation factor, and, for binary strings, the $1/2$-approximation factor, have been slightly improved (\cite{hsss19} and~\cite{RS19-LCS}, respectively). These works give improved algorithm for the two extreme cases where the size of the alphabet is very small or very large. In comparison, our approximation guarantee depends on the solution size rather than the size of the alphabet. Also, for the special case of balanced strings, we improve upon the result of~\cite{RS19-LCS} by obtaining an $o(|\Sigma|)$ approximate solution in subquadratic time. 
There are a few fine-grained complexity results for approximate \textsf{LCS}, but they only hold against deterministic algorithms, and rely on very strong assumptions~\cite{ab17, ar18, cglrr18}.

\subsection{Our Results}\label{sec:our_results}

For simplicity, we use $\lcs(A,B)$ to denote the size (not the whole sequence) of the longest common subsequence for two strings $A$ and $B$. Similarly, we use $\ED(A,B)$ to denote the edit distance and $\lis(A)$ for the size of the longest common subsequence. We sometimes normalize the solution by the length of the strings so that the size of the solution remains in the interval $[0,1]$. We refer to the normalized solutions by $\normal{\lcs(A,B)} = \lcs(A,B)/n$ and $\normal{\ED(A,B)} = \ED(A,B)/2n$ (here both strings have equal length $n$), and $\normal{\lis(A)} = \lis(A)/n$. In this way, $\normal{\ED(A,B)} + \normal{\lcs(A,B)} = 1$ (assuming both strings have equal length).

As mentioned earlier, recent developments for edit distance are based on a simple but rather useful observation. Edit distance satisfies triangle inequality, or in other words, given three strings $A_1, A_2, A_3$ of length $n$ such that $\normal{\ED(A_1,A_2)} \leq \delta$ and $ \normal{\ED(A_2,A_3)} \leq \delta$ hold, we can easily imply that $\normal{\ED(A_1,A_3)}\leq 2\delta$. 
While $\lcs$ does not satisfy the triangle inequality in any meaningful way, it does, {\em on average}, satisfy the following birthday-paradox-like property that we call \textit{birthday triangle inequality}. 
\begin{property}[birthday triangle inequality] 
Consider three equal-length strings $A_1$, $A_2$, and $A_3$ %of length $n$ 
such that $\normal{\lcs(A_1,A_2)} \geq \lambda$ and $\normal{\lcs(A_2,A_3)} \geq \lambda$. 
If the common subsequences correspond to random indices of each string, we expect that $\normal{\lcs(A_1,A_3)} \geq \lambda^2$.
\end{property}

Of course, this is not necessarily the case in general. More precisely, it is easy to construct examples%
\footnote{For example, $A_1 = 0^{n/2}0^{n/2}, A_2 = 0^{n/2}1^{n/2}, A_3 = 1^{n/2}1^{n/2}$.} in which $\normal{\lcs(A_1,A_2)} = 1/2$ and $\normal{\lcs(A_2,A_3)} = 1/2$, but $\normal{\lcs(A_1,A_3)} = 0$. 
Our main result shows that while it only holds on average, we can algorithmically replace the triangle inequality for edit distance with the birthday triangle inequality {\em on worst case inputs}.

\begin{theorem}[Main Theorem, formally stated as Theorems~\ref{thm:formal_lcs} and ~\ref{thm:formal_lcs_quadratic}]
Given strings $A,B$ both of length $n$ such that $\normal{\LCS(A,B)} = \lambda$, we can approximate the length of the \textsf{LCS} between the two strings within an $\Omega(\lambda^3)$ factor in subquadratic time. The approximation factor improves to $(1-\epsilon)\lambda^2$ when $1/\lambda$ is constant.
 \end{theorem}

We remark  that our algorithm is actually able to output the whole sequence of the solution, but we only focus on estimating the size of the solution for simplicity. We begin by comparing our main theorem to previous work on edit distance. In this case, $1/\lambda$ is constant w.l.o.g.%
\footnote{When we use our solution to approximate edit distance, we can safely assume that $\normal{\lcs(A,B)} = \Omega(1)$ since otherwise the edit distance of the two strings is very close to $2n$.}  
and therefore the approximation factor of our algorithm is $(1-\epsilon)\lambda^2$.
If $\delta = \normal{\ED(A,B)}$, then our \textsf{LCS} algorithm outputs a transformation from $A$ to $B$ using at most $2n  (1 -  (1-\epsilon)(1-\delta)^3 )$ operations.
Observe that when the strings are not overly close and $\delta= \Omega(1)$ by scaling $\epsilon$, we already recover a $(3+\epsilon')$-approximation for edit distance in truly subquadratic time. 
For mildly far strings, say $\delta = 0.1$, a more careful look at the expansion of $(1-\delta)^3$ reveals that we save an additive $\Theta(\delta^2)$ in the approximation factor. For example, with $\delta =0.1$ our approximation factor for edit distance is $2.71$ instead of $3$.

An interesting implication of our main result is for \textsf{LCS} over a large alphabet $\Sigma$, where the optimum $\normal{\LCS(A,B)}$ may be much smaller than $1$. This is believed to be the hardest regime for approximation algorithms (and indeed the only one for which we have any conditional hardness of approximation results~\cite{ab17, ar18, cglrr18}).
Here, we consider instances that satisfy a mild balance assumption: we assume that there is a character that appears with frequency at least $1/|\Sigma|$ in both strings%
\footnote{Note that in every instance in each string there is a character that appears with frequency at least $1/|\Sigma|$, but in general that may not be the same character.}. Then, our main theorem implies an $O(1/|\Sigma|^{3/4})$-approximate solution in truly subquadratic time (the first improvement over the trivial $1/|\Sigma|$ approximation in this regime). 

\begin{corollary}[\textsf{LCS}, formally stated as Corollary~\ref{cor:formal_lcs_sigma}]
Given a pair of strings $(A,B)$ of length $n$ over alphabet $\Sigma$ that satisfy the balance condition, we can approximate their \textsf{LCS} within an $O(|\Sigma|^{3/4})$ factor in truly subquadratic time.
\end{corollary}

Next, we show that a similar result can be obtained for \textsf{LIS}. Perhaps coincidentally, the approximation factor of our algorithm is also $\Omega(\lambda^3)$ which is same to  \textsf{LCS}, but the technique is completely different. Although \textsf{LIS} can be solved exactly in time $O(n \log n)$, there have been several attempts to approximate the size of  \textsf{LIS} and related problems in sublinear time~\cite{s61,f75,dglrrs99,ekkrv00,f04,accl07,ss10}. The best known solution is due to the work of Saks and Seshadhri~\cite{ss10} that obtains a $(1+\epsilon)$-approximate algorithm for \textsf{LIS} in polylogarithmic time, when the solution size is at least a constant fraction of the input size~\footnote{Their algorithm obtains an additive error of $\delta n$ in time $2^{ \tilde{O}( 1/ \delta ) }$.
 When the solution size is bounded by $\lambda n$, one needs to set $\delta < \lambda$ in order to guarantee a multiplicative factor approximation.}. In other words, if $\normal{\lis(A)} = \lambda$ and $1/\lambda$ is constant, their algorithm approximates $\lis(A)$ in polylogarithmic time. However, this only works if $1/\lambda$ is constant and even if $1/\lambda$ is logarithmically large, their method fails to run in sublinear time\footnote{There is a term $(1/\lambda)^{1/\lambda}$ in the running time.}. We complement the work of Saks and Seshadhri~\cite{ss10}  by presenting a result for \textsf{LIS} similar to our result for \textsf{LCS}. More precisely, we show that when $\normal{\lis(A)} = \lambda$, an $\Omega(\lambda^3)$ approximation of \textsf{LIS} can be obtained in truly sublinear time. Although our approximation factor is worse than that of ~\cite{ss10}, our result works for any (not necessarily constant) $\lambda$.

\begin{theorem}[\textsf{LIS}, formally stated as Theorem~\ref{theorem:lis}]
	Given an array $A$ of $n$ integer numbers such that $\normal{\lis(A)} = \lambda$. We can approximate the length of the \textsf{LIS} for $A$ in sublinear time within a factor $\Omega(\lambda^3)$.
\end{theorem}

If one favors the running time over the approximation factor, it is possible to improve the exponent of $n$ in the running time down to any constant $\kappa > 0$ at the expense of incurring a larger multiplicative factor to the approximation.

\subsection{Preliminaries}\label{sec:preli}
In \textsf{LCS} or edit distance, we are given two strings $A$ and $B$ as input. We assume for simplicity that the two strings have equal length and refer to that by $n$. In \textsf{LCS}, the goal is to find the largest subsequence of the characters which is shared between the two strings. In edit distance, the goal is to remove as few characters as possible from the two strings such that the remainders for the two strings are the same. We use $\lcs(A,B)$ and $\ED(A,B)$ to denote the size of the longest common subsequence and the edit distance of two strings $A$ and $B$.

In \textsf{LIS}, the input contains an array $A$ of $n$ integer numbers and the goal is to find a sequence of elements of $A$ whose values (strictly) increase as their indices increase. For \textsf{LIS}, we denote the solution size for an array $A$ by $\lis(A)$. We also use $\lis^{[\alpha,\beta]}(A)$ to denote the size of the longest increasing subsequence subject to elements whose values lie in range $[\alpha,\beta]$. Longest increasing subsequence is equivalent to \textsf{LCS} when the inputs are two permutations of $n$ distinct characters.  

Finally, we define a notation to  denote the normalized solution sizes. For \textsf{LCS}, we denote the normalized solution size by $\normal{\lcs(A,B)} = \lcs(A,B)/\sqrt{|A||B|}$ for $A$ and $B$ and we use $\normal{\ED(A,B)} = \ED(A,B)/(2\sqrt{|A||B|})$ for edit distance. Note that, when the two strings have equal length we have $\normal{\ED(A,B)} + \normal{\lcs(A,B)} = 1$. Similarly, for longest increasing subsequence, we denote by $\normal{\lis(A)} = \lis(A)/|A|$ the normalized solution size. We usually refer to the size of the input array by $n$.

Throughout this paper, we call an algorithm $f(\lambda)$-approximation for \textsf{LCS} if it is able to distinguish the following two cases: i) $\normal{\lcs(A,B)} \geq \lambda$ or ii) $\normal{\lcs(A,B)} < \lambda f(\lambda)$. A similar definition carries over to \textsf{LIS}. Once an $f(\lambda)$-approximation algorithm is provided for either \textsf{LCS} or \textsf{LIS}, one can turn it into an algorithm that outputs a solution with size $f(\lambda)(1-\epsilon)\lambda n$ provided that the optimal solution has a size $\lambda n$. The algorithm is not aware of the value of $\lambda$ but will start with $\lambda_0 = 1$ and iteratively multiply $\lambda_0$ by $1-\epsilon$ until a solution is found. 
% !TeX root = ../../main.tex
\subsection{Techniques Overview}\label{sec:tech} %old : \label{sec:results}
Our algorithm for \textsf{LCS} is inspired by the recent developments for edit distance~\cite{beghs18, cdgks18}. We begin by briefly explaining the previous ideas for approximating edit distance and then we show how we use these techniques to obtain a solution for \textsf{LCS}. Finally, in Section \ref{sec:tech_lcs} we outline our algorithm for \textsf{LIS}.
%%% 
\subsubsection{Summary of Previous \textsf{ED} Techniques}\label{sec:tech_ed}
 Indeed, edit distance is hard only if the two strings are far ($\normal{\ED(A,B)} = \delta$ and $\delta = n^{-o(1)}$) otherwise the $O(n+(n\delta)^2)$ algorithm of Landau \textit{et al.} \cite{lms98} computes the solution in truly subquadratic time. 
The algorithm of Chakraborty \textit{et al.} \cite{cdgks18} for edit distance has three main steps that we briefly discuss in the following.%  

\paragraph{Step 0 (window-compatible solutions):} In the first step, they construct a set of {\em windows}, or (contiguous) substrings, $W_A$ for string $A$, and $W_B$ for string $B$. Let $k$ denote the total number of windows of $W_A \cup W_B$. For simplicity, let all the windows have the same size $d$ and $n \simeq O(k d)$\footnote{The equality holds if we assume $\delta = \Omega(1)$.}. The construction features two key properties: 1) provided that the edit distances of the windows between $W_A$ and $W_B$ are available, one can recover a $1+\epsilon$ approximation of edit distance in time $\tilde O(n+k^2)$ via dynamic programming. 2) $k^2 \times d^2 \simeq O(n^2)$. % 
That is, if we naively compute the edit distance of every pair of windows, the overall running time would still  asymptotically be the same as that of the classic algorithm.

In order to obtain a solution for edit distance, it suffices to know the distances between the windows. However, Chakraborty \textit{et al.} \cite{cdgks18} show that knowing the distances between some of the window pairs is enough to obtain an approximately optimal solution for edit distance. % 
Step 1 provides estimates for the distances of the windows which is approximately correct except for $O(k^{2-\Omega(1)})$ many pairs and Step 2 shows how this can be used to obtain a solution for edit distance.
Discretization simplifies the problem substantially. For a fixed $0 \leq \delta \leq 1$, they introduce a graph $\graph{}_{\delta}$ where the nodes correspond to the windows and an edge between window $w_i \in W_A$ and window $w_j \in W_B$ means that $\normal{ \ED(w_i,w_j) } \leq \delta$. If we are able to construct $\graph{}_{\delta}$ for  logarithmically different choices of $\delta$, we can as well estimate the distances within a $1+\epsilon$ factor for the windows. Therefore the problem boils down to constructing $\graph{}_{\delta}$ for a fixed given $\delta$ without computing the edit distance between all pairs of windows. % 
\paragraph{Step 1 (sparsification via triangle inequality):} This step is the heart of the algorithm. 
Suppose we  choose a high-degree vertex $v$ from $\graph{}_{\delta}$ and discover all its incident edges by computing its edit distance to the rest of the windows. Triangle inequality implies that every pair of windows in $N(v)$ has a distance bounded by $2\delta$. Therefore by losing a factor 2 in the approximation, one can put all these edges in $\graph{}_{\delta}$ and not compute the edit distances explicitly. Although this does save some running time, in order to make sure the running time is truly subquadratic, we need to make a similar argument for paths of length 3 and thereby lose a factor 3 in the approximation. This method sparsifies the graph and what remains is to discover the edges of a sparse graph. % 

\paragraph{Step 2 (discovering the edges of the sparse graph):} Step 1 uses triangle inequality and discovers many edges between the vertices of $\graph{}_{\delta}$. However, it may not discover all the edges completely. When in the remainder graph, the degrees are small (and hence the graph is sparse) triangle inequality does not offer an improvement and thus a different approach is required. % 
Roughly speaking, Chakraborty \textit{et al.} \cite{cdgks18} subsample the windows of $W_A$ into a smaller set $S$ and discover all pairs of windows $w_i \in S$ and $w_j \in W_B$ such that edge $(i,j)$ is not discovered in Step 1. Next, they compute the edit distance of each pair of windows $(w_i,w_j), w_i \in W_A, w_j \in W_B$ such that there exist two nearby windows $(w_a, w_b)$ satisfying $w_a \in S, w_b \in W_B$ and the edge between $w_a$ and $w_b$ was missed in Step 1. The key observation is that even though this procedure does not discover all the edges, the approximated distances lead to an approximate solution for edit distance.

\subsubsection{\textsf{LCS}}\label{sec:tech_lcs}

\begin{figure}[t!]
\centering
\includegraphics[width = 0.7\textwidth]{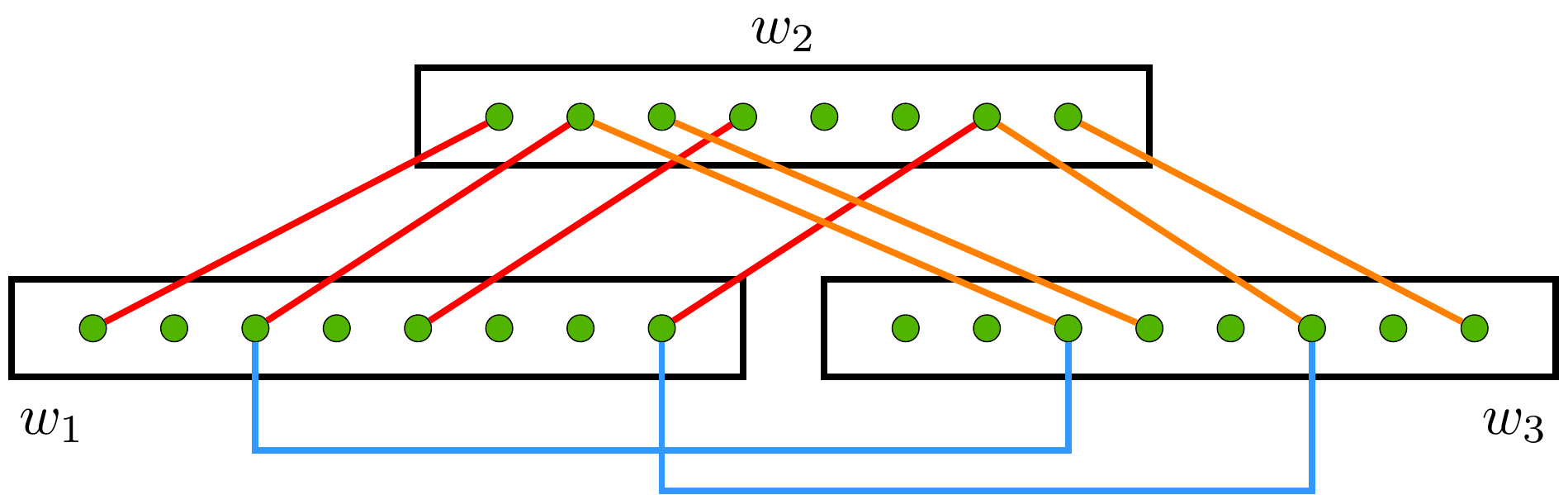}
\caption{Birthday paradox for triangle inequality: let $w_1, w_2, w_3$ be three windows of length $d = 8$ and assume $\lambda =1/2$. The \textsf{LCS} between $w_1$ and $w_2$ is $\lambda d = 4$ and the \textsf{LCS} between $w_2$ and $w_3$ is $\lambda d = 4$. Finally due to birthday paradox, we expect that the \textsf{LCS} between $w_1$ and $w_3$ is $\lambda^2 d = 2$. 
}\label{fig:birthday}
\end{figure}

Our algorithm for \textsf{LCS} mimics the same guideline. In addition to this, Steps 0 and 2 of our algorithm are \textsf{LCS} analogues of the ones used by Chakraborty \textit{et al.} \cite{cdgks18}. The main novelty of our algorithm is Step 1 which is a replacement for triangle inequality. Recall that unlike edit distance, triangle inequality does not hold for \textsf{LCS}.
\begin{challenge}
	How can we introduce a notion similar to triangle inequality to a non-metric setting such as \textsf{LCS}?
\end{challenge}
We introduce the notion of birthday triangle inequality to overcome the above difficulty. % 
Given windows $w_1$, $w_2$, and $w_3$ of size $d$ such that $\normal{\lcs(w_1,w_2)} \geq \lambda$ and $\normal{\lcs(w_2,w_3)} \geq \lambda$ hold, what can we say about the \textsf{LCS} of $w_1$ and $w_3$? In general, nothing! $\normal{\lcs(w_1,w_3)}$ could be as small as $0$. However, let us add some  randomness to the setting. Think of the \textsf{LCS} of $w_1$ and $w_2$  as a matching from the characters of $w_1$ to $w_2$ and similarly the \textsf{LCS} for $w_2$ and $w_3$ as another matching between characters of $w_2$ and $w_3$. Assume (for the sake of the thought experiment) that the characters of $w_2$ appear randomly in each matching. Since $\normal{\lcs(w_1,w_2)} \geq \lambda$,  each character of $w_2$ appears with probability at least $\lambda$ in the matching between $w_1$ and $w_2$. A similar argument implies that each character of $w_2$ appears with probability $\lambda $ in the matching of $w_2$ and $w_3$. Thus, (assuming independence), each character of $w_2$ appears in both matchings with probability $\lambda^2 $. This means that in expectation, there are $\lambda^2 d$ paths of length 2 between $w_1$ and $w_3$ which suggests $\normal{\lcs(w_1,w_3)} \geq \lambda^2$ as shown in Figure~\ref{fig:birthday}. This is basically birthday paradox used for the sake of triangle inequality.

Replacing triangle inequality by birthday triangle inequality is particularly challenging since birthday triangle inequality only holds on average. In contrast, triangle inequality holds for any tuple of arbitrary strings. Most of our technical discussions is dedicated to proving that we can algorithmically use birthday triangle inequality to obtain a solution for the worst-case scenarios. The most inconvenient step of our analysis is to show that our algorithm estimates the \textsf{LCS} of most window pairs in the sparsification phase. While this is straightforward for edit distance, birthday triangle inequality requires a deeper analysis of the underlying graph. In particular, we need to prove that if the undiscovered edges are too many, then birthday triangle inequality can be applied to certain neighborhoods of the graph.

There are two difficulties that we face here. On one hand, in order to apply birthday triangle inequality to a subgraph, we need to have enough structure for that subgraph to show the implication can be made. On the other hand, our assumptions cannot be too strong, otherwise such neighborhoods may not cover the edges of the graph. Therefore, the first challenge that we need to overcome is characterizing subgraphs in which birthday triangle inequality is guaranteed to be applicable. Our suggestion is the \textit{bi-cliques} structure. Although combinatorial techniques seem unlikely to prove this, we use the Blakley-Roy inequality to show that in a large enough bi-clique, we can use birthday triangle inequality to imply a bound on the \textsf{LCS} of certain pairs. The second challenge is to prove that if the underlying graph is dense enough, the graph contains many bi-cliques that cover almost all the edges that we plan to discover. This is again a challenging graph theoretic problem. We leverage extremal graph theory tools such as Turan's theorem for cliques and bi-cliques to obtain this bound.

Similar to edit distance, we construct a set $W = W_A \cup W_B$ of $k$ windows in Step 0 and aim to sparsify the edges of the $\lcs$-graph in Step 1. 
 Our construction ensures that $k d \simeq \Theta(n)$ and that knowing the \textsf{LCS} of the window pairs suffices to approximate the $\textsf{LCS}$ of the two strings. For a threshold $0 \leq \lambda \leq 1$, define a matrix $\ooutput{}: [k] \times [k] \rightarrow \{0,1\}$ to be a matrix which identifies whether $\normal{\lcs(w_i,w_j)} \geq \lambda$. In other words, $\ooutput{}[i][j] = 1 \Longleftrightarrow \normal{\lcs(w_i,w_j)} \geq \lambda$. For an $0 < \alpha \leq 1$, we call a matrix $\ooutput{}_{\alpha}$ an $\alpha$ approximation of $\ooutput{}$ if it meets the following two conditions:

\begin{align*}
\ooutput{}_{\alpha}[i][j] = 0 \Longrightarrow \| \lcs(w_i,w_j) \| < \lambda \hspace{1cm} \text{and}  \hspace{1cm}  \ooutput{}_{\alpha}[i][j] = 1 \Longrightarrow \| \lcs(w_i,w_j) \| \geq \alpha \cdot \lambda
\end{align*}

Notice that $\ooutput{}_{\alpha}$ gives more flexibility than $\ooutput{}$ for the cases that $\lambda \alpha \leq \lcs(w_i,w_j) < \lambda$. That is, both $0$ and $1$ are acceptable in these cases. Indeed an $\alpha$ approximation algorithm for the above problem is enough to obtain an $\alpha$ approximation algorithm for \textsf{LCS}. However, this is not necessary as Step 2 allows for incorrect approximation for up to $k^{2-\Omega(1)}$ many window pairs. Therefore, the problem of approximating \textsf{LCS} essentially boils down to approximating $\ooutput{}$ for a given basket of windows $W = W_a \cup W_b$ and a fixed $\lambda$ by allowing sufficiently small error in the output. A naive solution is to iterate over all pairs $w_i$ and $w_j$ and compute $\lcs(w_i,w_j)$ in time $O(d^2)$ and determine $\ooutput{}$ accordingly. However, this amounts to a total running time of $O(k^2 d^2)$ which is quadratic and not desirable. In order to save time, we need to compute the \textsf{LCS} of fewer than $k^2$ pairs of windows. To make this possible, we allow our algorithm to miss up to $O(k^{2-\Omega(1)})$ edges of the graph. Step 2 ensures that this does not hurt the approximation factor significantly.

We construct a graph from the windows wherein each vertex corresponds to a window and each edge identifies a pair with a large \textsf{LCS} (in terms of $\lambda$). Let us call this graph the $\lcs$-graph and denote it by $\graph{}_{\lambda}$. The goal is to detect the edges of the graph by allowing false-positive. 
As we discussed earlier, the hard instances of the problem are the cases where the $\lcs$-graph is dense for which we need a sparsifier. Roughly speaking, in our sparsification technique, our algorithm constructs another graph $\tgraph{}_{\lambda}$  such that $\tgraph{}_{\lambda}$ is valid in the sense that the edges of $\tgraph{}_{\lambda}$ correspond to pairs of windows with large enough \textsf{LCS}.
In addition to this, our algorithm guarantees that after the removal of the edges of $\tgraph{}_{\lambda}$ from $\graph{}_{\lambda}$ the remainder is sparse. In other words, $|E(\graph{}_{\lambda}) \setminus E(\tgraph{}_{\lambda})| = O(|V(\graph{}_{\lambda})|^{2-\Omega(1)})$. Of course, if the overall running time of the sparsification phase is truly subquadratic, the error of undiscovered edges can be addressed by the techniques of ~\cite{cdgks18} in Step 2. 
Below, we bring a formal definition for sparsification.
\begin{center}% 
	\noindent\framebox{\begin{minipage}{5.50in}
			\textsf{sparsification}\\
			\textsf{input}: Windows $w_1, w_2, \ldots, w_k$, parameters $\lambda$, and $\alpha$. \\
			\textsf{solution}: A matrix $\hat{\ooutput{}}_{\alpha} \in \{0,1\}^{k \times k}$ such that:
			{
				
			}
			\begin{itemize}
				\item $\hat{\ooutput{}}_{\alpha}[i][j] = 1 \Longrightarrow \normal{ \lcs(w_i,w_j) } \geq \alpha \cdot \lambda$
				\item $\left| \left\{ (i,j) ~|~ \normal{\lcs(w_i,w_j)} \geq \lambda \text{ and }\hat{\ooutput{}}_{\alpha} [i][j] = 0 \right\} \right| = k^{2 - \Omega(1)}$ % 
			\end{itemize}
	\end{minipage}}
\end{center}

We present two sparsification techniques for \textsf{LCS}. The first one (Section \ref{sec:sp1}), has an approximation factor of $( 1 - \epsilon ) \cdot \lambda^2$. In Section \ref{sec:sp2} we present another sparsification technique that has a worse approximation factor $\Omega(\lambda^3)$ but leaves fewer edges behind. Although the second sparsification technique has a worse approximation factor, it has the advantage that the number of edges that remain in the sparse graph is truly subquadratic regardless of the value of $\lambda$ and therefore it extends our solution to the case that $\lambda = o(1)$ (see Section \ref{sec:sp2} for a detailed discussion).%\\[0.05cm]

Let us note one last algorithmic challenge to keep in mind before we begin to describe our sparsification techniques. 
For edit distance, if window pairs $(w_1,w_2)$ and $(w_2,w_3)$ are close, we are {\em guaranteed} that $w_1$ and $w_3$ are also close; for longest common subsequence, we will argue that $(w_1,w_3)$ are {\em likely} to be have a long \textsf{LCS} (for a ``random'' choice of $(w_1,w_3)$). Nonetheless, in order to add $(w_1,w_3)$ as an edge to our graph we  have to {\em verify} that their \textsf{LCS} is indeed long. If we were to verify an edge naively, we would need as much time as computing the \textsf{LCS} between $(w_1,w_3)$ from scratch!

\paragraph{Sparsification 1, $(1-\epsilon) \lambda^2$-approximation}

\begin{figure}
\centering
\includegraphics[width=0.7\textwidth]{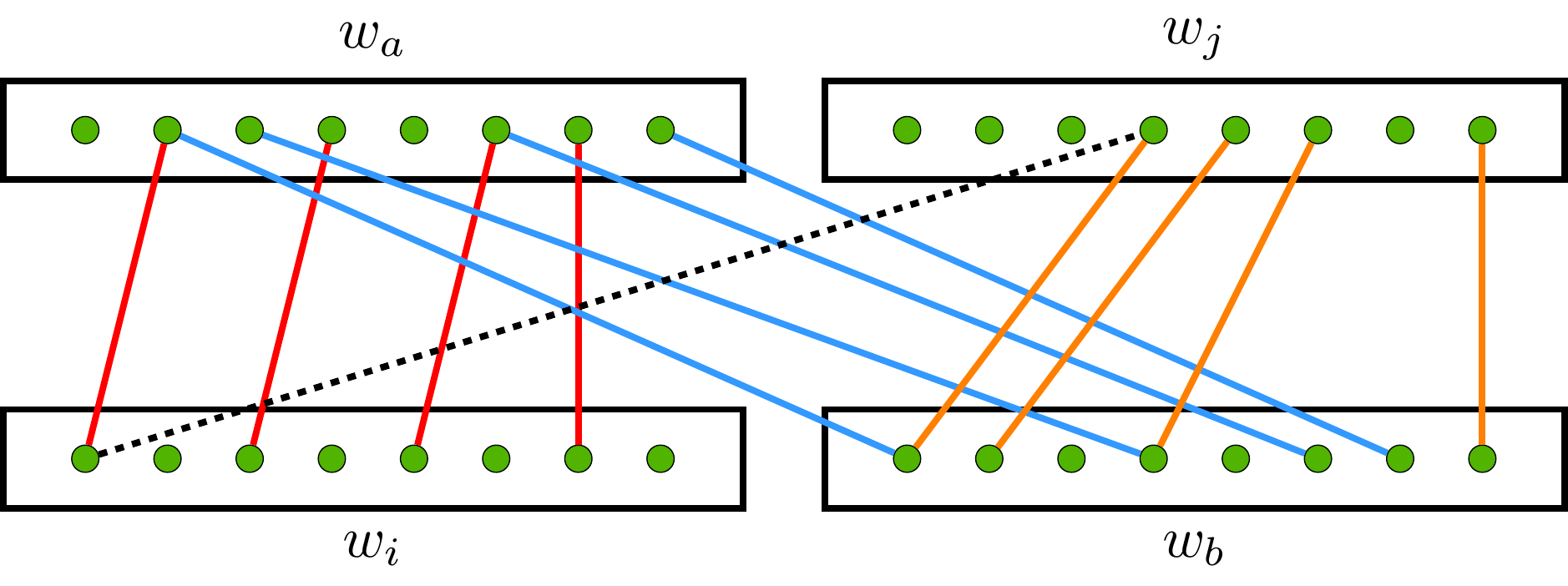}
\caption{Let $w_i, w_a, w_b$ and $w_j$ denote four windows and each of them has length $d=8$. This figure shows how the intersection of the edges of three windows are taken in order to construct a solution for the \textsf{LCS} of $w_i$ and $w_j$. If the size of the intersection is large, then such a tuple is called constructive. The solid lines represent \textsf{LCS} between two strings, and the dashed line represents the intersection of the three \textsf{LCS}s.}\label{fig:sparsification_1}
\end{figure}

Similar to edit distance, applying the birthday variant of triangle inequality to paths of length 2 for \textsf{LCS} does not improve the running time significantly. Therefore, we need to use birthday triangle inequality for paths of length $3$. 
To this end, we define the notion of constructive tuples as follows: a tuple  $\langle w_i, w_a, w_b, w_j\rangle$ is an $(\epsilon,\lambda)$-constructive tuple, if we have $\normal{\lcs(w_i,w_a)} \geq \lambda$, $\normal{\lcs(w_i,w_j)} \geq \lambda$, $\normal{\lcs(w_b,w_j)} \geq \lambda$ and by taking the intersection of the three \textsf{LCS} matchings, we are able to imply $\normal{\lcs(w_a,w_b)} \geq (1-\epsilon) \lambda^3$ (see Figure~\ref{fig:sparsification_1} for an example). Taking the intersection of the matchings can be done in linear time which is faster than computing the \textsf{LCS}.

Our sparsification technique here is simple but the analysis is very intricate. We subsample a set $S$ of windows and compute the \textsf{LCS} of every window in $S$ and all other windows. We set $|S| = k^{\gamma} \log k$, where $\gamma \in (0,1)$. % 
At this point, for some pairs, we already know their \textsf{LCS}. However, if neither $w_i$ nor $w_j$ is in $S$, 
we do not know if $\normal{\lcs(w_i,w_j)} \geq \lambda$ or not. Therefore, for such pairs, we try to find windows $w_a,w_b \in S$ such that $\langle w_i, w_a, w_b, w_j\rangle$ is constructive. If such a constructive tuple is found for a pair of windows, then we conclude that their normalized \textsf{LCS} is at least $(1-\epsilon) \lambda^3$.

All that remains is to argue that this method discovers almost all the edges of the $\lcs$-graph $\graph{}_\lambda$ and the number of undiscovered edges is $k^{2-\Omega(1)}$. % 
This is the most difficult part of the analysis. We note that even proving the existence of \textbf{only one} constructive tuple is already non-trivial \textbf{even when $\graph{}_\lambda$ is complete}. However, our goal is to show that almost all the edges are discovered via constructive tuples when $\graph{}_\lambda$ is dense (and of course not necessarily complete).

Define a pair of windows $w_i$ and $w_j$ to be \textit{well-connected}, if there are at least $k^{2-\gamma}$ different $(w_a,w_b)$ pairs such that $\langle w_i,w_a,w_b,w_j \rangle$ is $(\epsilon,\lambda)$-constructive. Since each window appears in $S$ with probability $k^{\gamma-1} \log k$, for each well-connected pair we find one constructive tuple via our algorithm with high probablity. Therefore, we need to prove that the total number of pairs $(w_i,w_j)$ such that $(w_i,w_j)$ is not well-connected but $\normal{\lcs(w_i,w_j)} \geq \lambda$ is subquadratic. Let us put these edges in a new graph $\gdelta$ whose vertices are all the windows. 
 
	We first leverage the Blakley-Roy inequality and a double counting technique to prove that if $\gdelta$ has a large complete bipartite subgraph, then there is one constructive tuple which includes only the vertices of this subgraph (Lemma \ref{lemma:clique}). 
	Next, we apply the Turan's theorem to show that if $\gdelta$ is dense, then it has a lot of large complete bipartite subgraphs. 
	 Finally, we use a probabilistic method  to conclude that $\gdelta$ cannot be too dense otherwise there are a lot of constructive tuples in the graph which implies that at least one edge $(w_i,w_j)$ in $\gdelta $ is well-connected. This is not possible since all the well-connected pairs are detected in our sparsification algorithm with high probability.

The above argument proves that if we sparsify our graph using our sparsification algorithm, the remainder graph would have a subquadratic number of edges. Therefore after plugging Step 2 into the algorithm, the running time remains subquadratic. However, since Turan theorem gives us a weak bound, the running time of the algorithm using this sparsification is $O(n^{2-\Omega(\lambda)})$ and is only truly subquadratic when $1/\lambda$ is constant.

\paragraph{Sparsification 2, $\Omega(\lambda^3)$-approximation} In Section \ref{sec:sp1}, we present another sparsification method that although gives us a slightly worse approximation factor $\Omega(\lambda^3)$ it always leaves a truly subquadratic number of edges behind and therefore the running time of the algorithm would be truly subquadratic regardless of the parameter $\lambda$. This sparsification is based on a novel data structure. We assume for simplicity in the following that all the windows are of the same length even though this does not hold in general and having windows of different length adds more complication to the algorithm. We discuss the details in Section~\ref{sec:sp1}.

Let $\opt_{i,a}$ denote the longest common subsequence of $w_i$ and $w_a$ (with some fixed tie-breaking rule, e.g.~lexicographically first). Define $\lcs_{w_a}(w_i,w_j)$ to be the size of the longest common subsequence between $\opt_{i,a}$ and $w_j$. Notice that this definition is no longer symmetric. Let $\normal{\lcs_{w_a}(w_i,w_j)}$ denote the relative value, i.e., $\normal{\lcs_{w_a}(w_i,w_j)} = \lcs_{w_a}(w_i,w_j) / \sqrt{|w_i| \cdot |w_j|} $. %  
 The first ingredient of the algorithm is a data-structure, namely $\lcsdata$. %  
 After a preprocess of time $O(|w_a| \sum_{i \in S} |w_i|)$, $\lcsdata$ is able to answer queries of the following type in time $O(|w_i| + |w_j|)$:
\begin{itemize}
\item ``for a $0 \leq \wt{\lambda} \leq 1$ either certify that $ \| \lcs_{w_a}(w_i,w_j) \| \geq \Omega( \wt{\lambda}^2)$ or report that $$\| \lcs_{w_a}(w_i,w_j) \| < O(\wt{\lambda})"$$.
\end{itemize}

In our sparsification, we repeat the following procedure $k^{\gamma}$ times, where $\gamma \in (0,1)$. We sample a window $w_a$ uniformly at random and construct $\lcsdata(w_a,S)$ for $S = \{w_i | i \neq a \text{ and } |w_i| \geq |w_a|\}$. After the preprocessing step, we make a query for every pair of windows $(w_i,w_j)$ such that $w_i,w_j \in S$ and determine if $\lcs_{w_a}(w_i, w_j)$ is at least $\Omega(\lambda^4)$ or upper bounded by $O(\lambda^2)$ (here $\wt{\lambda} = \lambda^2/2$). If their \textsf{LCS} is at least $\Omega(\lambda^4)$ we report this pair as an edge in our $\lcs$-graph. Finally, we use the Turan theorem to prove that the number of remaining edges in our graph is small. % 

	 To be more precise, we first construct a graph $\gdelta$ that reflects the edges that are not detected via our sparsification.
	 If $\gdelta$ is dense enough, then there is one vertex $v$ in $\gdelta$ with a large enough degree. We use the neighbors of $v$ to construct another graph $\fdelta$ with vertex set $N(v)$. An edge exists in $\fdelta$ if $\max\{ \|\lcs_{w_v(w_i,w_j)} \|, \|\lcs_{w_v(w_j,w_i)} \| \} \geq \Omega(\lambda^2)$.
	 We prove that $\fdelta$ has no large independent set. In other words, if we select a large enough set of vertices in $\fdelta$, then there is at least one edges between them.
	 Next, we apply the Turan theorem to prove that $\fdelta$ is dense.
	 Finally, we imply that since $\fdelta$ is dense, there is one vertex $u$ in the neighbors of $v$ such that there are a lot of 2-paths between $v$ and $u$. This implies that the edge $(u,v)$ should have been detected in our sparsification and therefore must not exist in $\gdelta$. This contradiction  implies that $\gdelta$ is sparse in the first place.

\subsubsection{\textsf{LIS}}\label{sec:tech_lis} %% 
In this section, we present our result for longest increasing subsequence. More precisely, we show that when the solution size is lower bounded by $n \lambda$  ($\lambda \in [0,1]$), one can approximate the solution within a factor $\Omega(\lambda^3)$ in time $\wt{O}(\sqrt{n} / \lambda^7)$. This married with a simple sampling algorithm for the cases that $\lambda < n^{-\Omega(1)}$, provides an $\Omega(\lambda^3)$-approximate algorithm with running time of $\tilde O(n^{0.85})$ (without further dependence on $\lambda$). We further extend this result to reduce the running time to $\wt{O}(n^{\kappa} \poly(1/\lambda))$ for any $\kappa > 0$ by imposing a multiplicative factor of $\poly(1/\lambda)$\footnote{The exponent of $1/\lambda$ depends exponentially on $1/\kappa$.} to the approximation.

Our algorithm heavily relies on sampling random elements of the array for which longest increasing subsequence is desired. Denote the input sequence by $A = \langle a_1, a_2, \ldots, a_n\rangle$. A naive approach to approximate the solution is to randomly subsample the elements of $A$ to obtain a smaller array $B$ and then compute the longest increasing subsequence of $B$ to estimate the solution size for $A$. Let us first show why this approach alone fails to provide a decent approximation factor. First, consider an array $A = \langle 1, 2, \dots, n\rangle$ which is strictly increasing. Based on $A$, we construct two inputs $A'$ and $A''$ in the following way:
\begin{itemize}
	\item $A'$ is exactly equal to $A$ except that a $p$ fraction of the elements in $A'$ are replaced by $0$.
	\item $A''$ is exactly equal to $A$ except that every block of length $\sqrt{n}$ is reversed in $A''$. In other words, $A'' = \langle \sqrt{n}, \sqrt{n}-1, \sqrt{n}-2, \ldots, 1, 2\sqrt{n}, 2\sqrt{n}-1, \ldots, \sqrt{n}+1, \ldots, n, n-1, n-2, \ldots, n-\sqrt{n}+1\rangle$.
\end{itemize}

We subsample the two arrays $A'$ and $A''$ with a rate of $1/\sqrt{n}$ to obtain two smaller arrays $B'$ and $B''$ of size roughly $O(\sqrt{n})$. It is easy to prove that $ \lis(B')  = \Omega(\sqrt{n})$ and $ \lis(B'')  = \Omega(\sqrt{n})$, yet $\lis(A') = \Omega(n)$ but $\lis(A'') = O(\sqrt{n})$. By setting $p = 1/e$\footnote{$e \simeq 2.7182$.} we can also make sure that $\lis(B')$ and $\lis(B'')$ are within a small multiplicative range even though the gap between $\lis(A')$ and $\lis(A'')$ is substantial. 

The above observation shows that the problem is very elusive when random sampling is involved. We bring a remedy to this issue in the following. Divide the input array into $\sqrt{n}$ subarrays of size $\sqrt{n}$. We denote the subarrays by $\sa_1, \sa_2, \ldots, \sa_{\sqrt{n}}$ and fix an optimal solution $\opt$ for the longest increasing subsequence of $A$. Define $\ssmall(\sa_i)$ to be the smallest number in $\sa_i$ that contributes to $\opt$\ and $\llarge(\sa_i)$ to be the largest number in $\sa_i$ that contributes to $\opt$. Moreover, define $\lis^{[\ell,r]}$ to be the longest increasing subsequence of an array subject to the elements whose values lie within the interval $[\ell,r]$. This immediately implies 
\begin{align*}
\lis(A) = \sum_{i=1}^{\sqrt{n}} {\lis^{[\ssmall(\sa_i),\llarge(\sa_i)]}(\sa_i)}.
\end{align*}
Another observation that we make here is that since we assume $\normal{\lis(A)} \geq \lambda$ and the size of each subarray is bounded by $\sqrt{n}$, then we have
\begin{align*}
\frac{\lis(A)}{ \max_i {\lis^{[\ssmall(\sa_i),\llarge(\sa_i)]}(\sa_i)}} \geq \sqrt{n} \lambda
\end{align*}
which  means that in order to approximate $\lis(A)$ it  suffices to compute 
$ % 
\lis^{[\ssmall(\sa_i),\llarge(\sa_i)]}(\sa_i)
$ % 
for $\wt{O}(1/\lambda)$ many randomly sampled subarrays. This is quite helpful since this shows that we only need to sample $\wt{O}(1/\lambda)$ many subarrays and solve the problem for them. However, we do not know the values of $\ssmall(\sa_i)$ and $\llarge(\sa_i)$ in advance. Therefore, the main challenge is to predict the values of $\ssmall(\sa_i)$ and $\llarge(\sa_i)$ before we sample the subarrays.

Indeed, one needs to read the entire array to correctly compute $\ssmall(\sa_i)$ and $\llarge(\sa_i)$ for each of the subarrays. However, we devise a  method to approximately guess these values without losing too much in the size of the solution. Roughly speaking, we show that if we sample $k = O(1 / ( \lambda \epsilon ) )$ different elements  from a subarray $\sa_i$ for some constant $\epsilon $ and denote them by $a_{j_1}, a_{j_2}, \ldots, a_{j_k}$,  then for at least one pair $(\alpha,\beta)$, $[a_{j_\alpha}, a_{j_\beta}]$ is approximately close to $[\ssmall(\sa_i), \llarge(\sa_i)]$ up to a $(1-\epsilon)$ factor.

\begin{figure}[t!]
\begin{center}
\includegraphics[width=0.9\textwidth]{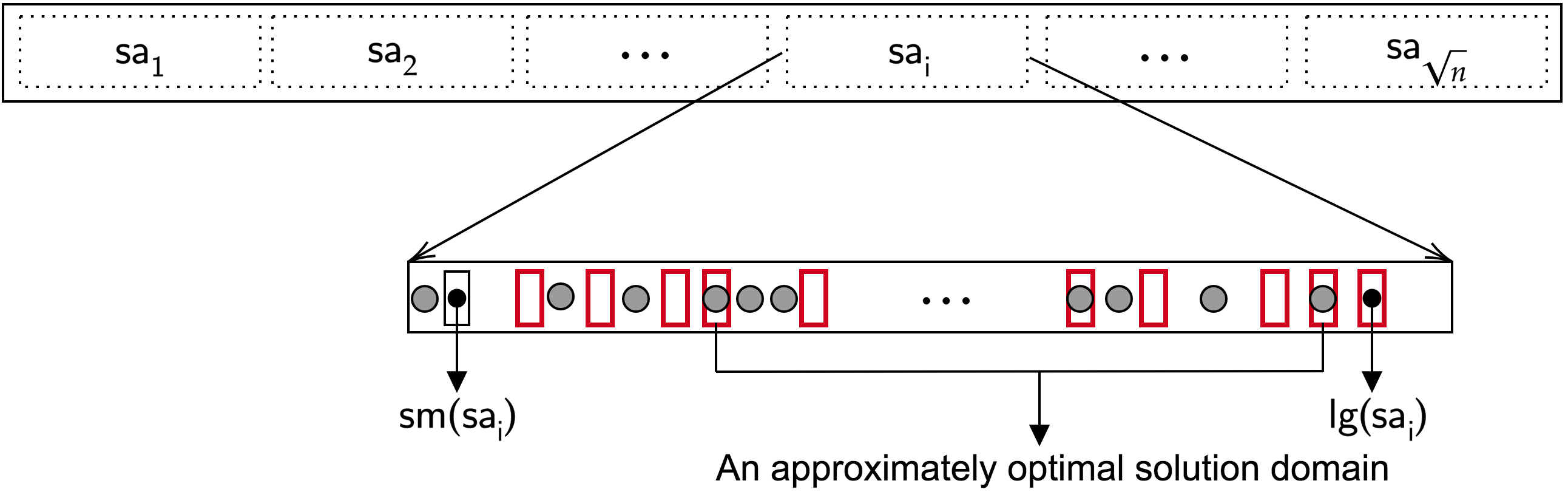}
\end{center}
\caption{Red rectangles show the elements of $\sa_i$ that contribute to $\lis(A)$ and gray circles show the elements of $\sa$ that are sampled via our algorithm.}\label{fig:fig1}
\end{figure}

The above argument provides  $O(( 1 / ( \lambda \epsilon ) )^2)$ candidate domain intervals for each $\sa_i$.  This does not provide a solution since we do not know which candidate domain interval approximates $[\ssmall(\sa_i), \llarge(\sa_i)]$ for each $\sa_i$. Of course, if we were to randomly choose one candidate interval for every subarray, we would make a correct guess for at least $O(\sqrt{n} ( \lambda \epsilon)^2)$ subarrays which provides an approximation guarantee of $\Omega(\lambda^2)$ for our algorithm. However, our assignments have to be monotone too. More precisely, let $[\widetilde{\ssmall}(\sa_i), \widetilde{\llarge}(\sa_i)]$ be the guesses that our algorithm makes, then we should have 

$$\widetilde{\ssmall}(\sa_1) \leq \widetilde{\llarge}(\sa_1) \leq \widetilde{\ssmall}(\sa_2) \leq \widetilde{\llarge}(\sa_2) \leq \ldots \leq \widetilde{\ssmall}(\sa_{\sqrt{n}}) \leq \widetilde{\llarge}(\sa_{\sqrt{n}}).$$

Random sampling does not guarantee that the sampled intervals are monotone. To address this issue, we introduce the notion of \textit{pseudo-solutions}. A pseudo-solution is an assignment of monotone intervals to subarrays in order to approximate $\ssmall(\sa_i)$ and $\llarge(\sa_i)$. The \textit{quality} of a pseudo-solution with intervals $[\ell_1, r_1], [\ell_2, r_2], \ldots, [\ell_{\sqrt{n}}, r_{\sqrt{n}}]$ is equal to $\sum_i \lis^{[\ell_i, r_i]}(\sa_i)$. For a fixed pseudo-solution, this can be easily approximated via random sampling. Thus, our goal is to construct a pseudo-solution whose quality is at least an $\Omega(\lambda^3)$ approximation of the size of the optimal solution. To this end, we present a greedy method in Section \ref{sec:lis_pseudo_solution} to construct the desired pseudo-solution.

Finally, in Section \ref{sec:lis_extension}, we show how the above ideas can be generalized to improve the running time down to $\wt{O}(n^{\kappa} \poly(1 / \lambda ) )$ for any arbitrarily small $\kappa > 0$ by imposing a factor $\poly(1 / \lambda )$\footnote{The exponent of $1/\lambda$ exponentially depends on $1/\kappa$.} to the approximation guarantee.  

\newpage
% !TeX root = ../../main.tex

\section{Organization of the Paper}\label{sec:step3}
Our algorithm for \textsf{LCS} is explained in Section~\ref{sec:window} (Step 0), Section~\ref{sec:step1} (Step 1), and Section~\ref{sec:step2} (Step 2). % 

In both our results for \textsf{LCS} and \textsf{LIS}, we assume that the goal is to find approximate solutions, provided that the solution size is at least $\lambda_0 n$. After the algorithms terminate, if the output is smaller than what we expect, we realize that the solution is smaller than $\lambda_0 n$. Therefore, we begin by setting $\lambda_0 = 1$ and iteratively multiply $\lambda_0$ by a $1-\epsilon$ factor until we obtain a solution. This only adds a multiplicative factor of $\log 1/\lambda$ to the running time and a multiplicative factor of $1-\epsilon$ to the approximation. Since we present two different sparsification techniques, we obtain two theorems: one is Theorem~\ref{thm:formal_lcs} and the other is Theorem~\ref{thm:formal_lcs_quadratic}.

\begin{theorem}\label{thm:formal_lcs}
Given strings $A,B$ of length $|A|=|B|=n$ with $\normal{\LCS(A,B)} = \lambda$, we can approximate the length of the \textsf{LCS} between the two strings within a factor $\Omega(\lambda^3)$ in time $\tilde O(n^{39/20})$.
\end{theorem}
\begin{proof}
Fix a sufficiently small constant $\epsilon$ (e.g.~$\epsilon = 1/10000$). Since we do not know the value of $\lambda$, we start with $\lambda = 1$ and iteratively try to solve the problem within a factor $\Omega(\lambda^3)$. Each time we are not able to find a solution, we multiply the value of $\lambda$ by $1-\epsilon$ and proceed. This imposes a $1-\epsilon$ factor to the approximation and a logarithmic factor to the runtime that can be hidden in the $\tilde O$ and $\Omega$ notations. Thus, in what follows, we assume that we fix a $\lambda$ and we know that the solution size is $\lambda n$. Also, we define $\kappa = n^{-1/140}$.

If $\lambda \leq \kappa$ we run the following algorithm: we choose $n{\lambda^3}$ characters of $A$ uniformly at random to obtain a string $A'$. With high probability, the $\textsf{LCS}$ of $A'$ and $B$ is at least $(1-\epsilon) \lambda^4 n$. Therefore, we set our aim to find such a solution. To this end, we spend a preprocessing time of $O(n \log n)$ on string $B$ and then we will be able to find such a solution in time $O(|A'| \lambda^4 n \log n)$ (see Theorem~\ref{theorem:smalllcs}). Thus, the overall runtime would be bounded by \begin{gather*}\tilde O(|A'| \lambda^4 n) \leq \tilde O(n \lambda^7 n) \leq \tilde O(n^{2-7/140}) = \tilde O(n^{39/20}).\end{gather*}
Moreover, the approximation factor is also $\Omega(\lambda^3)$ as desired. 

When $\lambda \geq \kappa$ we run the three steps of our algorithm (Step 0 stated in Fact~\ref{fact:parameters-cubic}, Step 1 stated in Theorem~\ref{thm:sp2}, and Step 2 stated in Lemma~\ref{lemma:focs}) by setting $d = \sqrt{n}\lambda$, $\gamma = 2/3$, $\wmax = \sqrt{n}$ and $k = \tilde O(\sqrt{n}/\lambda) \leq \tilde O(n^{71/140})$. After constructing the windows in Step 0 (Fact~\ref{fact:parameters-cubic}), we run the algorithm of Theorem~\ref{thm:sp2} (Step 1) for every $\lambda' \in \{\epsilon \lambda, \epsilon (1+\epsilon) \lambda, \epsilon (1+\epsilon)^2 \lambda, \ldots, 1\}$. If for a pair of windows $w_i,w_j$ our algorithm in Step 1 detects an edge at $\lambda'$ then we update the solution size for such a pair to $\max\{|w_i|,|w_j|\} \lambda'^4/16$. We then run the algorithm of Step 2 (Lemma~\ref{lemma:focs}) to find a solution. In what follows, we bound the approximation factor and the runtime of the algorithm.

\textbf{Approximation factor:} For now, we only consider the multiplicative and additive approximation losses that are incurred in Steps 0, 1, and 2 but we assume that for each $\lambda$, the output of Step 1 is without any errors. We then incorporate those errors to bound the overall approximation factor. These errors are listed below:
\begin{itemize}
	\item We lose a  multiplicative constant factor in Step 0 followed by an additive error of $\epsilon \lambda n$ (see Lemma~\ref{lem:improved_structrual_lemma}).
	\item We lose a multiplicative factor $1-2\epsilon$ in Step 1 due to the fact that we ignore window pairs whose \textsf{LCS} sizes drops below a threshold $\epsilon \lambda$ times the maximum window size.
	\item We also lose a multiplicative factor $1-\epsilon$ in Step 3 (Lemma~\ref{lemma:focs}).
\end{itemize}  

Since $\epsilon$ is very small and all the multiplicative loss factors are constant, we can assume that all the above errors amount to an overall $1/c$ for a \textbf{constant} factor $c$. Now we are ready to discuss the  loss in the approximation incurred in Step 1. If Step 1 did not incur any error, we would find a solution of size $1/c \lambda n$ for the two strings. Suppose that $(w_1, w'_1), (w_2, w'_2), \ldots$ is a such a window-compatible solution that provides a solution of total size $1/c \lambda n$ if the estimations were correct. We put these pairs in two sets based on whether the $A$ side window is larger or the $B$ side window is larger. At least one set gives us a solution of size $1/(2c) \lambda n$. Without loss of generality, we assume that these are the pairs whose $B$ side is at least as large as their $A$ side and in the rest of the proof we restrict ourselves to such pairs.

For each pair $(w_i, w'_i)$, define $\lambda_i$ to be the ratio of the actual \textsf{LCS} of $w_i$ and $w'_i$ divided by $|w'_i|$. Since $\sum |w'_i| \leq n$ and $\sum \lambda_i |w'_i| \geq 1/(2c) \lambda n$ then it follows from Lemma~\ref{lemma:math} that

\begin{gather*}\sum |w'_i| \lambda^4_i \geq (\frac{1/(2c) n \lambda}{\sum |w'_i|})^4 \sum |w'_i| = (1/(2c) \lambda)^4 n (n/\sum |w'_i|)^3 \geq (1/(2c) \lambda)^4 n,\end{gather*}
and therefore $\sum |w'_i| \lambda^4_i/16 \geq \frac{\lambda^4 n}{256 c^4} = \Omega(\lambda^4 n)$. Thus, via the estimations we find in Step 1, we would be able to find a solution that loses a factor of at most $\Omega(\lambda^3)$.

\textbf{Runtime:} The runtime of Step 0 is $\tilde O(k) = \tilde O(\sqrt{n}/\lambda) = \tilde O(n^{71/140})$ which is negligible. By Theorem~\ref{thm:sp2}, the runtime of Step 1 is equal to 
\begin{gather*}\tilde O(k^{1+\gamma} \wmax^2 + k^{2+\gamma} \wmax) \leq \tilde O(n^{1+71/140(5/3)} + n^{1/2 + 71/140 (8/3)}) \leq \tilde O(n^{1.86}).\end{gather*} 
Moreover, the number of remaining edges in the graph is bounded by (see Theorem~\ref{thm:sp2}) \begin{gather*}\tilde O(k^{2-\gamma} / \lambda) \leq  \tilde O(k^{2-2/3+1/70}) \simeq \tilde O(k^{1.35}).\end{gather*} This implies a runtime of (see Theorem~\ref{lemma:focs1}) \begin{gather*}\tilde O(k^{2-(2-1.35)/2}\wmax^2/\lambda^6) = \tilde O(k^{1.675}\wmax^2/\lambda^6) \leq \tilde O(n^{1+6/140+71/140(1.675)}) \leq \tilde O(n^{1.9})\end{gather*} for Step 2.
\end{proof}

\begin{theorem}\label{thm:formal_lcs_quadratic}
Given strings $A,B$ of length $|A|=|B|=n$ with $\normal{\LCS(A,B)} = \lambda$ where $\lambda$ is constant, we can approximate the length of the \textsf{LCS} between the two strings within a factor $(1-\epsilon)\lambda^2$ in $n^{2-\Omega_{\epsilon \lambda}(1)}$ time for any constant $0 < \epsilon < 1$.
\end{theorem}
\begin{proof}
Let $\epsilon' = \epsilon / 200$.
Since we do not know the value of $\lambda$, we start with $\lambda = 1$ and iteratively try to solve the problem within a factor $(1-\epsilon')\lambda^2$. Each time we are not able to find a solution, we multiply the value of $\lambda$ by $1-\epsilon'$ and proceed. This imposes a $1 - \epsilon'$ factor to the approximation and a constant factor to the runtime. Thus, in what follows, we assume that we fix a $\lambda$ and we know that the solution size is at least $\lambda n$. 

We run the three steps of our algorithm (Step 0 stated in Fact~\ref{fact:parameters-quadratic}, Step 1 stated in Theorem~\ref{thm:sp1}, and Step 2 stated in Lemma~\ref{lemma:focs}) by setting $d = \sqrt{n}$, $\wmax = \sqrt{n}$ and $k = \tilde O(\sqrt{n})$ (Notice that here $\lambda$ is constant). After constructing the windows in Step 0 (Fact~\ref{fact:parameters-quadratic}), we run the algorithm of Theorem~\ref{thm:sp1} (Step 1) for every $\lambda' \in \{ \epsilon \lambda, \epsilon (1+\epsilon') \lambda, \epsilon (1+\epsilon')^2 \lambda, \ldots, 1\}$. If for a pair of windows $w_i,w_j$ our algorithm in Step 1 detects an edge at $\lambda'$ then we update the solution size for such a pair to $\sqrt{|w_i||w_j|} (1-\epsilon')\lambda'^3$. We then run the algorithm of Step 2 (Lemma~\ref{lemma:focs}) to find a solution. In what follows, we bound the approximation factor and the runtime of the algorithm.

\textbf{Approximation factor:} For now, we only consider the multiplicative and additive approximation losses that are incurred in Steps 0, 1, and 2 but we assume that for each $\lambda$, the output of Step 1 is without any errors. We then incorporate those errors to bound the overall approximation factor. These errors are listed below:
\begin{itemize}
	\item We lose an additive error of $8 \epsilon' \lambda n$ in Step 0 (see Lemma~\ref{lem:improved_structrual_lemma_constant}).
	\item We lose a multiplicative factor $1-\epsilon'$ in Step 1 due to the fact that we ignore window pairs whose normalized \textsf{LCS} sizes drops below a threshold $\epsilon'\lambda$.
	\item We also lose a multiplicative factor $1-\epsilon'$ in Step 3 (Lemma~\ref{lemma:focs}).
\end{itemize}  
Since $\epsilon'  = \epsilon / 200 < 1/ 200$, we can assume that all the above errors amount to an overall $1 - 20\epsilon'$ factor. Now we are ready to discuss the  loss in the approximation incurred in Step 1. If Step 1 did not incur any error, we would find a solution of size $(1-20 \epsilon')\lambda n$ for the two strings. Suppose that $(w_1, w'_1), (w_2, w'_2), \ldots$ is a such a window-compatible solution that provides a solution of total size $(1-20\epsilon') \lambda n$ if the estimations were correct. % 
For each pair $(w_i, w'_i)$, define $\lambda_i $ to be $ \normal{ \lcs(w_i,w_i') } $. Since $\sum |w_i| \leq n$, $\sum |w'_i| \leq n$ and $\sum \lambda_i \sqrt{|w_i| \cdot |w'_i|} \geq (1-20\epsilon') \lambda n$, it follows from Lemma~\ref{lemma:math}  and Cauchy–Schwarz inequality that

\begin{align*} \sum  \sqrt{|w_i| \cdot |w'_i|}  (1 - \epsilon') \lambda_i^3 
& \geq (1 - \epsilon') \sum \sqrt{|w_i| \cdot |w'_i|}  \left(\frac{\sum \lambda_i \sqrt{|w_i| \cdot |w'_i|}}{\sum \sqrt{|w_i| \cdot |w'_i|}} \right)^3 \\  
& \geq (1 - \epsilon')\frac{\left((1-20\epsilon')\lambda n \right)^3}{\left(\sum \sqrt{|w_i| \cdot |w'_i|} \right)^2 } \\
& \geq (1 - \epsilon')\frac{\left((1-20\epsilon')\lambda n \right)^3}{n^2} \\
& \geq (1 - \epsilon')(1-20\epsilon')^3 \lambda^3 n \\ 
& \geq (1 - \epsilon')(1-60\epsilon') \lambda^3 n.\end{align*}
Thus, via the estimations we find in Step 1, we would be able to find a solution that loses a factor of at most $(1-\epsilon')(1-60\epsilon') \lambda^3 \geq (1-\epsilon)\lambda^3$.

\textbf{Runtime:}  The runtime of Step 0 is $\tilde O(k) = \tilde O(\sqrt{n}) $ which is negligible. By Theorem~\ref{thm:sp1} and setting $\gamma = 1/10$, the runtime of Step 1 is equal to 
\begin{gather*}\tilde O(\wmax^2 k^{1.1}  + \wmax k^{2.2} ) = \tilde O(n^{1.6}).\end{gather*} 
Moreover, the number of remaining edges in the graph is bounded by (see Theorem~\ref{thm:sp1}) 
\begin{gather*}\tilde O(k^{2- \epsilon \lambda^3 / ( 800 \wlayers \wgap )} ) = \tilde O(k^{2 - \Omega(\epsilon \lambda^6 / \log (1/\lambda))}) = \tilde O(k^{2 - \Omega(\epsilon \lambda^7)})\end{gather*} This implies a runtime of (see Theorem~\ref{lemma:focs2}) 
\begin{gather*}\tilde O(k^{2-\Omega(\epsilon \lambda^7)}\wmax^2/\lambda^6) = \tilde O(n^{2 - \Omega(\epsilon \lambda^7 / 2)}) =  n^{2 - \Omega_{\lambda,\epsilon}(1)}\end{gather*} for Step 2.
\end{proof}

As an immediate corollary of Theorem \ref{thm:formal_lcs}, we present an algorithm that beats the $1/|\Sigma|$ approximation factor in truly subquadratic time, when the strings are balanced.
\begin{corollary}\label{cor:formal_lcs_sigma}
Given a pair of strings $(A,B)$ of length $n$ over alphabet $\Sigma$ that satisfy the balance condition, we can approximate their \textsf{LCS} within an $O(|\Sigma|^{3/4})$ factor in time $O(n^{39/20})$.
\end{corollary}
\begin{proof}
Since $A$ and $B$ are balanced, there is a character $\sigma \in \Sigma$ that appears at least $n/|\Sigma|$ times in both strings. Indeed, finding a solution of size $n/|\Sigma|$ by restricting our attention to only character $\sigma$ can be done in time $O(n)$. If $\lcs(A,B) \leq n/|\Sigma|^{1/4}$ this already gives us an $O(|\Sigma|^{3/4})$ approximate solution. Otherwise, $\normal{\lcs(A,B)} > 1/ |\Sigma|^{1/4}$ and the approximation factor of our $\Omega(\lambda^3)$-approximation algorithm would be bounded by $O(|\Sigma|^{3/4})$.
\end{proof}

Finally, we bring our results for \textsf{LIS} in Section \ref{sec:lis}. 
We show that 
\begin{theorem}
Given a length-$n$ sequence $A$ with $  \lis(A)  = n \lambda $. %  
We can approximate the length of the \textsf{LIS}  within a factor of $\Omega(\lambda^3)$ in time $
\wt  O(n^{17/20} ) $.
\end{theorem}
\begin{proof}
If $\lambda < n^{-1/20}$ we sample the array with a rate of $n^{-3/20}$ and compute the \textsf{LIS} for the sampled array. 
The running time of the algorithm is $\wt O(n^{17/20})$. The approximation factor is $O(n^{-3/20}) \geq \Omega(\lambda^3)$.
Otherwise, by Theorem \ref{theorem:lis}, 
we estimate the size of \textsf{LIS} up to an $\Omega(\lambda^3)$ approximation factor in time $\wt \Omega(\lambda^{-7} \sqrt{n} ) \leq \wt O(n^{17/20})$. 
\end{proof}

 %%% 
% !TeX root = ../../main.tex

\newpage
\section{\textsf{LCS} Step 1: Sparsification via Birthday Triangle Inequality}\label{sec:step1}
Recall that we are given two sets of windows $W_A$ and $W_B$ for the strings and our goal is to approximate the \textsf{LCS} of all but a few pairs of windows from $W_A \times W_B$. For simplicity, we put all the windows in the same basket $W = W_A \cup W_B$ and denote the windows by $w_1, w_2, \ldots, w_k$ where $k$ is the total number of windows. Since the windows have different lengths, we define $\wmax = \max_{i \in [k]} |w_i|$ to be the maximum length of the windows. Similarly, we also define $\wmin = \min_{i\in [k]} |w_i|$ to be the minimum length of the windows. Let $\wgap = \wmax / \wmin$.
Let $\wlayers$ denote the number of different window sizes. 
Notations $\wgap$ and $\wlayers$ will be used in the later analysis.

In order to approximate the \textsf{LCS}'s we fix a $\lambda \in \{\epsilon \lambda_0, (1+\epsilon) \epsilon \lambda_0, (1+\epsilon)^2 \epsilon \lambda_0, \ldots, 1\}$ and sparsify graph $\graph{}_\lambda$. In Section~\ref{sec:sp1}, we present a sparsification algorithm (Algorithm~\ref{alg:sp1}) which provides $(1-\epsilon)\lambda^2$-approximation when $\lambda$ is constant. The formal guarantee of the algorithm is provided in Theorem~\ref{thm:sp1}. In Section~\ref{sec:sp2}, we present a sparsification which provides $\Omega(\lambda^3)$-approximation for any (potentially sub-constant) $\lambda$.

\newpage %%

% !TeX root = ../../main.tex

\subsection{Sparsification for constant $\lambda$ (Step 1 of Theorem~\ref{thm:formal_lcs_quadratic})}\label{sec:sp1}
Fix an arbitrary \textsf{LCS} for every pair of windows and refer to that as $\opt_{i,j}$ for two windows $w_i$ and $w_j$. Note that we do not explicitly compute $\opt_{i,j}$ in our algorithm. Let us for simplicity, think of each $\opt_{i,j}$ as a matching between the characters of the two windows. Also, denote by $\big(\opt_{i,a} \cap \opt_{a,b} \cap \opt_{b,j}\big)$ a solution which is constructed for windows $w_i$ and $w_j$ by taking pairs of characters $(x,y)$ such that:
\begin{equation*}
\begin{split}
(x,y) \in \big(\opt_{i,a} \cap \opt_{a,b} \cap \opt_{b,j}\big)   \Longleftrightarrow &\exists x', y' \text{ such that }\\
& (x,y') \in \opt_{i,a} \text{ and }\\
& (y',x') \in \opt_{a,b} \text{ and }\\
& (x',y) \in \opt_{b,j}.
\end{split}
\end{equation*}

Let $\left\|\big(\opt_{i,a} \cap \opt_{a,b} \cap \opt_{b,j}\big)\right\| = \frac{\left|\big(\opt_{i,a} \cap \opt_{a,b} \cap \opt_{b,j}\big)\right|}{\sqrt{|w_i| |w_j|}}$.

Our analysis is based on a notion which roughly reflects ``in how many ways a desirable solution can be made for a pair of windows $(w_i,w_j)$ by taking the intersection of the \textsf{LCS} for other pairs". Below, we provide a definition for this notion.
\begin{definition}[$(\epsilon,\lambda)$-constructive]\label{def:epsilon_delta_constructive}
 Let $0 < \epsilon, \lambda < 1$ be fixed values. We call a tuple $\langle w_i, w_a, w_b, w_j \rangle$ ($w_i \neq w_a \neq w_b \neq w_j$) an $(\epsilon, \lambda)$-\textit{constructive} tuple, if $$\left\|\big(\opt_{i,a} \cap \opt_{a,b} \cap \opt_{b,j}\big)\right\| \geq (1-\epsilon) \lambda^3.$$
 \end{definition}
The advantage of a constructive tuple is that if $\opt_{i,a}, \opt_{a,b},$ and $\opt_{b,j}$ are provided, one can construct a desirable solution for $\opt_{i,j}$ in linear time by taking the intersection of the given matchings. % 

We parametrize our algorithm by a value $0 < \gamma < 1$ to be set later. One may optimize the runtime of the algorithm by setting the value of $\gamma$ in terms of the number of windows and the length of the windows. We first sample a set $S$ of $O(k^{\gamma} \log k)$ windows. Next, we compute $\opt_{i,j}$ of every window $w_i \in S$ and every other window $w_j$ (not necessarily in $S$). Finally, we find all tuples $\langle w_i, w_a, w_b, w_j \rangle$ such that $w_a,w_b \in S$ and they satisfy the following property:
$$\left\|\big(\opt_{i,a} \cap \opt_{a,b} \cap \opt_{b,j}\big)\right\| \geq (1-\epsilon) \lambda^3.$$
Recall that we call such tuples $(\epsilon,\lambda)$-\textit{constructive} and update $\hat{{\sf O}}_{\lambda^2}[i][j], \hat{{\sf O}}_{\lambda^2}[j][i] \leftarrow 1$ accordingly. This is shown in Algorithm \ref{alg:sp1}.

\begin{algorithm}[!t]
\begin{algorithmic}[1]\caption{Sparsification for constant $\lambda$ (Step 1 of Theorem~\ref{thm:formal_lcs_quadratic})}\label{alg:sp1}
	\Procedure{\textsc{QuadraticSparsification}}{$w_1, w_2, \ldots, w_k, \lambda, \epsilon$} \Comment{Theorem~\ref{thm:sp1}}
	\State $S \leftarrow 40 k^{\gamma } \log k $ i.i.d. samples of $[k]$
	\State $\hat{{\sf O}}_{\lambda^2} \leftarrow \{ 0 \}^{k \times k} $ 
	\For{$w_i \in S$} \Comment{Takes $k |S|  \wmax^2$ time}
		\For{$j \leftarrow 1$ to $k$}
			\State $\opt_{i,j}, \opt_{j,i} \leftarrow \lcs(w_i,w_j)$
		\EndFor
	\EndFor
	\For{$w_i \in S$} \Comment{Takes $k |S| \wmax$ time}
		\For{$j \leftarrow 1$ to $k$}
			\If{$||\opt_{i,j}|| \geq \lambda$}
				\State $\hat{{\sf O}}_{\lambda^2} [i][j] \leftarrow 1$
				\State $\hat{{\sf O}}_{\lambda^2} [j][i] \leftarrow 1$
			\EndIf
		\EndFor
	\EndFor
	\For{$i \leftarrow 1$ to $k$} \Comment{Takes $k^2 |S|^2 \wmax$ time}
	\For{$j \leftarrow 1$ to $k$}
		\For{$w_a \in S$}
		\For{$w_b \in S$}
			\If{$\left\|\big(\opt_{i,a} \cap \opt_{a,b} \cap \opt_{b,j}\big)\right\| \geq ( 1 - \epsilon ) \lambda^3$} %\Comment{$\lambda^3$ or $(1-\epsilon) \lambda^3$?}
				\State $\hat{{\sf O}}_{\lambda^2} [i][j] \leftarrow 1$
				\State $\hat{{\sf O}}_{\lambda^2} [j][i] \leftarrow 1$
			\EndIf
		\EndFor
		\EndFor
	\EndFor
	\EndFor
	\State \Return $\hat{{\sf O}}_{\lambda^2}$
	\EndProcedure
\end{algorithmic}
\end{algorithm}

 The running time of our algorithm is equal to $O(k|S| \wmax^2 + k^2|S|^2 \wmax )$. The rest of this section is dedicated to proving that what remains in the $\lcs$-graph is sparse; this is formalized in Lemma~\ref{lemma:edges_upper_bound}.

\begin{definition}[$\gdelta$]\label{def:graph}
 Define a graph $\gdelta$ with $k$ vertices and the following edges:
\begin{align*}
E(\gdelta) = \left\{ (i,j) ~\bigg|~  \normal{ \lcs(w_i,w_j) } \geq \lambda \mathrm{~~~and~~~} \hat{{\sf O}}_{\lambda^2}[i][j] = 0 \right\}.
\end{align*}
\end{definition}
In other words, $\gdelta$ contains all of the edges that are not detected in our algorithm. We extend the notion of constructive tuples to the undetected edges in our algorithm. We call such tuples \textit{undetected-constructive}.
\begin{definition}[undetected-constructive tuple]
	We say a tuple $\langle w_i, w_a, w_b, w_j \rangle$ is \\$(\epsilon,\lambda)$-undetected-constructive if it is $(\epsilon,\lambda)$-constructive and also 
	\begin{align*}
	(i,a), (a,b), (b,j), (i,j) \in E(\gdelta).
	\end{align*}
\end{definition}

\begin{lemma}\label{lemma:edges_upper_bound}
With probability at least $1-2/k^3$, $\gdelta$ has at most $k^{2-\gamma \epsilon \lambda^3 / ( 80 \wlayers \wgap )}$ edges.
\end{lemma}

Instead of arguing directly about $\gdelta$, our proof uses another graph $\fdelta$, which is sub-sampled from $\gdelta$:
\begin{definition}[$\fdelta$]\label{def:graph-f}
Let $\fdelta$ be the graph constructed in the following way: for each node $v \in \gdelta$ we keep $v$ in $\fdelta$ with probability $$p := k^{-1 + \gamma /4}/4.$$ 
For each $u,v \in \fdelta$, if  $(u,v) \in \gdelta$ then we also draw an edge $u,v$ in $\fdelta$. We identify between the vertices and edges of $\fdelta$ and the corresponding vertices and edges of $\gdelta$, i.e.~$V(\fdelta) \subseteq V( \gdelta )$ and $E ( \fdelta ) \subseteq E( \gdelta )$. 
\end{definition}

The proof has two main ingredients. 
The first part of the proof uses the details of our algorithm to show that, w.h.p.~over the algorithm's randomness, $\gdelta$ contains few undetected-constructive tuples (Lemma~\ref{lem:well_connected_probability}).
Whenever this is the case, w.h.p.~over the sub-sampling procedure, $\fdelta$ does not contain any undetected-constructive tuples (Claim~\ref{cla:fgraph_has_no_undetected-constructive_tuple}).

The second part of the proof assumes by contradiction that $\gdelta$ is dense. It uses this assumption to conclude that $\fdelta$ is also dense (w.h.p.~over the sub-sampling procedure, see Claim~\ref{cla:fgraph_is_dense}), and therefore by Turan's Theorem contains a large bi-clique (Claim~\ref{claim:exists-biclique}). Finally, we use the large bi-clique to show that $\fdelta$ does contain an undetected-constructive tuple (Lemma~\ref{lemma:clique}) - a contradiction!

\subsubsection{Proof of Lemma~\ref{lemma:edges_upper_bound}, part 1: $\fdelta$ does not contain an undetected-constructive tuple} \hfill

We first introduce the notion of ``well-connected pairs".
\begin{definition}[well-connected pair]\label{def:well_connected}
We say that a pair of windows $(w_i, w_j)$ is \textit{well-connected} if there are at least $k^{2 - \gamma}$  pairs of windows $(w_a,w_b)$ such that $\langle w_i, w_a, w_b, w_j \rangle$ is $(\epsilon,\lambda)$-constructive.
\end{definition}
We argue that well-connected pairs are detected in our algorithm with high probability.

\begin{lemma}\label{lem:well_connected_probability}
Let $\hat{{\sf O}}_{\lambda^2} \in \{0,1\}^{k \times k}$ denote the output of Algorithm~\ref{alg:sp1}. 
With  probability at least $1-2/k^3$, 
for all $(i,j) \in [k] \times [k]$ such that $(w_i,w_j)$ is a well-connected pair (Definition~\ref{def:well_connected}),  we have $\hat{{\sf O}}_{\lambda^2} [i][j] = 1$.
\end{lemma}
\begin{proof}
%\Zhao{Chernoff bound} %\color{red}\textsf{TODO: Chernoff bound!}\color{black}
We consider a fixed $(i,j)$ such that pair $(w_i,w_j)$ is well-connected. Let
\begin{align*}
Q_{i,j} = \left\{ (a,b) ~\bigg|~ \left\| \opt_{i,a} \cap \opt_{a,b} \cap \opt_{b,j} \right\| \geq (1-\epsilon) \lambda^3 \right\}.
\end{align*}
Conceptually, we divide the process of sampling $S$ into two phases: we  sample $20k^\gamma \log k$ windows in the first phase, and then we sample  $20k^\gamma \log k$ more windows in the second phase. % 

For each $a \in [k]$, let $Q_{i, j, a} = \{b: (a, b) \in Q_{i, j}\}$. 
Since $\sum_{a \in [k]} |Q_{i,j,a}| = |Q_{i,j}|  \geq k^{2-\gamma}$, there are at least 
$\frac{k^{1-\gamma}}{2}$
different number $a$'s in $[k]$ such that $|Q_{i, j, a}| \geq \frac{k^{1-\gamma} }{ 2}$.
Hence, in the first phase, there is a sampled number $q$ such that $|Q_{i, j, q}| \geq k^{1-\gamma} / 2$
with probability at least \[1 - \left(1 - \frac{k^{1-\gamma} / 2}{k}\right)^{20 k^{\gamma} \log k} > \frac{k^5-1}{k^5}. \]

We fix such a $q$. 
In the second phase, there is a sampled number $r$ such that $r \in Q_{i, j, q}$ with probability at least 
\[1 - \left(1 - \frac{k^{1-\gamma} / 2}{k}\right)^{20 k^{\gamma} \log k} > \frac{k^5-1}{k^5}. \]

Since the number of $(i, j)$ pairs is at most $k^2$, the lemma is obtained by a union bound on all the  well-connected pairs $(w_i, w_j)$. %$\hat{{\sf O}}_{\lambda^2} [i][j] = 1$.
\end{proof}

We complete the first part of the proof by showing that with probability at least $0.99$, $\fdelta$ does not have an undetected-constructive tuple.
\begin{claim}[$\fdelta$ has no undetected-constructive tuple]\label{cla:fgraph_has_no_undetected-constructive_tuple}
	With probability at least $0.99$, there is no undetected-constructive tuple in $\fdelta$.
\end{claim}
\begin{proof}
	By Lemma~\ref{lem:well_connected_probability}, we assume that no edge $(i,j)$  in $\gdelta$ is well-connected. In other words, for all pairs of windows $(w_i,w_j)$ that are connected in $\gdelta$, 
	there are at most $k^{2-\gamma}$ pairs of windows $(w_a,w_b)$ such that $\langle w_i , w_a, w_b , w_j \rangle$ is $(\epsilon,\lambda)$-constructive. 
	
	Recall that for a tuple $\langle w_i , w_a, w_b , w_j \rangle$ to be undetected-constructive, edge $(i, j)$ should belong to $\gdelta$. For a fixed $(i,a,b,j)$, the probability that we keep it in $\fdelta$ is $p^4$. Therefore, the expected number of undetected-constructive tuples in $\fdelta$ is bounded by
	\begin{align*}
		\mathbb{E}[ \# \text{undetected-constructive~tuple} ] \leq k^2 k^{2-\gamma} \cdot p^4 = 1/256
	\end{align*}
	where the last step follows from $p = k^{-1 + \gamma/4}/4$.
	
	By Markov's inequality, we have 
	\begin{align*}
		\Pr[ \# \text{undetected-constructive~tuple} \geq 1/2 ] \leq \frac{ \mathbb{E}[ \# \text{undetected-constructive~tuple} ]  }{1/2} < 1/100.
	\end{align*}
	Thus, with probability $0.99$, there is no undetected-constructive tuple in $\fdelta$.
\end{proof}

\subsubsection{Proof of Lemma~\ref{lemma:edges_upper_bound}, part 2: $\fdelta$ contains an undetected-constructive tuple}\hfill

Now we assume by contradiction that the $\gdelta$ is dense.
To simplify the notation, we introduce another parameter,
$$\beta :=\gamma \epsilon \lambda^3 / ( 80 \wlayers \wgap ).$$
Note that $\beta < \gamma - \Omega(1)$.

We are able to show that if $\gdelta$ is dense, so is $\fdelta$.
\begin{claim}[$\fdelta$ is a dense graph]\label{cla:fgraph_is_dense}
	If for some $0 < \beta< \gamma/4-\Omega(1)$, $\gdelta$ contains is at least $ k^{2-\beta}$ edges  then with probability at least $0.98$ we have
	\begin{align*}
		|E (\fdelta) | \geq  \frac{| V( \fdelta ) |^{2-4\beta/\gamma}}{128}.
	\end{align*}
\end{claim}
The proof of Claim~\ref{cla:fgraph_is_dense} uses the following elementary graph theory fact:
\begin{fact}\label{fact:deg^2-EV}
	In every graph $G = (V,E)$, we have
	$$ \sum_v \deg(v)^2 \le 2|V||E|.$$
\end{fact}
\begin{proof}
	\begin{gather*} \sum_v \deg(v)^2 \le |V| \sum_v \deg(v) \le 2|V||E|.
	\end{gather*}
\end{proof}

\begin{proof}[Proof of Claim~\ref{cla:fgraph_is_dense}]
	Based on the sampling rate, we know that the following holds in expectation:
	\begin{gather}\label{eq:V(NF)}	
		\mathbb{E}[ | V ( \fdelta ) | ] = p | V ( \gdelta ) | = k^{\gamma/4}/4, 	
	\end{gather}	
	and	
	\begin{align*}	
		\mathbb{E}[ | E ( \fdelta ) | ] & \geq p^2 | E ( \gdelta ) | 	
		 \geq k^{\gamma/2 - \beta}/16. %
	\end{align*}

	Using standard Chernoff bound, we have with probability $0.99$,
	\begin{align*}
		|V ( \fdelta ) | \leq 2 \mathbb{E}[ | V ( \fdelta ) | ] = k^{\gamma/4}/2.
	\end{align*}
	In the rest of this proof, we show that with probability $0.99$,
	\begin{align*}
		|E ( \fdelta ) | \geq k^{\gamma/2 - \beta}/128,
	\end{align*}
	and then the claim follows.

	It is known that any graph can be made bipartite by removing at most half of its edges~\cite{west2001introduction}. Thus, for the sake of this proof, we only consider a bipartite subgraph of $\gdelta$ with parts $\mathcal{P}$ and $\mathcal{Q}$ that contains at least $k^{2-\beta}/2$ edges. We refer to this graph by $\hat{\gdelta}$. 
	We consider a two-step construction of $\hat \fdelta$ based on sampling in $\hat \gdelta$, where we first only sub-sample the vertices on the $\mathcal{P}$-side, and then also sub-sample the vertices on the $\mathcal{Q}$-side. Let $\hat \fdelta(\mathcal{P})$ denote the graph after subsampling only the $\mathcal{P}$-side, and $V(\hat \fdelta(\mathcal{P}), \mathcal{P})$ denote the set of vertices in $\hat \fdelta$ on the $\mathcal{P}$-side.

	The number of edges in $\hat \fdelta(\mathcal{P})$ is the sum of degrees of surviving $\mathcal{P}$-vertices, aka a sum of i.i.d.~random variables bounded in $[0,k]$. 
	It's expectation is given by: 
	\begin{gather*}
	\mathbb{E} [| E ( \hat \fdelta(\mathcal{P}) ) | ]= p | E ( \hat \gdelta(\mathcal{P}) ) | \ge k^{2-\beta}/2,
	\end{gather*}
	and by Hoeffding's inequality this sum concentrates around its expectation with high probability:
	\begin{align*}
		\Pr\left[| E ( \hat \fdelta(\mathcal{P}) ) | \le  p k^{2-\beta} / 4\right] \leq & \Pr\left[| E ( \hat \fdelta(\mathcal{P}) ) | \le \mathbb{E}[| E (\hat \fdelta(\mathcal{P}) ) |] -  p k^{2-\beta} / 4\right]  \\
		\leq & \exp\left(- \frac{2|\mathcal{P}|^2 p^2 k^{4 - 2\beta} / 16}{|\mathcal{P}| k^2}\right) \\
		\leq & \exp\left(-\Theta\left(k^{\gamma / 2 - 2\beta}\right)\right). &&\text{By $|P| \ge 1$}
	\end{align*}

		By Chernoff bound, 
	\begin{align*}
		\Pr[|V(\hat \fdelta(\mathcal{P}), \mathcal{P})| > 2pk] \leq [|V(\hat \fdelta(\mathcal{P}), \mathcal{P})| > 2p|\mathcal{P}|] \leq \exp\left(-p|\mathcal{P}| / 3\right) \leq \exp(-\Theta(k^{\gamma / 4 - \beta})),
	\end{align*}
	where the last inequality is obtained by $|\mathcal{P}| \leq k$.
	We henceforth fix any realization of $\hat \fdelta(\mathcal{P})$ conditioned on \[| E ( \hat \fdelta(\mathcal{P}) ) | \geq  p k^{2-\beta} / 4 \text{ and } |V(\hat \fdelta(\mathcal{P}), \mathcal{P})| \leq 2pk.\] 
	
We now consider the second step of the sub-sampling, which transforms $ \hat \fdelta(\mathcal{P})$ to $\hat \fdelta$. 
The expected number of edges in $\hat \fdelta$ satisfies:
\begin{gather*}
\mathbb{E}[| E ( \hat \fdelta ) |] = p| E ( \hat \fdelta(\mathcal{P}) ) | \ge p^2 k^{2-\beta} / 4 = k^{\gamma / 2 - \beta} / 64.
\end{gather*}
We now argue about concentration. 
Notice that the degree of each vertex in $\hat \fdelta(\mathcal{Q})$ is at most  the number of vertices on $\mathcal{P}$-side in $\hat \fdelta(\mathcal{P})$ is at most $ |V(\hat \fdelta(\mathcal{P}), \mathcal{P})| \leq 2pk$. Therefore, by Hoeffding's inequality,% 
	\begin{align*}
 \Pr\left[| E ( \hat \fdelta ) | < k^{\gamma / 2 - \beta} / 128 \right] 
		\le & \Pr \left[| E ( \hat \fdelta ) | <\mathbb{E}[| E ( \hat \fdelta ) |]  - k^{\gamma / 2 - \beta} / 128   \right] \\
		\le & \exp\left(-\Theta\left(\frac{|\mathcal{Q}|^2 k^{\gamma - 2\beta}}{|\mathcal{Q}|p^2 k^2}\right)\right) \\
		\le & \exp\left(- \Theta\left(k^{\gamma / 2 - 2\beta}\right)\right) && \text{By $|Q| \ge 1$}.
 	\end{align*}

By taking a union bound on the low probability events and the fact that $E(\hat \fdelta)$ is a subset of $E( \fdelta)$, we have 	$|E ( \fdelta ) | \geq k^{\gamma/2 - \beta}/128$.
\end{proof}

We now use the fact that $\fdelta$ is dense to argue that it contains a large bi-clique.
\begin{claim}[$\fdelta$ contains a large bi-clique]\label{claim:exists-biclique}
With probability at least $0.98$ $\fdelta$, contains a complete bipartite subgraph $K_{\gamma/(5\beta),\gamma/(5\beta)}$. 
\end{claim}
\begin{proof}
Using Tur\'an's Theorem (Lemma~\ref{lem:turan}), we know, for any integer $s \geq 2$, that a graph $G$ with $n$ vertices and $n^{2-1/s}$ edges has at least one $K_{s,s}$ subgraph. We apply this to 
graph $\fdelta$. Since $\lambda$ is constant then $\beta/\gamma$ becomes constant and thus $|V(\fdelta)|^{\beta/\gamma}$ is super constant. It follows from Claim~\ref{cla:fgraph_is_dense} that \[|E(\fdelta)| \geq \frac{|V(\fdelta)|^{ 2 - 4 \beta / \gamma }}{128} > |V(\fdelta)|^{ 2 - 5 \beta / \gamma }\] holds with probability at least $0.98$. This in turn implies that with probability at least $0.98$ $\fdelta$ contains a complete bipartite subgraph $K_{\gamma/(5\beta),\gamma/(5\beta)}$. 
\end{proof}

The last step in the proof is to use the large bi-clique to exhibit an undetected-constructive tuple.
\begin{lemma}\label{lemma:clique}
Let $X$ and $Y$ be two sets of windows such that for every $w_i \in X$ and $w_j \in Y$ there is an  edge $(i,j)$ in $E(\gdelta)$. If $|X| \geq 16 \wlayers \wgap/(\epsilon \lambda^3) $ and $|Y| \geq 16 \wlayers \wgap/(\epsilon \lambda^3) $, 
then there exist $w_i, w_a, w_b, w_j \in X \cup Y$ such that $\langle w_i, w_a, w_b, w_j\rangle$ is $(\epsilon,\lambda)$-constructive, i.e.,
	\begin{align*}
		\left\| \opt_{i,a} \cap \opt_{a,b} \cap \opt_{b,j} \right\| \geq (1-\epsilon) \lambda^3
	\end{align*}
and either $w_i, w_b \in X$ and $w_a, w_j \in Y$ or $w_i, w_b \in Y$ and $w_a, w_j \in X$.
\end{lemma}
\begin{proof}
Let $\epsilon' = \epsilon / 4$. By assumption, we know that $|X|, |Y| \geq 4 \wlayers \wgap / (\epsilon' \lambda^3)$. Moreo\-ver the total number of different window sizes is bounded by $\wlayers$. Thus, there exist two sets $\hat{X} \subseteq X$ and $\hat{Y} \subseteq Y$ such that $|\hat{X}|,|\hat{Y}| \geq 4 \wgap / (\epsilon' \lambda^3)$ and the windows within $\hat{X}$ are of the same size and the windows within $\hat{Y}$ are of the same size (though the windows of $\hat{X}$ and $\hat{Y}$ may have different sizes).
Let $d_x$ denote the window size for each window in $\hat{X}$, and $d_y$ denote the window size for each window in $\hat{Y}$. 

We select $X' \subseteq \hat{X}$ and $Y'\subseteq \hat{Y}$ such that 
\begin{enumerate}
\item $|X'| \geq 4 / (\epsilon' \lambda^3)$.
\item $|Y'| \geq 4 / (\epsilon' \lambda^3)$.
\item $|X'| d_x \leq (1+\epsilon') |Y'| d_y $.
\item $|Y'| d_y \leq (1+\epsilon') |X'| d_x $.
\end{enumerate}
To do this, if $|\hat{X}|d_x >  (1+ \epsilon' ) |\hat{Y}| d_y$, then we set $Y' = \hat{Y}$ and select $X'$ as an arbitrary subset of $\hat{X}$ with size $\left\lceil\frac{|Y'|d_y}{d_x} \right\rceil$.
Otherwise, we set $X' = \hat{X}$ and select $Y'$ an arbitrary subset of $\hat{Y}$ with size $\left\lceil\frac{|X'|d_x}{d_y} \right\rceil$.

\begin{figure}[t!]
\begin{center}
\includegraphics[width=0.95\textwidth]{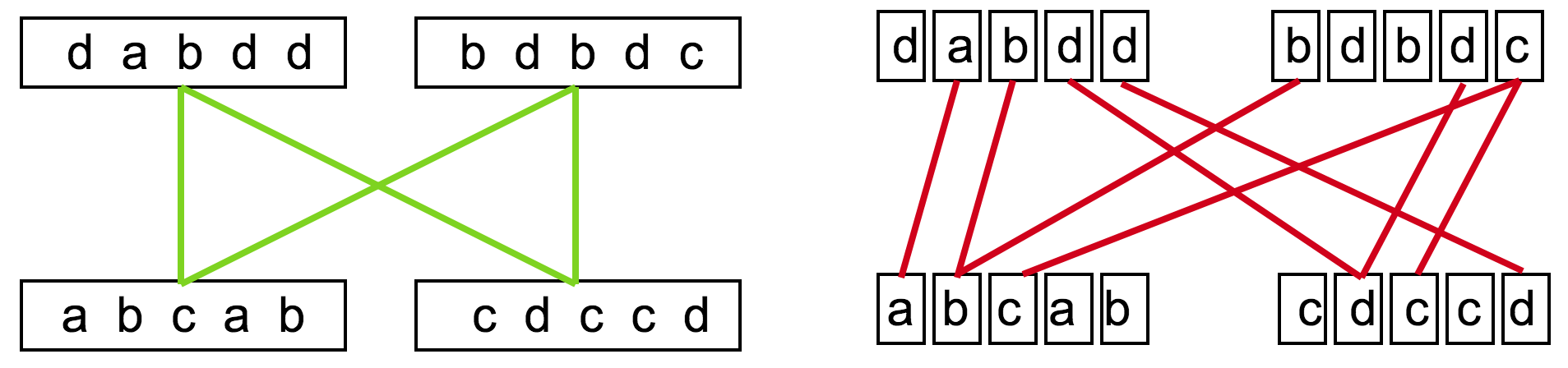}
\end{center}
\caption{The graph on the left is an example of the string-based graph and the graph on the right is an example of the character-based graph.}\label{fig:fig1}
\end{figure}

We define a character-based (bipartite) graph $G_C = (V_C, E_C)$ as follows: 
Each window of $X'$ has $d_x$ nodes in the character-based graph such that each node represents a character of the window. 
Similarly, each window of $Y'$ has $d_y$ nodes in the character-based graph. 
Two nodes $x, y$ in the character-based graph are adjacent if and only if 
$(x, y) \in \opt_{i, j}$ where  $w_i$ is the window containing character $x$ and $w_j$ is the window containing character $y$.
Let $l_x = |X'|$ and $l_y = |Y'|$.

The total number of nodes in the character-based graph is $|V_C| = l_x d_x + l_y d_y$. 
By the last two desiderata of definition of $X',Y'$ that, each side of the character based-graph has approximately the same number of nodes:
\begin{gather}l_xd_x ,  l_y d_y \approx_{\text{$(1\pm \eps')$-factor}} |V_C|/2. \label{eq:lxdx=lydy}
\end{gather}
Similarly, we also have that:
\begin{gather}  \max\{l_x, l_y\}\min\{d_x, d_y\}  \approx_{\text{$(1\pm \eps')$-factor}} |V_C|/2. \label{eq:lxdx=lydy2}
\end{gather}

The total number of edges in the character-based graph satisfies
\begin{gather} |E_C| \leq l_xl_y \min\{d_x, d_y\}, \label{eq:E_c-ub}
\end{gather}
and, by \eqref{eq:lxdx=lydy},
\begin{gather}
|E_C|  \geq l_xl_y\lambda\sqrt{d_xd_y}  
 \approx_{\text{$\sqrt{1\pm \eps'}$-factor}} \sqrt{l_x l_y} \lambda |V_C|/2.
\end{gather}
By Blakley-Roy inequality (Lemma~\ref{lem:blakley_roy_inequality}), 
the number of walks of length $3$ in the character-based graph is at least
\begin{align*}
\#\text{3-walks} & \geq  |V_C| \cdot  \left( 2 \frac{ |E_C|}{|V_C| } \right)^3 \\
&\approx_{\text{$(1\pm \eps')^{1.5}$-factor}}    \lambda^3 (l_x l_y)^{1.5} |V_C|.
\end{align*}

We are interested in the number of 3-walks that are not degenerate. Thus, we need to exclude such degenerate walks from the total count. The number of such walks is upper bounded by
\begin{align*}
\#\text{degenerate 3-walks} & \leq (\text{max degree of $G_C$}) \cdot 4 |E_C| \\
& \leq \max\{l_x, l_y\} \cdot 4 (l_x l_y \min\{d_x, d_y\}) && \text{(Eq.~\eqref{eq:E_c-ub})}\\
& \approx_{\text{$(1\pm \epsilon')$-factor}}  2 l_xl_y \cdot |V_C|&& \text{(Eq.~\eqref{eq:lxdx=lydy2})}\\
\end{align*}
Therefore the ratio of the number of degenerate 3-walks  and the total number of $3$-walks is bounded by
\begin{align*}
\frac{\#\text{degenerate 3-walks}}{\#\text{3-walks}}  & 
\approx_{\text{$(1\pm O(\epsilon'))$-factor}} 
\frac{2 l_xl_y \cdot |V_C|}{\lambda^3 (l_x l_y)^{1.5} |V_C|}\\
& =  \frac{2 }{\lambda^3 \sqrt{l_x l_y}}\\
& \leq \frac{2}{ 4/\epsilon'} && \text{(Def.~of $X',Y'$)}\\
& = \epsilon'/2.
\end{align*}
Hence, the total number of  $3$-paths (3-walks that are not degenerate) is at least
\begin{align*}
\#\text{$3$-paths}\gtrsim_{\text{$(1\pm \epsilon')^{2.5}$-factor}}   \lambda^3 (l_x l_y)^{1.5} |V_C|.
\end{align*}

On the other hand, the number of $4$-tuples of windows that contribute to those $3$-paths of characters is $8 { l_x \choose 2 } { l_y \choose 2 } \leq 2l_x^2 l_y^2$.
Thus, there must exist a $4$-tuple containing at least 
\begin{align*}
\frac{\#\text{$3$-paths}}{\#\text{$4$-tuples}} &\gtrsim_{\text{$(1\pm \epsilon')^{2.5}$-factor}}
 \frac{\lambda^3 (l_x l_y)^{1.5} |V_C|  }{2l_x^2 l_y^2}\\
&=\frac{ \lambda^3 |V_C| }{\sqrt{l_x l_y}} \\
& \approx_{\text{$\sqrt{1\pm \epsilon'}$-factor}} \frac{\lambda^3 \sqrt{l_xd_x l_y d_y}  }{\sqrt{l_x l_y}} && \text{(Eq.~\eqref{eq:lxdx=lydy})}\\
 &= \lambda^3\sqrt{d_x d_y}
\end{align*}
many $3$-walks. This means that such a $4$-tuple is $(\epsilon,\lambda)$-constructive.

\end{proof}

\subsubsection{Guarantees of the sparsification algorithm}

\begin{theorem}[Sparsification for constant $\lambda$ (Step 1 of Theorem~\ref{thm:formal_lcs_quadratic})]\label{thm:sp1}
Given $k$ windows $w_1, \cdots, w_k$. Let $\wmax = \max_{i \in [k] }|w_i|$, $\wmin = \min_{i\in [k]} |w_i|$ and $\wgap = \wmax / \wmin$. 
Let the number of different window sizes be $\wlayers$. 
For any constant $\lambda \in (0,1)$ and $0 < \epsilon <1$, there is a randomized algorithm (Algorithm~\ref{alg:sp1}) that runs in time 
\begin{align*}
O( \wmax^2 k^{1.1} \log k + \wmax k^{2.2} \log^2 k ),
\end{align*} 
outputs a table $\wh{\sf O}_{\lambda^2} \in \{ 0 , 1 \}^{k \times k}$ such that 
\begin{align*}
 \normal{ \lcs(w_i,w_j) } \geq (1-\epsilon) \lambda^3 , \mathrm{~if~} \wh{\sf O}_{\lambda^2} [i][j] = 1
\end{align*}
and
\begin{align*}
\left| \left\{ (i,j) ~\bigg|~ \normal{ \lcs(w_i,w_j) } \geq \lambda, \mathrm{~~~and~~~}  \wh{\sf O}_{\lambda^2} [i][j] = 0 \right\} \right| \leq k^{2- \epsilon \lambda^3 / ( 800 \wlayers \wgap )} .
\end{align*}
The algorithm has success probability at least $1-1/\poly(k)$.
\end{theorem}
\begin{proof}
We set $\gamma = 1/10$. The overall running time is
\begin{align*}
& ~ O(  k |S| \wmax^2 + k^2 |S|^2 \wmax ) \\
= & ~ O( k \cdot k^{\gamma} \log k  \cdot \wmax^2 + k^2 \cdot k^{2\gamma} \log^2 k \cdot \wmax  ) \\
= & ~  O( k^{1.1} \log k \cdot \wmax^2 + k^{2.2} \log^2 k \cdot \wmax )
\end{align*}
where the first step follows from $|S| = O(k^{\gamma} \log k)$, the second step follows from $\gamma = 0.1$.

The guarantee of table $\wh{\sf O}_{\lambda^2}$ follows from properties of graph $\gdelta$ (Algorithm~\ref{alg:sp1} provides the first property of table in Theorem statement, Lemma~\ref{lemma:edges_upper_bound} provides the second property of table in Theorem statement). 
\end{proof}
% !TeX root = ../../main.tex

\subsection{Sparsification for arbitrary $\lambda$ (Step 1 of Theorem~\ref{thm:formal_lcs})}\label{sec:sp2}
One shortcoming of Lemma \ref{lemma:edges_upper_bound} is that the number of remaining edges is only truly subquadratic if $\lambda$ is constant. As we discuss in Section \ref{sec:sp1}, the overall running time of the algorithm depends on the number of edges in the remaining graph; and in order for the running time to be truly subquadratic, we need to reduce the number of edges to truly subquadratic. In this section, we show how one can obtain this bound even when $\lambda$ is sub-constant. However, instead of losing a factor $\lambda^2$ in the approximation, our technique loses a factor of $\Omega(\lambda^3)$. 

From here on, we assume that all of the windows are of the same length and we show at the end of the section that this is almost without loss of generality. We begin by giving a definition:

\begin{definition}[$\lcs_{w_a}(w_i,w_j)$]
For two windows $w_i$ and $w_j$, and a window $w_a$, define $\lcs_{w_a}(w_i,w_j)$ as the length of the \textsf{LCS} of $\opt_{i,a}$ and $w_j$, where $\opt_{i,a}$ denotes a fixed \textsf{LCS} of $w_i$ and $w_a$.
\end{definition}

Notice that unlike $\lcs$, this new definition is not symmetric. That is, the size of $\lcs_{w_a}(w_i,w_j)$ may be different from the size of $\lcs_{w_a}(w_j,w_i)$. Throughout this section and Section~\ref{sec:application_of_lcsdata},  $\opt_{i,a}$ refers to a fixed (e.g., lexicographically smallest) longest common subsequence of $w_i$ and $w_a$. We moreover assume that $\opt_{i,a}$ and $\opt_{a,i}$ refer to the same matching.

Similar to $\lcs$, we also normalize the size of $\lcs_s$ by the geometric mean of the lengths of the two windows. Our assumption from here on, until the statement of Theorem~\ref{thm:sp2} is that all the windows are of the same length which implies that $$\normal{\lcs_{w_a}(w_i,w_j)} = \lcs_{w_a}(w_i,w_j) / \sqrt{|w_i||w_j|} = \lcs_{w_a}(w_i,w_j) / |w_i| = \lcs_{w_a}(w_i,w_j) / |w_j|.$$ We bring a reduction in Theorem~\ref{thm:sp2} to make our solution work for windows of arbitrary length.
In what follows, we first give an algorithm for detecting \textit{close pairs} (a notion that we introduce later in the section) of windows, and then prove that the number of remaining pairs whose normalized $\lcs$ is at least $\lambda$ is truly subquadratic.

Our algorithm is based on a data structure which we call $\lcsdata(w_a,S,\wt{\lambda})$. The reader can think of $\wt{\lambda} = \lambda^2 / 2$. Let us fix a threshold $\wt{\lambda}$ and a window $w_a$. $\lcsdata(w_a,S,\wt{\lambda})$ receives a set $S$ of windows as input and preprocesses the windows in time $\wt{O}(|w_a|^2 |S|)$. Next, $\lcsdata(w_a,S,\wt{\lambda})$ would be able to answer each query of the following form in almost linear time ($O(\wmax)$):\\
\begin{quote}
	Given two windows $w_i, w_j \in S$, either certify that $\normal{ \lcs_{w_a}(w_i,w_j) } < \wt{\lambda}$ or find a solution for $\normal{ \lcs_{w_a}(w_i,w_j) }$ of size at least $\wt{\lambda}^2 / 4$.
\end{quote}
\;\\
We first show in Section \ref{sec:application_of_lcsdata}, how $\lcsdata$ gives us a sparsification in truly subquadratic time and then discuss the algorithm for $\lcsdata$ in Section \ref{sec:implemenation_of_lcsdata}.

\subsubsection{$\Omega(\lambda^3)$ Sparsification using $\lcsdata$}\label{sec:application_of_lcsdata}

We use a constant parameter $\gamma$ in our algorithm, and in the end we adjust $\gamma$ to minimize the total running time. 
Our algorithm repeats the following procedure $ O(k^{\gamma} \log k)$ times: sample a window $w_a$ uniformly at random and let $S$ be the set of all the other windows. Next, by setting $\wt \lambda = \lambda^2/2$, we obtain $\lcsdata(w_a,S,\lambda^2/2)$ via running the preprocessing step. Finally, for each pair of windows $w_i, w_j \in S$, we make a query to $\lcsdata$ to verify one of the following two possibilities:
\begin{itemize}
	\item $\normal{ \lcs_{w_a}(w_i,w_j) } < \lambda^2/2$;
	\item $\normal{ \lcs_{w_a}(w_i,w_j) } \geq \lambda^4/16$.
\end{itemize} 
If the latter is verified we set $\hat{{\sf O}}_{\lambda^3}[i][j]$ and $\hat{{\sf O}}_{\lambda^3}[j][i]$ to 1 otherwise we take no action. In what follows, we prove that after the above sparsification, the number of edges in the remaining graph is truly subquadratic.

\begin{algorithm}
	\begin{algorithmic}[1]\caption{Sparsification for arbitrary $\lambda$ (Step 1 of Theorem~\ref{thm:formal_lcs})}\label{alg:sp2}
		\Procedure{\textsc{CubicSparsification}}{$w_1, w_2, \ldots, w_k, \lambda$} \Comment{Theorem~\ref{thm:sp2}}
		\State $\wt{\lambda} \leftarrow \lambda^2/2$
		\For{$ \text{counter} = 1 \to  10 k^{\gamma} \log k $}
		\State Sample $a \sim [k]$ uniformly at random
		\State $S \leftarrow \emptyset$
		\For{$i = 1 \to k$}
		\If{$i \neq a$}
			\State $S \leftarrow S \cup \{w_i\}$
		\EndIf
		\EndFor
		\State$\lcsdata.\textsc{Initial}(w_a,S,\wt{\lambda})$ \Comment{Algorithm~\ref{alg:lcs_cmp}, Lemma~\ref{lem:initial_lcs_cmp}}
		\For{$w_i \in S$}
		\For{$w_j \in S$} %\Zhao{I'm not sure it is accept or reject...}  \Xiaorui{correct to me}
		\If{$\lcsdata.\textsc{Query}(w_i,w_j)$ outputs accept} \Comment{Alg.~\ref{alg:lcs_cmp}, Lemma~\ref{lem:query_lcs_cmp}}
		\State $\hat{{\sf O}}_{\lambda^3} [i][j] \leftarrow 1$ \Comment{$\normal{ \lcs_{w_a}(w_i,w_j) } \geq \wt{\lambda}^2/8$}
		\EndIf
		\EndFor
		\EndFor
		\EndFor
		\State \Return $\hat{{\sf O}}_{\lambda^3}$
		\EndProcedure
	\end{algorithmic}
\end{algorithm}

 Our goal for the rest of this section is to prove (Lemma~\ref{lem:upper_bound_on_E_G_delta_Delta}) an upper bound on the number of edges that remain in the following graph:
 
 \begin{definition}[$\gdelta$]\label{def:graph-2}
 Define a graph $\gdelta$ with $k$ vertices and the following edges:
\begin{align*}
E(\gdelta) = \left\{ (i,j) ~\bigg|~  \normal{ \lcs(w_i,w_j) } \geq \lambda \mathrm{~~~and~~~} \hat{{\sf O}}_{\lambda^3}[i][j] = 0 \right\}.
\end{align*}
\end{definition}

We define a notation called ``close'' which is similar to the notion of ``well-connected'' vertices in Section~\ref{sec:sp1}. %
\begin{definition}[close]\label{def:close}
Let $\gamma \in (0,1)$ and $\lambda \in (0,1)$. We say a pair $(w_i,w_j)$ of windows is \textit{close}, if there are at least $k^{1-\gamma}$ windows $w_a$ such that $\normal{ \lcs_{w_a}(w_i,w_j) } \geq \lambda^2/2$. 
\end{definition}

Our first observation is that Algorithm \ref{alg:sp2} detects all the close pairs with high probability.

\begin{lemma}\label{lem:cubic_table_is_good}
Let $\hat{{\sf O}}_{\lambda^3} \in \{0,1\}^{k \times k}$ be the output of Algorithm \ref{alg:sp2}. For each $(i,j) \in [k] \times [k]$, if $(w_i,w_j)$ is close (Definition~\ref{def:close}), then $\hat{{\sf O}}_{\lambda^3} [i][j] = 1$ holds with probability at least $1-1/k^3$.
\end{lemma}
\begin{proof}
We consider a fixed $(i,j)$ such that $(w_i,w_j)$ is close. 
By Definition~\ref{def:close}, there are at least $k^{1-\gamma}$ windows $w_a$ such that $\normal{ \lcs_{w_a}(w_i,w_j) } \geq \lambda^2/2$. If such a $w_a$ window is sampled in our algorithm, then we detect the edge between the close pair.
The probability that none of these windows is sampled   
is at most %  
\begin{align*}
\left( 1 - \frac{ k^{1-\gamma} }{ k } \right)^{10 k^{\gamma} \log k}  = \left( 1 - \frac{ 1 }{ k^{\gamma} } \right)^{10 k^{\gamma} \log k} \leq \frac{1}{k^5}.
\end{align*}
Taking a union over at most $k^2$ pairs completes the proof. 
\end{proof}

Before we proceed to Lemma \ref{lem:upper_bound_on_E_G_delta_Delta}, we bring Lemma \ref{lem:existing_a_correlated_pair} as an auxiliary observation.

\begin{lemma}[existence of a correlated pair]\label{lem:existing_a_correlated_pair}
Let $\lambda \in (0,1)$. Given a window $w_a$ and a set $T$ containing at least $2 / \lambda$ windows. If for each $w_i \in T$ we have $\normal{ \lcs(w_a,w_i) } \geq \lambda$, then there exist two windows $w_i, w_{j} \in T$ such that both $\normal{ \lcs_{w_a}( w_i, w_{j} ) } \geq \lambda^2/2$ and $\normal{ \lcs_{w_a}( w_j, w_{i} ) } \geq \lambda^2/2$ hold. 
\end{lemma}
\begin{proof}
Let $d$ be the size of the windows.  We consider %
a character-based bipartite graph: On one side, it has $d$ nodes, and on the other side it has $d|T|$ nodes. 
Two nodes $x, y$ in the character-based graph are adjacent iff 
$(x, y) \in \opt_{a, i}$ where  $w_a$ is the window containing character $x$ and $w_i$ is the window containing character $y$.
Since $\normal{ \lcs(w_a,w_i) } \geq \lambda$  for every $w_i \in T$,
the total number of edges in character-based bipartite graph is at least $\lambda |T| d$.

For $\ell$-th character in window $w_a$, we use $D_{\ell}$ to denote the degree of the corresponding node in the character-based bipartite graph. The number of $2$-walks between pairs of nodes on the side with $d|T|$ nodes is at least
\begin{align*}
  \sum_{\ell=1}^{d} D_{\ell} (D_{\ell}-1) 
= & ~  \left( \sum_{\ell=1}^{d} D_{\ell}^2 - \sum_{\ell=1}^{d} D_{\ell} \right) \\
\geq & ~  \left( \frac{1}{d} ( \sum_{\ell=1}^{d} D_{\ell} )^2 -  \sum_{\ell=1}^{d} D_{\ell} \right) & \text{~by~Cauchy-Schwarz~inequality} \\
\geq & ~  \left( \frac{1}{d} (\lambda |T| d )^2 - (\lambda |T| d ) \right) & \text{~by~Eq.~\eqref{eq:existing_a_correlated_pair_1} explained below}  \\
= & ~  \left( \lambda^2 d |T|^2 - \lambda |T| d \right)  \\
\geq & ~  1/2 \lambda^2 d |T|^2  & \text{~by~} |T| \geq 2/\lambda
\end{align*}

The number of $3$-tuples $(w_i, w_a, w_j)$ is at most $|T|^2$. Thus, there must exist a pair $(w_i,w_j)$ such that
\begin{align*}
\normal{ \lcs_{w_a} (w_i, w_j) } \geq  \lambda^2/2.
\end{align*}
This also means that 
\begin{align*}
\normal{ \lcs_{w_a} (w_j, w_i) } \geq \lambda^2/2
\end{align*}
which is the other conclusion of our lemma.

It remains to show Eq.~\eqref{eq:existing_a_correlated_pair_1}. 
Since the number of edges in the character-based bipartite graph is at least $\lambda |T| d$, we have
\begin{align}\label{eq:existing_a_correlated_pair_2} 
\sum_{\ell=1}^{d} D_{\ell} & \geq \lambda |T| d \nonumber \\
& \geq \lambda (2/\lambda) d > d. && \text{($|T| \geq 2/\lambda$ by lemma premise)}
\end{align}

We also require the following simple fact: 
\begin{align}\label{eq:existing_a_correlated_pair_3}
 \forall x \geq y \geq z \geq 0,\;\;\; x(x-z) \geq y(y-z).
\end{align}
Combining the last two inequalities, we finally have:
\begin{align}\label{eq:existing_a_correlated_pair_1}
 \frac{1}{d} ( \sum_{\ell=1}^{d} D_{\ell} )^2 -  \sum_{\ell=1}^{d} D_{\ell}  
& = \frac{\sum_{\ell=1}^{d} D_{\ell} }{d}( \sum_{\ell=1}^{d} D_{\ell} - d)  \nonumber \\
& \geq  \frac{1}{d} (\lambda |T| d )^2 - (\lambda |T| d ) && \text{(Ineq.~\eqref{eq:existing_a_correlated_pair_2} and~\eqref{eq:existing_a_correlated_pair_3})}.
\end{align}

\end{proof}

Now, we are ready to prove that the remaining graph is sparse.

\begin{lemma}[upper bound on $|E(\gdelta)|$]\label{lem:upper_bound_on_E_G_delta_Delta}

\begin{align*}
| E( \gdelta ) | \leq  \frac{2 k^{2-\gamma}}{\lambda}
\end{align*}
holds with probability at least $1-1/k^3$.
\end{lemma}

In the proof of this lemma we will construct an auxiliary graph $\fdelta$ in the following way:
\begin{definition}[$\fdelta$]\label{def:graph-3}
Let $a$ denote the node in $V(\gdelta)$ that has the highest degree. The vertices of $\fdelta$ would be the set of neighbors of $a$ in $\gdelta$.%, that is $V(\fdelta)= N(a)$. 
We add an edge between vertices $(i,j)$ in $\fdelta$ if both $\normal{ \lcs_{w_a}(w_i,w_{j}) } \geq \lambda^2/2$ and $\normal{ \lcs_{w_a}(w_j,w_{i}) } \geq \lambda^2/2$ hold. 
\end{definition}

\begin{proof}[Proof of Lemma~\ref{lem:upper_bound_on_E_G_delta_Delta}]

We prove the lemma by contradiction. % 
Suppose 
\begin{align*}
|E ( \gdelta ) | > \frac{2 k^{2-\gamma}}{\lambda}.
\end{align*}
Since the number of vertices in $\fdelta$ is equal to the maximum degree of $\gdelta$  we have
\begin{gather}\label{eq:V(delta)}
|V(\fdelta)| > \frac{2 k^{1-\gamma}}{\lambda}.
\end{gather}

 Using Lemma \ref{lem:existing_a_correlated_pair}, we have for each set $T \subseteq V(\fdelta)$ with $|T| \geq  2/\lambda$, there exist two nodes $u$ and $v$ in $T$ such that $ (u,v)$ is an edge of $\fdelta$.

 If we look at the complement of graph $\fdelta$, we know there is no clique $K_{r}$ where $r = 2/\lambda$.  Using Turan's theorem (Lemma~\ref{lem:turan41}) we know that the complement of graph $\fdelta$ has  at most $(1-\frac{1}{r-1})\frac{q^2}{2} < (1-\frac{1}{r})\frac{q^2}{2}$ edges, where $q := | V( \fdelta ) |$. Then  we have 
 \[|E(\fdelta) | > \frac{q(q-1)}{2} - (1-\frac{1}{r}) \frac{q^2}{2} = \frac{q^2}{ 2r} - \frac{q} { 2} .\]
This implies that there exists a vertex $b$ whose  degree in $\fdelta$ is more than  
\begin{gather*} \frac{|V(\fdelta)| }{r } - 1   =  \frac{|V(\fdelta)| }{2/\lambda} - 1  
 \ge k^{1-\gamma}-1, \end{gather*}
 where the inequality follows by Eq.~\eqref{eq:V(delta)}. Thus, the degree of $b$ is at least $k^{1-\gamma}$. For each vertex $c$ of $\fdelta$ which is adjacent to $b$ we have % 
\begin{gather*}
\normal{\textsf{LCS}_{w_c}(w_a,w_b)} = \normal{\textsf{LCS}_{w_a}(w_c,w_b)} \geq \lambda^2/2.
\end{gather*}
Since the  degree of $b$ in $\fdelta$ is at least $k^{1-\gamma}$, this means that pair $(w_a,w_b)$ is a close pair and the edge $(w_a,w_b)$ should not have existed between the vertices in $\gdelta$ in the first place.	
\end{proof}
Now, we are ready to bring our main theorem of this section. In Theorem~\ref{thm:sp2} we assume that the windows may have different length but the total number of distinct window sizes is bounded by $\wlayers$.
\begin{theorem}[Sparsification for arbitrary $\lambda$ (Step 1 of Theorem~\ref{thm:formal_lcs})]\label{thm:sp2}
Given $k$ windows $w_1, \cdots, w_k$. Let $\wlayers$ denote the number of different sizes for windows. For any $\lambda \in (0,1)$ and  $\gamma \in (0,1)$, there is a randomized algorithm (Algorithm~\ref{alg:sp2}) that runs in time
\begin{align*}
 O( \wlayers^2 (k^{1+\gamma} \wmax^2 \log^2 k + k^{2+\gamma} \wmax \log k) )
\end{align*}
and outputs a table $\wh{\sf O}_{\lambda^3} \in \{ 0 , 1 \}^{k \times k}$ such that
\begin{align*}
 \lcs(w_i,w_j) /\max\{|w_i|,|w_j|\} \geq \lambda^4 / 16, \text{~if~} \wh{\sf O}_{\lambda^3} [i][j] = 1
\end{align*}
and
\begin{align*}
\left| \left\{ (i,j) ~|~ \lcs(w_i,w_j) / \max\{|w_i|,|w_j|\} \geq \lambda, \mathrm{~~~and~~~} \wh{\sf O}_{\lambda^3}[i][j] = 0 \right\} \right| = O( k^{2-\gamma} \wlayers^2 / \lambda ) .
\end{align*}
The algorithm has success probability $1-1/k^3$.
\end{theorem}
\begin{proof}

We run the algorithm explained above $\wlayers + \binom{\wlayers}{2}$ times. Each of the first $\wlayers$ runs considers one set of windows with equal sizes. Each of the next $\binom{\wlayers}{2}$ runs on one pair of window sizes. For such runs, we pad the smaller windows by dummy characters to make sure all windows have equal lengths. Our estimation for each pair of windows is obtained in one of the runs (which focuses on those particular lengths). Thus, both the runtime and the number of false-negatives of our algorithm are multiplied by a factor of $O(\wlayers^2)$.

The running time of each round of Algorithm~\ref{alg:sp2} is
\begin{align*}
= & ~ O( \lcsdata.\textsc{Initial} \text{~time} + |S|^2 \cdot ( \lcsdata.\textsc{Query} \text{~time})  ) \\
= & ~ O(  |S| \wmax^2 \log k +   |S|^2 \wmax ) \\
= & ~ O(  k \wmax^2 \log k + k^2 \wmax )
\end{align*}

Since the algorithm repeats the sampling procedure $O(k^{\gamma} \log k)$ rounds, thus the overall running time is
\begin{align*}
O(k^{\gamma} \log k) \cdot O( k \wmax^2 \log k + k^2 \wmax ) = O( k^{1+\gamma} \wmax^2 \log^2 k + k^{2+\gamma} \wmax \log k ).
\end{align*}
\end{proof}

% !TeX root = ../../main.tex

\subsubsection{Implementation of $\lcsdata$}\label{sec:implemenation_of_lcsdata}
The goal of this section is to design a data-structure that will be used in our subquadratic time algorithm for \textsf{LCS}. The data structure receives a window $w_a$, a threshold $\wt\lambda$, and a set $S$ of windows 
$w_1, w_2, \dots, w_s$. (To avoid confusing, one can assume $a=0$ and $w_a = w_0$ but we refer to this special window by $w_a$ to be consistent with the previous sections.) All the windows $w_a, w_1, w_2, \ldots, w_s$ have equal sizes. We present an $O(s|w_a|^2 \log n)$ time preprocessing algorithm and an $O(|w_a|)$ time query algorithm for $\lcsdata$  such that for any $i, j \in [s]$
\begin{enumerate}
\item if $\|\lcs_{w_a}(w_i, w_j)\| \geq \wt \lambda $, then the query algorithm outputs accept;
\item if $\|\lcs_{w_a}(w_i, w_j)\| < {\wt \lambda}^2 / 4$, then the query algorithm outputs reject.
\end{enumerate}

We bring the high level idea in the following (see also Algorithm~\ref{alg:lcs_cmp} for pseudocode).
We first compute $\opt_{a, i}$ for every $i \in [s]$. Next, for each window $w_i$, we find a set of at most $2 /\wt \lambda$ common subsequences between $w_a$ and $w_i$. Let us put the corresponding indices in the $w_a$ side of all such common subsequences in a set $Y_{a,i}$ for each window $w_i$. Our algorithm maintains the property that if we remove all characters corresponding to $Y_{a,i}$ from $w_a$, the \textsf{LCS} of the remainder with $w_i$ is smaller than $\frac{|w_a| \wt \lambda}{2}$. For each pair $w_i,w_j \in S$, if $\textsf{LCS}_{w_a}(w_i,w_j) \geq |w_a| \wt \lambda$ then at least half of the characters that contribute to such a solution lie in $Y_{a,j}$. Thus, the intersection of at least one common subsequence in $Y_{a,j}$ and $\opt_{a,i}$ should have size $\frac{|w_a| \wt \lambda}{2}/(2/\wt \lambda) = |w_a|\wt \lambda^2 /4$. 

In what follows, we discuss the details.

\begin{algorithm}[!t]
	\begin{algorithmic}[1]\caption{$\lcsdata$ data structure}\label{alg:lcs_cmp}
	\State {\bf data structure} $\lcsdata$
	\State
	\State {\bf members}
		\State\hspace{4mm} $\opt_{a,1}, \cdots, \opt_{a,s}$
		\State\hspace{4mm} $Y_{a,1}, \cdots, Y_{a,s}$
		\State\hspace{4mm} $\wt{\lambda} \in (0,1)$
	\State {\bf end members}
	\State
		\Procedure{\textsc{Initial}}{$w_a, \{ w_i \}_{i\in [s]}, \wt\lambda$} \Comment{Lemma~\ref{lem:initial_lcs_cmp}}
		\For{$i \in [s]$}
			\State compute $\opt_{a, i}$, a fixed (e.g. lexicographically smallest) \textsf{LCS} between $w_a$ and $w_i$
		\EndFor
		\For{$i \in [s]$}
			\State $Y_{a, i} \leftarrow \emptyset$
			\State $w'_a \leftarrow w_a$
			\While {$\textsf{LCS}(w'_a,w_i) \geq |w_a|\wt \lambda/2$}
				\State $Y_{a,i} \leftarrow Y_{a,i} \cup$ characters in the \textsf{LCS} of $w'_a$ and $w_i$
				\State $w'_a \leftarrow w'_a \setminus $ characters in the \textsf{LCS} of $w'_a$ and $w_i$
			\EndWhile
		\EndFor
		\State \Return $\{ \opt_{a, i} \}_{i\in [s]}$, $\{ Y_{a, i} \}_{i \in [s]}$ 
		\EndProcedure
		\State
		\Procedure{\textsc{Query}}{$w_i, w_j$} \Comment{Lemma~\ref{lem:query_lcs_cmp}}
		\If {$|\opt_{a, i} \cap Y_{a, j}| > \wt \lambda |w_a| / 2$}
			\State \Return accept
		\Else
		\State \Return reject
		\EndIf
		\EndProcedure
		\State
	\State {\bf end data structure}
	\end{algorithmic}
\end{algorithm}

\begin{lemma}[\textsc{Initial}]\label{lem:initial_lcs_cmp}
Given a parameter $\wt{\lambda}$, a window $w_a$, and a set of windows $w_1, \cdots,$  $w_s$, the \textsc{Initial} procedure of data structure $\lcsdata$ (Algorithm~\ref{alg:lcs_cmp}) takes $O(s|w_a|^2 \log n )$ time, and outputs $\{ \opt_{a,i} \}_{i \in [s]}$ and $\{ Y_{a,i} \}_{i \in [s]}$ such that the following hold.
\begin{enumerate}
\item $\opt_{a, i}$ corresponds to the $w_a$ side indices of a fixed longest common subsequence between $w_a$ and $w_i$ for every $i \in [s]$.
\item $Y_{a, i}$ corresponds to the $w_a$ side indices of at most $2 / \wt \lambda$ common subsequence between $w_a$ and $w_i$ for every $i \in [s]$.
\item For any common subsequence between $w_a$ and $w_i$, if none of its $w_a$ side indices are in $Y_{a, i}$, then the length of this common subsequence is less than $\wt \lambda |w_a|/2$.
\end{enumerate}
\end{lemma}
\begin{proof}

In order to construct set $Y_{a,i}$ for each window $w_i$ we run the following algorithm: We start by setting $Y_{a,i}$ to an empty set. Next, we set $w'_a = w_a$ and iteratively compute the \textsf{LCS} of $w'_a$ and $w_i$. Each time we find a solution, we compare its size to $\wt \lambda |w_a|/2$. If the solution size is smaller, we terminate the algorithm. Otherwise, we add the original positions of the common subsequence in $w_a$ to set $Y_{a,i}$ and continue by removing these characters from $w'_a$. We stop when the solution size drops below $\wt \lambda |w_a|/2$. It immediately follows that since each time the size of $Y_{a,i}$ is increased by at least $\wt \lambda |w_a|/2$ then we repeat this procedure at most $2/\wt \lambda$ times.

To bound the runtime, we state the following fact: when the size of the \textsf{LCS} between a string $Q$ and a string $R$ is bounded by $\ell$, one can preprocess one of the strings ($R$ in this case) in time $O(|R|\log n)$ and then compute the \textsf{LCS} in time $O(|Q| \ell \log n)$ (see Theorem~\ref{theorem:smalllcs}). Notice that in order to construct $Y_{a,i}$ for a window $w_i$, the total size of the \textsf{LCS}'s that we find is bounded by $|w_a|$. Therefore, for a fixed window $w_i$,  if we preprocess $w_i$ once and use it for all computations, the total runtime is bounded by $$O(|w_a| |Y_{a,i}| \log n + |w_i| \log n) = O(|w_a|^2 \log n + |w_i| \log n).$$ Thus, the total preprocessing time of our algorithm over all windows is $O(s|w_a|^2 \log n)$.  
\end{proof}

\begin{lemma}[\textsc{Query}]\label{lem:query_lcs_cmp}
For any $(i,j) \in [s] \times [s]$, the \textsc{Query} of data structure $\lcsdata$ (Algorithm~\ref{alg:lcs_cmp}) runs in time $O(|w_a|)$ with the following properties:
\begin{enumerate}
\item If $\lcs_{w_a}(w_i, w_j) \geq \wt \lambda |w_a|$, then it outputs accept.
\item If $\lcs_{w_a}(w_i, w_j) < {\wt \lambda}^2 |w_a| / 4$, then it outputs reject.
\end{enumerate}
\end{lemma}
\begin{proof}
Our algorithm takes the intersection of the characters that contribute to $\opt_{a,i}$ with $Y_{a,j}$. As explained earlier, if $\lcs_{w_a}(w_i, w_j) \geq \wt \lambda |w_a|$ then at least half of these characters belong to $Y_{a,j}$ and thus the intersection of one of the common subsequences contributing to $Y_{a,j}$ with $\opt_{a,i}$ is at least $\wt \lambda^2|w_a| / 4$. Indeed, this can only happen if $\lcs_{w_a}(w_i, w_j) \geq  {\wt \lambda}^2 |w_a| / 4$ in the first place. Otherwise we certainly output reject.
 \end{proof}

% !TeX root = ../../main.tex
\section{\textsf{LCS} Step 0: Window-compatible Solutions}\label{sec:window}
The goal of this section is to show that window-compatible solutions provide very accurate estimates for \textsf{LCS}. Roughly speaking, in this section, we construct a set of windows for $A$ and a set of windows for $B$ and we show that it suffices to only compute the \textsf{LCS} of pairs of windows. If that information is available, then we can estimate the \textsf{LCS} of the original two strings very accurately.

We use the same definition of window-compatible transformation as \cite{beghs18}, but we will use slightly different sets of windows. Every window is a (contiguous) substring of each string. Notice that windows may not have the same length. We use $W_A$ to denote the set of windows constructed for $A$ and $W_B$ for the set of the windows that we construct for $B$.

Let us first state our definition of window-compatible solutions for \textsf{LCS}.
\begin{definition}[Window-compatible common subsequence]
\begin{itemize}
\item Let $S = \langle w_1,  \cdots, w_{\alpha} \rangle$ and $S' = \langle w_1',  \cdots w_{\alpha}' \rangle$ be two sequences of non-overlapping windows from $W_A$ and $W_B$, respectively. 
 We call a common subsequence of $(A, B)$ {\em window-compatible with respect to $S$ and $S'$}, if it is a union of $k$ common subsequences of $({w_1}, {w_1'}),\dots,({w_{\alpha}}, {w_{\alpha}'})$ % 
\item
We call a common subsequence {\em window-compatible}, if it is window-compatible with respect to some pair of sequences of non-overlapping windows from $W_A$ and $W_B$, ordered by their respective starting indices.
\end{itemize}
\end{definition}

$W_A$ will simply be a partitioning of $A$ into disjoint windows of length $d$ (for parameter $d$ to be finalized later).
Setting $d = n^{x}$ leads to a truly subquadratic time solution for any $0 < x < 1$, however, one can play with the value of $d$ to optimize the running time.
Below we consider two constructions of windows $W_B$ for string $B$.

\subsection{Window construction for constant $\lambda$ (Step 0 of Theorem~\ref{thm:formal_lcs_quadratic})}\label{sec:LCS_generate_multiple_levels}
For clarity, we use $B^{(j,l)}$ to denote a length-$l$ substring of $B$ which starts at index $j$ and ends at index $j+l-1$. Our construction has multiple layers, where each layer contains windows with equal lengths.
For parameter $\epsilon_0 \in (0,1)$ to be defined later,
let $$f := \lceil\log_{1+\eps_0}(1/\eps_0)\rceil+1 = \Theta\left(\frac{1}{\eps_0}\log(\frac{1}{\eps_0})\right).$$

For each $i \in \{-f, \cdots, f \}$, we define $d_i = d ( 1 + \epsilon_0  )^i$, $g_i = \epsilon_0 d_i$, and $t_i = n/d_i$. For each $i$, let $W_B^i$ denote the set of all the windows in this layer. In the $i$-th layer, all the windows have the same size which is $d_i$. $g_i$ is the shift size, i.e.~for each window (except the leftmost and rightmost windows), if we shift by $\pm g_i$
we get another window in $W_B^i$. $t_i$ is the number of windows in the $i$-th layer.

\begin{definition}[Window construction for constant $\lambda$]\label{def:multiple_levels_window}
	\begin{align*}
	W_B^i :=& \big\{ B^{(\mathrm{left},\mathrm{len})} ~\big|~ \mathrm{left} = x\cdot g_i + y\cdot d_i, \mathrm{len} = d_i ,\\
	& \forall x \in \{ 0,1,\cdots, d_i/g_i - 1\}, \forall y \in \{0,1,\cdots,t_i-1\} \big\} 
	\end{align*}
	where we use $B^{(\mathrm{left},\mathrm{len})}$ denotes a (contiguous) substring of string $B$ that starts at $\mathrm{left}$ and has length $\mathrm{len}$. Finally $W_B$ is $\cup_{i \in \{-f,\cdots,f\} } W_B^i$. 
\end{definition}

\begin{algorithm}
	\begin{algorithmic}[1]\caption{Window construction algorithm for constant $\lambda$ }\label{alg:levels}
		\Procedure{\textsc{QuadraticWindows}}{$A,B,n,d,\epsilon_0$} % 
		\State \Comment{Generate $W_A$:}
		\State ${W_A} \leftarrow \emptyset$
		\State $t \leftarrow n/d$
		\For{$y = 0 \to t -1$}
		\State $\text{left} \leftarrow y \cdot d $
		\State $ W_A \leftarrow W_A \cup \left\{A^{( \text{left},d ) }\right\}$
		\EndFor
		\State \Comment{Generate $W_B$:}		
		\State $f \leftarrow \lceil\log_{1+\eps_0}(1/\eps_0)\rceil+1$ \Comment{$d \cdot (1+\epsilon_0)^{f} = \Theta(d/\epsilon_0)$}
		\State $W_B \leftarrow \emptyset$
		\For{$i = -f \to f$}
		\State $W_B^i \leftarrow \emptyset$
		\State $d_i \leftarrow d (1+\epsilon_0)^i$, $t_i \leftarrow n/d_i$, $g_i \leftarrow \epsilon_0 d_i$ \Comment{$d_i$ is the length, $g_i$ is the shift}
		\For{$x = 0 \to d_i/g_i-1$}
		\For{$y = 0 \to t_i -1$}
		\State $\text{left} \leftarrow x \cdot g_i + y \cdot d_i $
		\State $\text{len} \leftarrow d_i $
		\State $W_B^i \leftarrow W_B^i \cup \left\{B^{( \text{left} , \text{len} ) }\right\}$
		\EndFor
		\EndFor
		\State $W_B \leftarrow W_B \cup W_B^i$
		\EndFor
		\State \Return $W_A, W_B, f$
		\EndProcedure
	\end{algorithmic}
\end{algorithm}

\begin{fact} [Parameters for constant $\lambda$ (Theorem~\ref{thm:formal_lcs_quadratic})] \label{fact:parameters-quadratic}\hfill

Recall that $\epsilon$ is a small constant and $\lambda$ is the relative length of the true LCS between $A$ and $B$.	In order to make sure the loss in the approximation is negligible, we use $$\epsilon_0 := \epsilon \lambda.$$ 

	\begin{itemize}
		\item The total number of windows is equal to $k = |W_A|+|W_B| = O((1/\lambda)^3 n/d)$.
		\item The maximum size of the windows is equal to $\wmax = \Theta(d / \lambda)$.
		\item The minimum size of the windows is equal to $\wmin = \Theta(d \lambda)$.
		\item The ratio of the maximum window size over the minimum window size is bounded by $\wgap = \wmax / \wmin = \Theta(1/\lambda^2)$.
	    \item The number of different layers which is equal to the number of different window sizes is equal to $\wlayers = O( (1/\lambda)  \log (1/\lambda) )$% 
\end{itemize}
\end{fact}
\begin{proof}
We derive the number of windows (the rest of the parameters follow immediately from the construction). 
In each layer of $W_B$, the number of windows is 
$$|W_B^i|=O(t_i d_i/g_i) = O(n/(d_i \eps_0)).$$ 
Thus the total number of windows is given by 
\begin{align*}|W_B| & = \sum_{i=-f}^{f}|W_B^i| \\
& =  O\left(n/\eps_0 \sum_{i=-f}^{f} 1/d_i\right) && \text{(Def.~\ref{def:multiple_levels_window})} \\
& = O\Big(n/(d\eps_0) \underbrace{\sum_{i=-f}^{f} 1/(1+\eps_0)^i}_{=O(1/\eps_0^2)}\Big) \\
& = O(n/(d\eps_0^3)) && \text{(Geometric sequence)}.
\end{align*}
\end{proof}

% !TeX root = main.tex
\subsubsection{Near-optimality of window-compatible common subsequence}\label{sec:optimality}
We present a \\structural lemma for window-compatible common subsequences. This essentially, shows that the problem of computing \textsf{LCS} between the two strings reduces to the problem of computing the \textsf{LCS}'s between the windows.

\begin{lemma}[Window-compatible structural lemma (constant $\lambda$)]\label{lem:improved_structrual_lemma_constant} \hfill

For any two strings $A,B$ of length $n$:
If $\normal{\LCS(A,B)} \geq \lambda$, there exists a window-compatible common subsequence of $A,B$ of length at least $ \lambda n - 8\epsilon_0 n$ for the window sets constructed in Definition~\ref{def:multiple_levels_window}.
\end{lemma}
\begin{proof}

Fix a LCS of $A,B$.

Let $A_j$ be the $j$-th window of $A$, and let $\wt{B}_j$ be the minimal substring of $B$ containing all characters common to $A_j$ and the LCS. We consider three cases:
\begin{description}
\item[{Case 1: $|\wt{B}_j| \in [\epsilon_0 d  , d/\epsilon_0]$:}] In this case we can keep this pair. % 

\item [Case 2: $|\wt{B}_j| \ge d/\epsilon_0$:] Then we throw out this pair. We throw out at least $d/\epsilon_0$ characters, but we only decrease the entire $\LCS$ by at most $d$.

\item[Case 3: $|\wt{B}_j| \le d \epsilon_0$]: Then we also throw out this pair. We throw out at least $d$ characters, but we only decrease the entire $\LCS$ by at most $\epsilon_0 d$.
\end{description}

Let $Z$ denote the set of $A$-windows that we don't throw out. 
Whenever we throw out a pair, the amount of $\LCS$ got decreased is always at most $\epsilon_0$ fraction of the characters we throw out. There are $2n$ characters in total. Then we decrease the $\LCS$  by at most $2\epsilon_0 n$, i.e.,
\begin{align}\label{eq:improved_LCS_sum_i_in_Z_at_least_LCS_AB_minus_2_epsilon0_n}
\sum_{j\in Z} \LCS(\wt{A}_j, \wt{B}_j) \geq  \LCS(A,B)  - 2 \epsilon_0 n,
\end{align}

We now derive, for each $j \in Z$, a {\em window} $B_j \subseteq \wt{B}_j$ of approximately the same length.
Let $i \in \{-f+1,\dots,f\}$ be the layer such that $|\wt{B}_j| \in (d(1+\eps_0)^i, d(1+\eps_0)^{i+1})$.
Similarly, $\wt{B}_j$'s starting index is in $\left((y-1) \cdot \eps_0d(1+\eps_0)^{j-1}, y \cdot \eps_0d(1+\eps_0)^{j-1} \right]$ for some $y$.
Let $B_j \in W_B$ be the window of length $d(1+\eps_0)^{j-1}$ with starting index in $y \cdot \eps_0d(1+\eps_0)^{j-1}$. 
Notice that $B_i$ ends at 
\begin{align*}
y \cdot \eps_0d(1+\eps_0)^{j-1} + d(1+\eps_0)^{j-1} & =   y \cdot \eps_0d(1+\eps_0)^{j-1} + d(1+\eps_0)^{j} - \eps_0d(1+\eps_0)^{j-1}\\
& = (y-1) \cdot \eps_0d(1+\eps_0)^{j-1} + d(1+\eps_0)^{j},
\end{align*}
which is a lower bound on $\wt{B}_j$'s end index, hence indeed $B_j \subseteq \wt{B}_j$.
Finally, notice that

\begin{align}\label{eq:window-loss-2}
| \wt{B}_j \setminus B_j | \leq & ~ d(1+\epsilon_0)^{i+1} - d(1+\epsilon_0)^{i-1} \nonumber \\
\leq & ~ d(1+\epsilon_0)^{i-1} 3 \epsilon_0 \nonumber \\
< & ~ 3 \epsilon_0 |\wt{B}_j|.
\end{align}
Therefore, in total the number of characters we lose by restricting to windows $B_j$ is bounded by $3\eps_0 n$:
\begin{align*}
\sum_{i \in Z} |\LCS(A_j,B_j)|
\geq  & ~ \sum_{j \in Z}  ( | \LCS(A_j,\wt{B}_j) | - 3 \epsilon_0 |\wt{B}_j| ) && \text{(Eq.~\eqref{eq:window-loss-2})}\\
\geq & ~   \sum_{j\in Z} | \LCS(A_j, \wt{B}_j) | - 3 \epsilon_0 n   && \text{($\wt{B}_j$'s are disjoint)}\\
\geq & ~  | \LCS(A,B) | - 8 \epsilon_0 n &&\text{(Eq.~\eqref{eq:improved_LCS_sum_i_in_Z_at_least_LCS_AB_minus_2_epsilon0_n})}.
\end{align*}

\end{proof}

\subsection{Window construction for arbitrary $\lambda$ (Step 0 of Theorem~\ref{thm:formal_lcs})}\label{sec:LCS_generate_multiple_levels_cubic}

Our construction for this case is similar to Section~\ref{sec:LCS_generate_multiple_levels}, with the exception that when we're OK with losing constant factors in approximation, we can afford less shifts and less layers, which will improve the running time.

Let $$f := \log(1/\eps_0).$$
For each $i \in \{0,\cdots,f\}$, we define $d_i = d \cdot 2^i$, and $t_i = n/d_i$. For each $i$, let $W_B^i$ denote the set of all the windows in this layer. We consider window set $W_B^i$, in $i$-th layer, all the window have sizes that are multiples of $d$, and in the range $(d_i/2, d_i]$. The shift for each layer will be $d_i$.

\begin{definition}[Window construction for arbitrary $\lambda$]\label{def:cubic_levels_window}
	\begin{align*}
	W_A, W_B^0 :=& \big\{ B^{(\mathrm{left},\mathrm{len})} ~\big|~ \mathrm{left} = y\cdot d,\\
	&  \forall y \in \{0,1,\cdots,t_0-1\} \big\} 
	\end{align*}
	\begin{align*}
	W_B^i :=& \big\{ B^{(\mathrm{left},\mathrm{len})} ~\big|~ \mathrm{left} = y\cdot d_i, \mathrm{len} = d_i/2 + x \cdot d ,\\
	& \forall x \in \{ 1,\cdots, d_i/(2d)\} \forall y \in \{0,1,\cdots,t_i-1\} \big\} 
	\end{align*}
	where we use $B^{(\mathrm{left},\mathrm{len})}$ to denote a (contiguous) substring of string $B$ that starts at $\mathrm{left}$ and has length $\mathrm{len}$. Finally $W_B$ is $\cup_{i \in \{0,1,\cdots,f\} } W_B^i$. 
\end{definition}

In other words, the window construction in Definition~\ref{def:cubic_levels_window} considers all the substrings that start at any multiple of $d \cdot 2^i$ and whose length is a multiple of $d$, and at most $d \cdot 2^i-1$.

\begin{algorithm}
	\begin{algorithmic}[1]\caption{Window construction algorithm for arbitrary $\lambda$ }\label{alg:cubic-levels}
		\Procedure{\textsc{CubicWindows}}{$A,B,n,d,\epsilon_0$} % 
		\State \Comment{Generate $W_A$ and $W_B^0$:}
		\State $W_A,W_B^0  \leftarrow \emptyset$
		\State $t \leftarrow n/d$
		\For{$y = 0 \to t -1$}
		\State $\text{left} \leftarrow y \cdot d $
		\State $ W_A \leftarrow W_A \cup \left\{A^{( \text{left},d ) }\right\}$
		\State $ W_B^0 \leftarrow W_B^0 \cup \left\{B^{( \text{left},d ) }\right\}$
		\EndFor
		\State \Comment{Generate $W_B$:}		
		\State $f \leftarrow \log(1/\eps_0)$ \Comment{$d \cdot 2^f = d/\epsilon_0$}
		\State $W_B \leftarrow W_B^0$
		\For{$i = 1 \to f$}
		\State $W_B^i \leftarrow \emptyset$
		\State $d_i \leftarrow d \cdot 2^i$, $t_i \leftarrow n/d_i$ %\Comment{$d_i$ is the length, $g_i$ is the shift}
		\For{$x = 1 \to d_i/(2d)$}
		\For{$y = 0 \to t_i -1$}
		\State $\text{left} \leftarrow y \cdot d_i $
		\State $\text{len} \leftarrow d_i/2 + x \cdot d $
		\State $W_B^i \leftarrow W_B^i \cup \left\{B^{( \text{left} , \text{len} ) }\right\}$
		\EndFor
		\EndFor
		\State $W_B \leftarrow W_B \cup W_B^i$
		\EndFor
		\State \Return $W_A, W_B, f$
		\EndProcedure
	\end{algorithmic}
\end{algorithm}

\begin{fact} [Parameters for arbitrary $\lambda$ (Theorem~\ref{thm:formal_lcs})]\label{fact:parameters-cubic} \hfill

Let $\epsilon>0$ be a small constant and $\lambda$ is the relative length of the true LCS between $A$ and $B$.	
In order to make sure the loss in the approximation is small, we use $$\epsilon_0 := \epsilon \lambda.$$ 

	\begin{itemize}
		\item The total number of windows is equal to $k = |W_A|+|W_B| = \tO(n/d)$.
		\item The maximum size of the windows is equal to $\wmax = \Theta(d / \lambda)$.
		\item The minimum size of the windows is equal to $\wmin = \Theta(d)$.
		\item The ratio of the maximum window size over the minimum window size is bounded by $\wgap = \wmax / \wmin = \Theta(1/\lambda)$.
	    \item The number of different layers which is equal to the number of different window sizes is equal to $\wlayers = O(\log (1/\lambda) )$.
\end{itemize}
\end{fact}
\begin{proof}
We derive the number of windows (the rest of the parameters follow immediately from the construction). 
In each layer of $W_B$, the number of windows is 
\begin{gather*}|W_B^i|=O(t_i \cdot (d_i/d)) = O(n/d).\end{gather*} 
Therefore the total number of windows is \begin{gather*} |W_B| = O(fn/d) = \tO(n/d).\end{gather*} 
\end{proof}

\subsubsection{Up-to-constant-factor-optimality of window-compatible common subsequence}\label{sec:optimality-cubic}
We present another structural lemma for window-compatible common subsequences. 
It is similar to Lemma~\ref{lem:improved_structrual_lemma}, but loses a constant factor in length of common subsequence.

\begin{lemma}[Window-compatible structural lemma (arbitrary $\lambda$)]\label{lem:improved_structrual_lemma} \hfill

For any two strings $X,Y$ of length $n$, it is possible to map them into $A,B$ in one of the following ways: 
\begin{itemize}
\item $A=X,B=Y$ 
\item $A=Y,B=X$
\item $A=\rev(X), B = \rev(Y)$, or
\item $A=\rev(Y), B = \rev(X)$,
\end{itemize}
 such that the following holds:
If $\normal{\LCS(A,B)} \geq \lambda$, there exists a window-compatible common subsequence of $A,B$ of length at least $ \Omega(\lambda n) - 2\epsilon_0 n$ for the window sets constructed in Defi\-nition~\ref{def:cubic_levels_window}.
\end{lemma}
\begin{proof}
Fix any LCS of $X,Y$, and suppose we first try $A=X,B=Y$. 
For each $j$, let $A_j$ be the $j$-th window of $A$, and let $\wt{B}_j$ be the minimal substring of $B$ containing all characters common to $A_j$ and the LCS.

We will analyze a few different cases of $\wt{B}_j$; the trickiest case is when $\wt{B}_j$ and $\wt{B}_{j+1}$ are both contained in the same $W_B^0$-window of length $d$. Denote this window by $\wh{B}_{j,j+1}$. We claim that in this case the opposite cannot also be true: The minimal substring of $A$ containing all characters common to $\wh{B}_{j,j+1}$ and the LCS is not contained in any $W_A$-window --- this is because it intersects both $A_j$ and $A_{j+1}$.
We will throw out all such $j$'s; wlog their contribution is at most half of the LCS (otherwise we can consider the reverse assignment $A=Y,B=X$).

We also throw out all the $\wt{B}_{j}$'s such that $|\wt{B}_{j}| > d/(2\eps_0)$. Throwing out each such $\wt{B}_{j}$ can decrease the LCS by at most $d$, but removes $d/(2\eps_0)$ characters; hence the total loss to LCS is at most $2\eps_0 \cdot n$.

Finally, we further throw out either all the $\wt{B}_{j}$'s that are contained in some $W_B^0$-window of length $d$, or all the ones that aren't; again this step loses at most half of the LCS.

\noindent \textbf{Case 1: we keep all the $\wt{B}_{j}$'s that are contained in some $W_B^0$-window of length $d$.}
By the previous paragraph, each of them is the only $\wt{B}_{j}$ contained in that window, so we can just match $A_j$ to that window.

\noindent \textbf{Case 2: we keep $\wt{B}_{j}$'s that are not contained in any $W_B^0$-window of length $d$.}
For each $\wt{B}_{j}$, consider the index that is a multiple of $d \cdot 2^i$ for the largest possible $i$ (notice that this is unique). 
We truncate $\wt{B}_{j}$ to the left of this index; wlog the total contribution of truncated characters to the LCS is at most half (otherwise we reverse both strings). Each $\wt{B}_{j}$ is now contained in a disjoint $W_B^i$-window, so we can match $A_j$ to that window.

\end{proof}

\subsection{Dynamic Programming for window-compatible \textsf{LCS}}\label{sec:LCS_main_result}

\begin{algorithm}[!t]
	\begin{algorithmic}[1]\caption{Dynamic programming algorithm for block-based $\LCS$ problem}\label{alg:dp_LCS}
		\Procedure{\textsc{DP-LCS}}{$W_A,W_B,M$}\Comment{Lemma~\ref{lem:dp_LCS}}
		\State Note that $M$ is table s.t.~$M(k_1, k_2) \le \lcs(A_{k_1},B_{k_2})$.
		\State Note that $W_A$ and $W_B$ are sets of windows.
		\State
		\State Let $S_1$ denote the sorted list of right-indices of $W_A$-windows.
		\State Let $S_2$ denote the sorted list of right-indices of $W_B$-windows.
		\For{$i_1 \in S_1$}
		\For{$i_2 \in S_2$}
		\State $C[i_1][i_2] \leftarrow 0$
		\State $x_1 \leftarrow 0$
		\For{$A_{k_1}, B_{k_2}$ have right index $i_1,i_2$} \Comment{Case 1}
		\State $\text{left}_1 \leftarrow $ left index of $A_{k_1}$
		\State $\text{left}_2 \leftarrow $ left index of $B_{k_2}$
		\State $\text{tmp} \leftarrow C[\text{left}_1][\text{left}_2]  + M(A_{k_1}, B_{k_2})$
		\If{$\text{tmp} > x_1$}
		\State $x_1 \leftarrow \text{tmp}$
		\EndIf
		\EndFor
		\State Let $\text{prev}(i_1)$ denote the first index $\in S_1$ that is earlier than $i_1$
		\State Let $\text{prev}(i_2)$ denote the first index $\in S_2$ that is earlier than $i_2$
		\State $x_2 \leftarrow C[\text{prev}(i_1)][i_2] $ \Comment{Case 2}
		\State $x_3 \leftarrow C[i_1][\text{prev}(i_2)] $ \Comment{Case 3}
		\State $C[i_1][i_2] \leftarrow \max(x_1,x_2,x_3)$
		\EndFor
		\EndFor
		\EndProcedure
	\end{algorithmic}
\end{algorithm}

\begin{lemma}[Dynamic Programming for Longest Common Sequence]\label{lem:dp_LCS}
	Given two strings $A,B$ of length $n$, respective sets of windows $W_A, W_B$, and estimates-table $M$ on the pairwise LCS, Algorithm~\ref{alg:dp_LCS} computes the longest common subsequence of $A,B$ that is both:
(i) window-compatible, and (ii) on each pair of windows uses the common subsequence length from $M$.
	
	The algorithm runs in time $|W_A||W_B|$.
\end{lemma}

\begin{proof}
	The proof is based on a classic DP. Define $D[i][j]$ which stores the size of the solution ending at windows $w_i \in A$ and $w_j \in B$. We then can update the size of the solution for each pair in time $O(1)$ which gives us a solution in time $O(|W_A||W_B|)$.
\end{proof}

 %%% this is step 0
% !TeX root = ../../main.tex

\newpage
\section{\textsf{LCS} Step 2: Nearby Searching}\label{sec:step2}
The arguments of this section follow from the work of~\cite{cdgks18}. However, for the sake of completeness, we restate the theorems here.
This section is dedicated to presenting a method $\nonmetric(W_A,W_B)$ to be used for estimating the longest common subsequence. The output of this method is a matrix $\hat{M}: [|W_A|] \times [|W_B|] \rightarrow \mathbb{Z}^+$ where $\hat{M}[i][j]$ estimates the \textsf{LCS} of windows $w_i \in W_A$ and $w_j \in W_B$. We allow for errors in a few elements of $\hat{M}$ but we prove at the end of this section that with high probability, the error does not affect the solution significantly.

Our algorithm is pretty simple: We call a pair $(w_i, w_j)$ of windows \textit{underestimated} if their \textsf{LCS} is large enough (at least $\epsilon \lambda$ when normalizing the \textsf{LCS} size) but it is not computed within the desired approximation factor in Step 1.

 We are guaranteed by the bounds of Step 1 that the total number of underestimated pairs of windows is sublinear in the number of pairs. Thus, we show their count by $k^{2-\eta}$. Define $\mathcal{W} =  k^{\eta/2} \wmax$ and let two windows of the $A$ side or two windows of the $B$ side be \textit{nearby} if and only if the distance of the starting indices of the windows is bounded by $\mathcal{W}$. Fix $\epsilon_{\textsf{nbs}}$ to be a small error rate. In our algorithm, we sample the windows of the $A$ side with a rate $p = (10 \log n) k^{-\eta/2}/\epsilon_{\mathsf{nbs}}$ and discover all the underestimated pairs whose $A$ side is subsampled via a naive brute-force. We then recompute the \textsf{LCS} of each pair of windows $(w_i, w_j)$ such that $w_i \in W_A, w_j \in W_B$ and we detect an underestimated pair $(w_{i'}, w_{j'})$ such that $w_i$ and $w_{i'}$ are nearby and $w_j$ and $w_{j'}$ are also nearby. Finally, starting from the distance matrix provided in Step 1, we update the solution for each pair of windows that we compute their \textsf{LCS} from scratch and output it as matrix $\hat{M}$. We prove in Lemma~\ref{lemma:focs} that the error of our estimations is bounded with high probability.

\begin{algorithm}
	\begin{algorithmic}[1]\caption{Nearby Searching}\label{alg:nbs}
%		\Procedure{\textsc{QuadraticWindows}}{$A,B,n,d,\epsilon_0$} %\Comment{Lemma~\ref{lem:levels}}
%		\State \Comment{Generate $W_A$:}
		\State $p \leftarrow (10 \log n) k^{-\eta/2}/\epsilon_{\mathsf{nbs}}ß$
		\State $\mathcal{W} \leftarrow k^{\eta/2} \wmax$
		\State $\hat{M} \leftarrow M$
		\For{$w \in W_A$}
		\State \text{Skip the following commands with probability } $1-p$
		\For{$w' \in W_B$}
		\State Compute the \textsf{LCS} of $w$ and $w'$
		\If{$(w,w')$ is underestimated in $M$}
		\For{any window $w_1$ in $W_A$ which is nearby to $w$}
				\For{any window $w_2$ in $W_B$ which is nearby to $w'$}
				\State Update $\hat{M}$ by computing the \textsf{LCS} of $w_1$ and $w_2$ from scratch.
		\EndFor
		\EndFor
		\EndIf
		\EndFor
		\EndFor
		\State \Return $\hat{M}$
	\end{algorithmic}
\end{algorithm}

Let $M$ be the output of Step 1 of our algorithm and $\hat{M}$ be the output of Step 2. We refer to the size of the optimal \textsf{LCS} made by matrices $M$ and $\hat{M}$ by $\mathsf{LCS}(M)$ and $\mathsf{LCS}(M')$ respectively. We moreover, define a matrix $M'$ which is the same as $M$ except that the \textsf{LCS} values of all the underestimated pairs are corrected in $M'$. Similarly, we denote by $\mathsf{LCS}(M')$ the size of the longest common subsequence made by running DP on matrix $M'$. We prove below that the error between $\mathsf{LCS}(M')$ and $\mathsf{LCS}(\hat{M})$ is bounded.

\begin{lemma}[follows from \cite{cdgks18}]\label{lemma:focs}
	After running Algorithm~\ref{alg:nbs}, $\mathsf{LCS}(\hat{M}) \geq  \mathsf{LCS}(M') -  2n \epsilon_{\mathsf{nbs}}$ holds with probability at least $1-1/n^4$.
\end{lemma}
\begin{proof}	
It follows from the definition that the error can only come from the underestimated pairs of windows; for all other pairs, the corresponding entries of matrix $\hat{M}$ are at least as large as the respective entries of $M'$. We prove below that the error from underestimated pairs is bounded.

To this end, fix an optimal window-compatible solution with respect to $M'$ and let 
\begin{align*}
(x_1, y_1), (x_2, y_2), \ldots 
\end{align*}
be the first indices of the underestimated pairs of windows that contribute to this optimal solution. Let the pairs be sorted by their index. That is $x_1 < x_2 < x_3 < \ldots$ and $y_1 < y_2 < y_3 < \ldots$ hold. The proof is based on the following intuitive argument: with high probability, all but very few pairs of underestimated windows that contribute to the optimal window-compatible solution of $M'$ are detected in Algorithm~\ref{alg:nbs} and thus their values are corrected.

To see this, consider one pair $(x_i, y_i)$. This means that there is a window $w \in W_A$ and a window $w' \in W_B$ such that the starting index of $w$ is equal to $x_i$ and the starting index of $w'$ is equal to $y_i$. Moreover, the value of the pair $(w,w')$ is underestimated in $M$. Let $X$ be the set of all $(x_j,y_j)$'s in $(x_1, y_1), (x_2, y_2), \ldots$ such that $|x_j - x_i| \leq \mathcal{W}$ and $|y_j - y_i| \leq \mathcal{W}$. Notice that each elements of $X$ in addition to being an underestimated pair of windows also contributes to the optimal window-compatible solution of $M'$.  If  $|X| \geq \epsilon_{\mathsf{nbs}} k^{\eta/2}$, then with high probability the $A$-side window of one of these underestimated pairs will be sampled in Algorithm~\ref{alg:nbs} and therefore after detecting the underestimated pair, we compute the \textsf{LCS} of $(x_i, y_i)$ from scratch.  If $|X| < \epsilon_{\mathsf{nbs}} k^{\eta/2}$ then at least one of the following two constraints should hold: 
\begin{itemize}
	\item The number of $x_j$'s in $(x_1, y_1), (x_2, y_2), \ldots$ such that $x_i - \mathcal{W} \leq x_j \leq x_i$ is smaller than $\epsilon_{\mathsf{nbs}} k^{\eta/2}$.
	\item The number of $y_j$'s in $(x_1, y_1), (x_2, y_2), \ldots$ such that $y_i - \mathcal{W} \leq y_j \leq y_i$ is smaller than $\epsilon_{\mathsf{nbs}} k^{\eta/2}$.
\end{itemize}
Keep in mind that in the constrains above, each $(x_j, y_j)$ is an underestimated pair which contributes to the optimal solution of $M'$. If the former holds, we call $x_i$ \textit{lonely}. Similarly, if the latter holds, we call $y_i$ lonely. A simple argument implies that the number of lonely $x_i$'s (and similarly the number of lonely $y_i$'s) is bounded by $(\epsilon_{\mathsf{nbs}} k^{\eta/2})n/\mathcal{W}$: Divide the string into blocks of size $\mathcal{W}$. In each block, at most $\epsilon_{\mathsf{nbs}} k^{\eta/2}$ $x_j$'s are lonely. Thus, the total number of lonely $x_j$'s is bounded by $(\epsilon_{\mathsf{nbs}} k^{\eta/2})n/\mathcal{W}$. A similar argument also holds for the number of lonely $y_j$'s.

Therefore, the size of $|X|$ may be smaller than $\epsilon_{\mathsf{nbs}} k^{\eta/2}$ for at most $2(\epsilon_{\mathsf{nbs}} k^{\eta/2})n/\mathcal{W}$ underestimated pairs in $(x_1, y_1), (x_2, y_2), \ldots$. This implies that with high probability, the error of $\hat{M}$ in comparison to $M'$ is bounded by 

\begin{equation*}
\begin{split}
2\wmax(\epsilon_{\mathsf{nbs}} k^{\eta/2})n/\mathcal{W} & = 2\wmax(\epsilon_{\mathsf{nbs}} k^{\eta/2})n/(k^{\eta/2} \wmax)\\
&= 2n\epsilon_{\mathsf{nbs}}
\end{split}
\end{equation*}
\end{proof}

\begin{theorem}[nearby searching for $\Omega(\lambda^3)$ approximation]\label{lemma:focs1}
	Given that the number of underestimated pairs in Step 1 is bounded by $k^{2-\eta}$, for an arbitrary small constant $\epsilon > 0$, Step 2 takes time $\tilde O(\wmax^2 k^{2-\eta/2}/\lambda^6)$ and has approximation factor $1-\epsilon$ with probability at least $1-1/n^3$.
\end{theorem}
\begin{proof}
	To make sure the multiplicative factor in the approximation remains $1-\epsilon$, we set $\epsilon_{\mathsf{nbs}} = \Theta(\lambda^4)$. Thus, the runtime of the algorithm would be as follows: In the sampling process, we subsample each window with probability $p = (10 \log n) k^{-\eta/2}/\epsilon_{\mathsf{nbs}} = \tilde O(k^{-\eta/2} / \lambda^4)$. Therefore, the number of sampled windows would be $\tilde O(k^{1-\eta/2}/\lambda^4)$ with high probability and therefore the runtime for detecting the underestimated pairs of windows would be 
	\begin{align*}
		\tilde O(\wmax^2 k^{2-\eta/2}/\lambda^4).
	\end{align*}
	
	Since the total number of underestimated pairs is bounded by $k^{2-\eta}$ and we sample the $A$-side windows with probability $\tilde O(k^{-\eta/2} / \lambda^4)$, the expected number of underestimated pairs we detect in the first step is $\tilde O(k^{2-3\eta/2}/\lambda^4)$. Moreover, for each underestimated pair of windows that we detect, we recompute the \textsf{LCS} of all the nearby pairs of windows. The number of nearby windows on the $A$-side is $O(k^{\eta/2}/\lambda)$ and the number of nearby windows on the $B$ side is $\tilde O(k^{\eta/2}/\lambda)$. %\Saeed{This calculation needs a double-check.}
	This makes a total runtime of $\tilde O(\wmax^2 k^{2-\eta/2}/\lambda^6)$.
	
	Notice that the runtime of the algorithm is in expectation since it depends on the number of underestimated edges we detect in the sampling phase. To make the runtime fixed, we observe that with probability at least $1/2$, the algorithm terminates after $\tilde O(\wmax^2 k^{2-\eta/2}/\lambda^6)$ operations. Thus, we run the algorithm in parallel $10 \log n$ times and report the output of any instance that terminates before $\tilde O(\wmax^2 k^{2-\eta/2}/\lambda^6)$ operations. This comes with a constant factor overhead in the error rate of the algorithm which remains smaller than $1/n^3$.
\end{proof}

\begin{theorem}[nearby searching for $(1-\epsilon)\lambda^2$ approximation]\label{lemma:focs2}
	Given that the number of underestimated pairs in Step 1 is bounded by $k^{2-\eta}$, for an arbitrary small constant $\epsilon > 0$, Step 2 takes time $\tilde O(\wmax^2 k^{2-\eta/2}/\lambda^6)$ and has approximation factor $1-\epsilon$ with probability at least $1-1/n^3$.
\end{theorem}
\begin{proof}
	To make sure the multiplicative factor in the approximation remains $1-\epsilon$, we set $\epsilon_{\mathsf{nbs}} = \Theta(\lambda^3)$. Thus, the runtime of the algorithm would be as follows: In the sampling process, we subsample each window with probability $p = (10 \log n) k^{-\eta/2}/\epsilon_{\mathsf{nbs}} = \tilde O(k^{-\eta/2} / \lambda^3)$. Therefore, the number of sampled windows would be $\tilde O(k^{1-\eta/2}/\lambda^3)$ with high probability and therefore the runtime for detecting the underestimated pairs of windows would be 
	\begin{align*}
		\tilde O(\wmax^2 k^{2-\eta/2}/\lambda^3).
	\end{align*}
	
	Since the total number of underestimated pairs is bounded by $k^{2-\eta}$ and we sample the $A$-side windows with probability $\tilde O(k^{-\eta/2} / \lambda^3)$, the expected number of underestimated pairs we detect is $\tilde O(k^{2-3\eta/2}/\lambda^3)$. Moreover, for each underestimated pair of windows that we detect, we recompute the \textsf{LCS} of all the nearby pairs of windows. The number of nearby windows on the $A$-side is $O(k^{\eta/2}/\lambda)$ and the number of nearby windows on the $B$ side is $\tilde O(k^{\eta/2}/\lambda^2)$. %\Saeed{This calculation needs a double-check.}
	This makes a total runtime of 
	\begin{align*}
	\tilde O(\wmax^2 k^{2-\eta/2}/\lambda^6).
	\end{align*}
	
	Similar to Theorem~\ref{lemma:focs1}, one can make sure the runtime of the algorithm is 
	\begin{align*}
	\tilde O(\wmax^2 k^{2-\eta/2}/\lambda^6)
	\end{align*}
	and the error remains bounded by $1/n^3$.
\end{proof} %%% this is step 2
% !TeX root = ../../main.tex

\newpage
\section{Longest Increasing Subsequence}\label{sec:lis}
%\
We outlined the algorithm in Section \ref{sec:tech_lis}. Here, we bring the details for each step of the algorithm. In Section \ref{sec:domains} we discuss the solution domains and show how we construct them. Next, in Section \ref{sec:lis_pseudo_solution} we discuss the details of constructing pseudo-solutions and finally in Section \ref{sec:eval} we show how we can obtain an approximate solution from the pseudo-solutions. Also, in Section \ref{sec:lis_extension} we bring an improvement to the running time at the expense of having a larger approximation factor for the algorithm.
\subsection{Solution Domains}\label{sec:domains}
We assume from now on that $\lis(A) \geq n \lambda$ holds. As mentioned earlier, we divide the input array into $\sqrt{n}$ subarrays of size $\sqrt{n}$ and denote them by\\ $\sa_1, \sa_2, \ldots, \sa_{\sqrt{n}}$. For a fixed optimal solution $\opt$, our goal is to approximate the smallest and the largest number of each subarray that contribute to $\opt$. Let us refer to these numbers as the \textit{domain} of each subarray. Let $\epsilon := 1/1000$ be the accuracy parameter. For a subarray $\sa_i$, we sample $k$ (will be decided later) different elements and refer to them by $a_{j_1}, a_{j_2}, \ldots, a_{j_k}$.

\begin{algorithm}[h!]
\begin{algorithmic}[1]\caption{Constructing the candidate domains}\label{alg:construct_candidate_domains}
\Procedure{ConstructCandidateDomains}{$\sa_i$} \Comment{Lemma~\ref{lem:construct_candidate_domains}}
	\State \Comment{Given random access to a subarray $\sa_i$}
	\State $k \leftarrow 20 \log (1/\delta) / ( \lambda \epsilon^2 )$
	\State Sample $k$ elements from $\sa_i$, and denote the sampled elements by $a_{j_1}, a_{j_2}, \ldots, a_{j_k}$
	\For{ $\alpha$ in $[k]$ }
		\For{$\beta$ in $[k]$}
			\State If $a_{j_\alpha} \leq  a_{j_\beta}$, then construct a candidate domain $[a_{j_\alpha}, a_{j_\beta}]$
		\EndFor
	\EndFor
	\State \Return all the constructed candidate domains
\EndProcedure
\end{algorithmic}
\end{algorithm}

We first prove that,
\begin{lemma}[constructing candidate domains]\label{lem:construct_candidate_domains}
	Let $\lambda \in (0,1)$,  $\epsilon \in (0,1/2)$ and $\delta \in (0,1/10)$.
	Let $\sa_i$ be a length-$\sqrt{n}$ subarray whose contribution to the optimal solution is at least $\epsilon \sqrt{n} \lambda$, i.e., $\lis^{ [\ssmall(\sa_i) , \llarge (\sa_i) ] }( \sa_i ) \geq \epsilon \sqrt{n} \lambda$. If we uniformly sample $k = 20 \log(1/\delta) / ( \lambda \epsilon^2 ) $ elements $a_{j_1}, a_{j_2}, \ldots, a_{j_k}$ from $\sa_i$, then with probability at least $1-\delta$, there exists a pair $(\alpha,\beta) \in [k] \times [k]$ such that the following two conditions hold \\
	\begin{enumerate}
	\item  $\ssmall(\sa_i) \leq a_{j_\alpha} \leq a_{j_\beta} \leq \llarge(\sa_i)$,
	\item $\lis^{[a_{j_\alpha}, a_{j_\beta}]}(\sa_i) \geq (1-\epsilon) \lis^{[\ssmall(\sa_i), \llarge(\sa_i)]}(\sa_i)$. 
	\end{enumerate}
\end{lemma}
\begin{proof}
At least $\epsilon \sqrt{n} \lambda$ elements of $\sa_i$ appear in $\opt$. Let us put all these elements in an array $\mathsf{b}$ in the same order that they appear in $\sa_i$. Then it is obvious that $\mathsf{b}$ has at least $\epsilon \sqrt{n} \lambda $ elements. %
To prove the lemma, we bound the probability that none of the first $\epsilon/2$ fraction of the elements of $\mathsf{b}$ are sampled in our algorithm. % 
\begin{align*}
& ~ \Pr[\text{none of the elements in the first $\epsilon/2$ fraction of $\mathsf{b}$ is sampled}] \\ 
\leq & ~ \left( 1 - \frac{\epsilon}{2} \cdot \epsilon \sqrt{n} \lambda\cdot \frac{1}{\sqrt{n}} \right)^{k} \\
= & ~ \left( 1 - \frac{\epsilon^2 \lambda}{2}  \right)^{ \frac{2}{\epsilon^2   \lambda} \cdot 10 \log (1/\delta) } \\
\leq & ~ e^{ - 10 \log (1/\delta) } \\
\leq & ~ \delta/2,
\end{align*} 
where 
the first step follows from the fact that $\mathsf{b}$ contains at least $\epsilon \lambda \sqrt{n} $ elements,
the second step follows from $k = 20 \log(1/\delta) / ( \lambda \epsilon^2 )$, 
and the third step follows from the fact that $(1-1/x)^x \leq 1/e$ for $\forall x \geq 4$. % 
Hence, with probability at least $1-\delta/2$,  at least one of the elements in the first $\epsilon/2$ fraction of $\mathsf{b}$ are sampled.
	
With the same analysis, one can prove that with probability at least $1-\delta/2$ at least one of the elements in the last $\epsilon/2$ fraction of $\mathsf{b}$ are also sampled.

Taking a union bound of two events, with probability at least $1-\delta$, at least one of the elements in the first and at least one of the elements in the last $\epsilon/2$ fraction of $\mathsf{b}$ are sampled. 

 Therefore, the $\lis$ of $\sa_i$ subject to this interval is at least a $1-\epsilon$ fraction of\\ $\lis^{[\ssmall(\sa_i),\llarge(\sa_i)]}(\sa_i)$.
	
\end{proof}

Notice that the average contribution of each subarray to $\opt$ is $\sqrt{n}\lambda$ and Lemma \ref{lem:construct_candidate_domains} applies to a subarray if its contribution to $\opt$ is at least an $\epsilon$ fraction of this value. Therefore Lemma \ref{lem:construct_candidate_domains} implies that a considerable fraction of the solution is covered by the candidate domains.

\begin{corollary}[existence of a desirable solution]\label{cor:existence_of_good_solution}
Let $\lambda \in (0,1)$ such that $\lis(A) \geq n\lambda$ and $\eps \in (0, 1/4)$.
If we run Algorithm~\ref{alg:construct_candidate_domains}  with parameter $\delta = \eps$ on every subarray independently, then with probability at least $1-\exp(-\Omega( \epsilon^2 \sqrt{ n } \lambda ) )$,
there exists a set $T \subseteq [\sqrt{n}]$ and elements $\alpha_i$ and $\beta_i$ sampled from $\sa_i$ for each $i \in T$ such that the following conditions hold:
\begin{enumerate}
\item For any $i \in T$, $\alpha_i \leq \beta_i$.
\item For any $i, j \in T$ satisfying $i < j$, $\beta_i < \alpha_j$.
\item $\sum_{ i \in T } \lis^{[{\alpha_i},{\beta_i}]} (\sa_i) \geq (1-4\epsilon) \lis(A)$.
\end{enumerate}
\end{corollary}
\begin{proof}

Lemma \ref{lem:construct_candidate_domains} holds for all subarrays whose contribution to $\opt$ is at least $\epsilon \sqrt{n} \lambda$. Let $S \subseteq [\sqrt{n}]$ denote the set of coordinates such that for each $i \in S$
\begin{align*}
 \lis^{ [\ssmall (\sa_i) , \llarge(\sa_i) ]} ( \sa_i )  \geq \epsilon \sqrt{n} \lambda.
\end{align*}
Since $\sum_{i=1}^{\sqrt{n}}  \lis^{ [\ssmall (\sa_i) , \llarge(\sa_i)] } ( \sa_i ) %
= \lis(A) $ and $ \lis^{[ \ssmall (\sa_i) , \llarge(\sa_i) ]} ( \sa_i )  \leq \sqrt{n} $, we have

\begin{equation}\label{equ:solution_domain_2} \sum_{i \in S} \lis^{ [\ssmall (\sa_i) , \llarge(\sa_i)] } ( \sa_i ) \geq  \sum_{i=1}^{\sqrt{n}}  \lis^{ [\ssmall (\sa_i) , \llarge(\sa_i)] } ( \sa_i ) - \sqrt{n}\cdot \eps\sqrt{n}\lambda \geq \lis(A) - \eps n \lambda. \end{equation}

Let $T \subseteq S$ denote the set of coordinates such that for each $i \in T$, 
\begin{align*}
 \lis^{[\alpha_i,\beta_i]} (\sa_i)  \geq (1-\epsilon)  \lis^{[ \ssmall (\sa_i) , \llarge(\sa_i) ]} ( \sa_i ) , \text{~~~and~~~}  \lis^{[ \ssmall (\sa_i) , \llarge(\sa_i) ]} ( \sa_i )  \geq \epsilon \sqrt{n} \lambda.
\end{align*}
Now we show that with probability at least $1 -  \exp( - \Omega(\epsilon^2 \sqrt{n} \lambda ))$, 
\begin{equation}\label{equ:solution_domain_2a}
\sum_{i\in T} \lis^{[ \ssmall (\sa_i) , \llarge(\sa_i)]} (\sa_i)  \geq (1 - 2\epsilon) \sum_{i\in S} \lis^{[ \ssmall (\sa_i) , \llarge(\sa_i)]} (\sa_i).
\end{equation}
For each $i \in S$, let $X_i$ denote a random variable such that
\begin{align*}
X_i 
= 
\begin{cases}
\lis^{[ \ssmall (\sa_i) , \llarge(\sa_i)]} (\sa_i), & \text{~with~probability of } i \in T ; \\
0, & \text{~with~probability of } i \notin T,
\end{cases}
\end{align*}
and $X = \sum_{i \in S} X_i$.
By Lemma \ref{lem:construct_candidate_domains} (with $\delta = \eps$),
We have $$\E[X] \geq (1 - \epsilon) \sum_{i\in S} \lis^{[ \ssmall (\sa_i) , \llarge(\sa_i)]} (\sa_i).$$
By Hoeffding bound (Theorem~\ref{thm:hoeffding}),  
\begin{align*}
\Pr[  X - \E[X]  \geq \epsilon \E[X] ] \leq & ~ 2 \exp \left( - \frac{2 \epsilon^2 (\E[X])^2}{ \sum_{i \in S}  \left(\lis^{[ \ssmall (\sa_i) , \llarge(\sa_i)]}(\sa_i) \right)^2 } \right) \\
\leq & ~ 2 \exp\left(- \frac{2\epsilon^2 (1 - \epsilon)^2 \left(\sum_{i\in S} \lis^{[ \ssmall (\sa_i) , \llarge(\sa_i)]} (\sa_i)\right)^2}{ n^{3/2} }\right) \\
\leq & ~ 2 \exp( - 2 \epsilon^2 (1 - \epsilon)^4 \sqrt{n}\lambda) \\
\leq & ~ \exp( - \Omega(\epsilon^2 \sqrt{n} \lambda )).
\end{align*}
Hence, Equation~\eqref{equ:solution_domain_2a} holds with probability at least $1 -  \exp( - \Omega(\epsilon^2 \sqrt{n} \lambda ))$.

Conditioned on Equation~\eqref{equ:solution_domain_2a}, 
we have 
\begin{equation}\label{equ:solution_domain_3}
\begin{split}
 \sum_{i\in T}  \lis^{[\alpha_i,a_{\beta_i}]} (\sa_i)  
\geq & ~ \sum_{i\in T} (1-\epsilon)  \lis^{[ \ssmall (\sa_i) , \llarge(\sa_i) ]} ( \sa_i )  \\
\geq & ~ (1-\epsilon) (1-2\epsilon)  \sum_{i \in S}  \lis^{[ \ssmall (\sa_i) , \llarge(\sa_i) ]} ( \sa_i ) \\
\geq & ~ (1-\epsilon)(1-2\epsilon)  ( \lis(A) -  \epsilon n \lambda ) \\
\geq & ~ (1-4\epsilon) \lis(A)
\end{split}
\end{equation}
where the first inequality follows from the definition of $T$, the second inequality follows from Equation~\eqref{equ:solution_domain_2a}, the third inequality follows from Equation~\eqref{equ:solution_domain_2} and the last inequality follows from $ \lis(A) \geq n  \lambda $.

Finally, by Equation~\eqref{equ:solution_domain_2a} and \eqref{equ:solution_domain_3}, we have 
\begin{align*}
\Pr \left[ \sum_{i\in T}  \lis^{[\alpha_i,a_{\beta_i}]} (\sa_i)  \geq (1-4\epsilon) \lis(A) \right] \geq 1 - \exp( - \Omega ( \epsilon^2 \sqrt{n} \lambda) ).
\end{align*}
\end{proof}

\subsection{Constructing Approximately Optimal Pseudo-solutions}\label{sec:lis_pseudo_solution}
We call a sequence of $\sqrt{n}$  intervals $[\ell_1, r_1], [\ell_2, r_2], \ldots, [\ell_{\sqrt{n}}, r_{\sqrt{n}}]$ a pseudo-solution if all of the intervals are monotone. That is $\ell_1 \leq r_1 < \ell_2 \leq r_2 < \ell_3 \leq r_3 \leq \ldots \leq \ell_{\sqrt{n}} \leq r_{\sqrt{n}}$. These intervals denote solution-domains for the subarrays. We also may decide not to assign any solution domain to a subarray in which case we show the corresponding interval by $\emptyset$. We define monotonicity of a pseudo-solution such that it is not violated by $\emptyset$.  The quality of a pseudo-solution is defined as $\sum_i \lis^{[\ell_i, r_i]} (\sa_i)$. We denote the quality of a pseudo-solution $\ps$ by $\quality(\ps)$.

Another way to interpret Corollary \ref{cor:existence_of_good_solution} is that one can construct a pseudo-solution using the sampled elements whose quality is at least a $1-\epsilon$ fraction of $\lis(A)$. In this section, we present an algorithm to construct a small set of pseudo-solutions with the promise that at least one of them has a quality of at least $\lis(A)/t$, where $t$ is the number of pseudo-solutions. Finally, in Section \ref{sec:eval}, we present a method to approximate the size of the optimal solution using pseudo-solutions.

We construct the pseudo-solutions via Algorithm \ref{alg:construct_pseudo_solutions}. 
The input of Algorithm \ref{alg:construct_pseudo_solutions} is the set of  candidate domain intervals obtained by Algorithm \ref{alg:construct_candidate_domains} on every subarray.
We first find an assignment of candidate solution domains to the subarrays which is monotone and has the largest number of candidate domain intervals. (This step can be implemented by dynamic programming or solved with an algorithm similar to activity selection algorithm.)  We make a pseudo-solution out of this assignment and update the set of candidate intervals by removing the ones which are used in our pseudo-solution. We then repeat the same procedure to construct the second pseudo-solution and update the candidate solution domains accordingly. We continue on, until the number of solution domains used in our pseudo-solution drops below $\epsilon \lambda \sqrt{n}$ in which case we stop.

\begin{algorithm}[t!]
\begin{algorithmic}[1]\caption{Constructing the pseudo solutions}\label{alg:construct_pseudo_solutions}
\Procedure{ConstructPseudoSolutions}{$\cdi_1, \ldots, \cdi_{\sqrt{n}}$} \Comment{Lemma~\ref{lem:construct_pseudo_solutions_size},\ref{lem:construct_pseudo_solutions_quality},\ref{lem:construct_pseudo_solutions_time}}
	\State \Comment{$\{ \cdi_{i} \}_{ i \in [ \sqrt{n} ] } $ is $\sqrt{n}$ sets of candidate domain intervals }
	\State $\textsf{pseudo-solutions} \leftarrow \emptyset$
	\While{ {\bf true} }
		 \State $\mathsf{assg} \leftarrow$ 
		  largest assigment of candidate domain intervals to subarrays which is monotone 
		 \If{$\mathsf{assg}$ contains less than $\epsilon \sqrt{n} \lambda$ non-empty candidate domain intervals}
		 	 \State \textbf{break}
		 \Else
			 \State Add $\mathsf{assg}$ to $\textsf{pseudo-solutions}$\;
			 \For{$i \leftarrow 1$ to $\sqrt{n}$}
			 	 \If{ $\mathsf{assg}$ contains a candidate domain interval for subarray $\sa_i$ }
			 	 	\State remove the corresponding candidate domain interval from $\cdi_i$
			 	 \EndIf
			 \EndFor
		\EndIf
	\EndWhile
	\State \Return pseudo-solutions $\ps_1, \ps_2, \ldots, \ps_{t} $
\EndProcedure
\end{algorithmic}
\end{algorithm}

We first prove in Lemma \ref{lem:construct_pseudo_solutions_size} that the number of pseudo-solutions constructed in Algorithm \ref{alg:construct_pseudo_solutions} is bounded by $O( k^2  / ( \lambda \epsilon ) )$. Next, we show in Lemma \ref{lem:construct_pseudo_solutions_quality} that at least one of the pseudo-solutions constructed by Algorithm \ref{alg:construct_pseudo_solutions} has a quality of at least $ \Omega( \lis(A) / t )$ where $t$ is the number of pseudo-solutions. Finally we prove in Lemma \ref{lem:construct_pseudo_solutions_time} that the running time of Algorithm \ref{alg:construct_pseudo_solutions} is $O(t k^2 \sqrt{n} \log n)$.

\begin{lemma}[number of pseudo-solutions]\label{lem:construct_pseudo_solutions_size}
 For each $i \in [\sqrt{n}]$, 
let $ \cdi_i $ be a set of at most $k^2$ candidate domain intervals. Let $t$ denote the number of pseudo-solutions constructed in Algorithm \ref{alg:construct_pseudo_solutions}. Then, we have $t \leq  k^2 / ( \lambda \epsilon ) $.
\end{lemma}
\begin{proof}
	Note that for each subarray we have at most $k^2$ candidate domain intervals. Since there are $\sqrt{n}$ subarrays, then in total we have $\sqrt{n} k^2$ candidate domain intervals. Each time we construct a pseudo-solution, the total number of the candidate domain intervals is decreased by at least $\epsilon \sqrt{n} \lambda$. Thus, the total number of pseudo-solutions $t$ can be upper bounded,
	\begin{align*}
	t \leq \frac{ \sqrt{n} k^2 }{ \epsilon \sqrt{n} \lambda }  \leq \frac{ k^2  }{  \lambda \epsilon }.
	\end{align*}
\end{proof}

\begin{lemma}[quality of pseudo-solutions]\label{lem:construct_pseudo_solutions_quality}
	Let $\ps_1, \ps_2, \ldots, \ps_{t}$ be the pseudo-solutions constructed by Algorithm \ref{alg:construct_pseudo_solutions}. If $\lis(A) \geq n \lambda$ holds, then with probability at least $1-\exp(-\Omega( \sqrt{ n } \lambda ) )$, there exists an $i \in [t]$ such that
	\begin{align*}
	\quality(\ps_i) \geq \frac{\lis(A)}{2 t}.
	\end{align*}
\end{lemma}
\begin{proof}
Let us focus again on the actual solution domains 
\begin{align*}
[\ssmall(\sa_1), \llarge(\sa_1)],[\ssmall(\sa_2), \llarge(\sa_2)], \ldots, [\ssmall(\sa_{\sqrt{n}}), \llarge(\sa_{\sqrt{n}})].
\end{align*}

We define set $S \subseteq [\sqrt{n}]$ such that
\begin{align*}
 \lis^{ [ \ssmall (\sa_i) , \llarge (\sa_i) ] }  \geq \epsilon \sqrt{n} \lambda, \forall i \in S.
\end{align*}

Using Corollary~\ref{cor:existence_of_good_solution} with $\eps \leq 1/10$, we know that there is a monotone pseudo-solution $[\alpha_i, \beta_i]_{i \in T}$ ($T \subseteq S$) such that $[\alpha_i, \beta_i]$ are candidate domain intervals and 
\begin{align*}
\sum_{i \in T}  \lis^{ [ \alpha_i , \beta_i ] } ( \sa_i )  \geq (1 - 4\epsilon)  \lis(A) .
\end{align*}
Denote this pseudo-solution as $\mathsf{sol}$.
At the time we terminate Algorithm \ref{alg:construct_pseudo_solutions}, 
there are at most $\epsilon \sqrt{n}\lambda$ candidate domain intervals of $\mathsf{sol}$ that do not belongs to any pseudo-solution of the pseudo-solution set. 
Also, since each candidate domain interval contributes at most $\sqrt{n}$ to the quality of the pseudo-solution containing the interval, we have  
\begin{align*}
\sum_{i=1}^t \quality  ( \ps_i )  \geq ( 1 -  4\epsilon )\lis(A) - \epsilon \sqrt{n} \lambda \cdot \sqrt{n} \geq ( 1 - 4 \epsilon )\lis(A) - \epsilon  \lis(A)  = ( 1 - 5 \epsilon ) \lis(A).
\end{align*} 
Thus, there exists an $i\in [t]$ such that
\begin{align*}
\quality  ( \ps_i ) \geq ( 1 - 5 \epsilon ) \lis(A) / t \geq  \lis(A)  / (2t).
\end{align*}
\end{proof}

\begin{lemma}[running time]\label{lem:construct_pseudo_solutions_time}
For each $i \in [\sqrt{n}]$, let $ \cdi_i $ be a set of at most $k^2$ candidate domain intervals.  Let $t$ denote the number of pseudo-solutions. The running time of Algorithm \ref{alg:construct_pseudo_solutions} is bounded by $O(t k^2 \sqrt{n} \log n)$.
\end{lemma}
\begin{proof}
	Lemma \ref{lem:construct_pseudo_solutions_size} states that Algorithm \ref{alg:construct_pseudo_solutions} terminates after constructing $t$\\ pseudo-solutions. Now we show that constructing each pseudo-solution takes time\\ $O(k^2 \sqrt{n} \log n)$.
	Our solution is based on a dynamic programming technique. Let $D: [\sqrt{n}] \times [k^2] \rightarrow \mathbb{N}$ be an array such that $D[i][j]$ stores the size of the largest monotone pseudo-solution for the first $i$ subarrays which ends with the $j$'th candidate domain interval of $\sa_i$. Using classic segment-tree data structure (this data structure can be found in many textbooks, e.g. \cite{clrs09}), %
	one can compute the value of $D[i][j]$ in time $O(\log n)$ from the previously computed elements of the array. 

	Thus, the total running is bounded by $O(t k^2 \sqrt{n} \log n)$.
\end{proof}

\subsection{Evaluating the Pseudo-solutions}\label{sec:eval}
We finally use a concentration bound to show that the quality of a pseudo-solution can be approximated well by sampling a small number of subarrays. Since a pseudo-solution specifies the range of the numbers used in every subarray, the quality of a pseudo-solution, or in other words, the size of the corresponding increasing subsequence of a pseudo-solution can be formulated as 
\begin{align*}
\quality(\ps) := \sum_{i=1}^{\sqrt{n}} \lis^{[\ell_i, r_i]}(\sa_i)
\end{align*}
where $[\ell_i, r_i]$ denotes the corresponding solution domain of $\ps$ for $\sa_i$.

\begin{algorithm}[t!]
\begin{algorithmic}[1]\caption{Evaluate the pseudo solutions}\label{alg:evaluate_pseudo_solutions}
\Procedure{EvaluatePseudoSolutions}{$\ps_1, \ldots, \ps_{t}$} \Comment{Lemma~\ref{lemma:estimate}}
	\State \Comment{$\{ \ps_{i} \}_{ i \in [t] } $ is a set of pseudo solutions }
	\State $p \leftarrow \frac{1000 t \log^4  n }{\epsilon^4 \lambda\sqrt{n}}$ 
	\State Randomly sample each $i \in [\sqrt{n}]$ with probability p, and put all the samples in a set $W$
	\For{each $\ps_j$} 
	\State $\tilde{\quality}(\ps_j) \leftarrow 0$
	\For{each interval $[\ell_i, r_i]$ in $\ps_j$}
		 \If{$ i \in W$}
		 	 \State $\tilde{\quality}(\ps_j)  \leftarrow \tilde{\quality}(\ps_j) + \lis^{[\ell_i, r_i]}(\sa_i) / p$
		\EndIf
	\EndFor \EndFor
	\State \Return largest $\tilde{\quality}(\ps_j)$ for all $j \in [t]$ %pseudo-solutions $\ps_1, \ps_2, \ldots, \ps_{t} $
\EndProcedure
\end{algorithmic}
\end{algorithm}

In Lemma \ref{lemma:estimate}, we prove that by sampling $O(\log^{O(1)} n  / \lambda^4)$ many subarrays and computing $\lis^{[\ell_i, r_i]}(\sa_i)$ for them, one can approximate the quality of a pseudo-solution pretty accurately.

\begin{lemma}[the quality of pseudo-solution]\label{lemma:estimate}
Let $\lambda \in (0,1)$ and $\eps$ be a constant in $(0, 1/100)$.
Let $\ps_1, \ps_2, \cdots, \ps_t$ be a set of $t$ pseudo-solutions. 
With probability at least $1- \exp(- \Omega(\log^2 n))$,
Algorithm~\ref{alg:evaluate_pseudo_solutions} runs in time $O( t^2 \sqrt{n} \log^{O(1)} n /\lambda)$  such that,
\begin{enumerate}
\item If there exists an $i \in [t]$, $\quality(\ps_i) \geq \frac{\lambda n }{ 2t} $, then the algorithm outputs an estimation at least $\frac{\lambda n}{4t}$.
\item If $\quality(\ps_i) < \frac{\lambda n }{ 8t} $ for all $i \in [t]$,  then the algorithm outputs an estimation smaller than $\frac{\lambda n}{4t}$.
\end{enumerate}
\end{lemma}
\begin{proof}
We iterate over all $t$ pseudosolutions, and for each one we estimate their quality separately. In the end we output the largest estimated value over all pseudosolutions. More precisely, we define $p = \frac{1000 t \log^4  n }{\epsilon^4 \lambda\sqrt{n}}$ and for each pseudosolution, we sample each of its subarrays with probability $p$. We then compute the \textsf{LIS} of the sampled subarrays subject to the range which corresponds to the pseudosolution. We then estimate the quality of the pseudosolution by scaling the sum of \textsf{LIS}'s by a factor $1/p$. In what follows, we prove that the estimation error is negligible.

By Chernoff bound, for each pseudosolution, with probability $1 - \exp(-\Omega(\log^3 n))$ at most $2p\sqrt{n} = \frac{2000t \log^4 n}{\epsilon^4 \lambda}$ subarrays are sampled. 
For each pseudo-solution, Algorithm~\ref{alg:evaluate_pseudo_solutions} computes the value of longest increasing subsequence subject to the corresponding range once for every sampled subarray.
Hence, the total running time of the algorithm is $O(t^2 \sqrt{n}\log^{O(1)} n / \lambda)$.

Consider an arbitrary pseudo-solution $\ps \in \{\ps_1, \dots, \ps_t\}$ and let $\tilde{\quality}(\ps)$ denote its estimated \textsf{LIS}. 
We show that %
\begin{align}\label{equ:estimation_lis}
\Pr\left[\left(1 - \frac{\epsilon}{4}\right)^2 \left( \quality(\ps) -\frac{ \epsilon\lambda^4 n}{100t}\right) \leq \tilde \quality(\ps)  \leq \left(1 + \frac{\epsilon}{4}\right)^2 \quality(\ps) + \frac{\epsilon \lambda^4 n}{100 t} 
\right]  \leq  1 - \exp(-\Omega(\log^3 n)).\end{align}
Then the lemma holds by a union bound on all the pseudo-solutions.

Let $T$ be the set of subarray indices such that $i \in T$ iff there is a non-empty interval $[\ell_i, r_i]$ of $\ps$ corresponding to subarray $\sa_i$. 
Let $p = \frac{1000 t \log^4  n }{\epsilon^4 \lambda\sqrt{n}}$ be  the probability of sampling a subarray.  For each $i \in T$, let $X_i$ denote a random variable such that
\begin{align*}
X_i 
= 
\begin{cases}
1, & \text{~with~prob.~} p ; \\
0, & \text{~with~prob.~} 1-p .
\end{cases}
\end{align*}
and \[X =\sum_{i \in T} \frac{1}{p} \lis^{[\ell_i, r_i]}(\sa_i) X_i.\] 
We have 
\begin{align*}
\E[X] = \E \left[ \sum_{i \in T}  \frac{1}{p} \lis^{[\ell_i, r_i]}(\sa_i) X_i \right] = \sum_{i \in T}  \lis^{[\ell_i, r_i]}(\sa_i) = \quality (\ps).
\end{align*}
Let \[T_j = \left\{i \in T: (1+\epsilon / 4)^{j-1} \leq \lis^{[\ell_i, r_i]}(\sa_i) < (1 + \epsilon / 4)^j\right\}\]
for integer $1 \leq j \in \left\lceil \log_{1+\epsilon} \sqrt{n} \right\rceil $.

Let $\Delta = \frac{\epsilon^2 \lambda \sqrt{n}}{1000 t \log n}$.
Consider $T_j$'s such that $|T_j| \geq \Delta$.
The contribution of $\quality(\ps)$ mostly comes from $T_j$ in the following sense 
\begin{equation}\label{equ:lis_evaluation_1}\sum_{j : |T_j| \geq  \Delta} \sum_{i \in T_j} \lis^{[\ell_i, r_i]}(\sa_i) \leq \quality(\ps)\end{equation}
and 
\begin{equation}\label{equ:lis_evaluation_2}\begin{split}
\sum_{j : |T_j| \geq  \Delta} \sum_{i \in T_j} \lis^{[\ell_i, r_i]}(\sa_i) = & \quality(\ps) -\sum_{j : |T_j| <  \Delta} \sum_{i \in T_j}  \lis^{[\ell_i, r_i]}(\sa_i)\\
\geq & \quality(\ps) - \Delta \sqrt{n} \cdot  \left\lceil \log_{1+\epsilon} \sqrt{n} \right\rceil \\
\geq & \quality(\ps)  - \frac{\epsilon \lambda n}{100t}.
\end{split}
\end{equation}
Now we bound the random variable $\sum_{j : |T_j| \geq  \Delta} \sum_{i \in T_j} \lis^{[\ell_i, r_i]}(\sa_i) X_i / p$.
By Chernoff bound, for each $j$ such that $|T_j| \geq \Delta$, 
we have 
\begin{equation}\label{equ:lis_evaluation_3}
\Pr\left[\left(1 - \frac{\epsilon}{4}\right) p|T_j|\leq \sum_{i \in T_j} X_i \leq \left(1 + \frac{\epsilon}{4}\right)p|T_j|\right] \geq 1 - \exp(-\Omega(\epsilon^2 p |T_j|)) = 1 - \exp(-\Omega(\log^3 n)).
\end{equation}
Notice that for set $T_j$, 
\begin{equation}\label{equ:lis_evaluation_4}  \frac{(1 + \epsilon/4)^{j - 1}}{p} \sum_{i \in T_j}X_i \leq \sum_{i \in T_j} \frac{1}{p} \lis^{[\ell_i, r_i]}(\sa_i) X_i \leq \frac{(1 + \epsilon/4)^j}{p} \sum_{i \in T_j}X_i.\end{equation}
We have 
\begin{equation}\label{equ:lis_evaluation_5} \begin{split}
& \Pr\left[ \left(1 - \frac{\epsilon}{4}\right)^2  \sum_{i \in T_j}  \lis^{[\ell_i, r_i]}(\sa_i) \leq \sum_{i \in T_j}  \frac{1}{p} \lis^{[\ell_i, r_i]}(\sa_i) X_i \leq \left(1 + \frac{\epsilon}{4}\right)^2  \sum_{i \in T_j}  \lis^{[\ell_i, r_i]}(\sa_i)\right]\\
 \geq & \Pr\left[
  \left(1 - \frac{\epsilon}{4}\right)^2  \sum_{i \in T_j}  \lis^{[\ell_i, r_i]}(\sa_i) \leq \frac{(1 + \epsilon/4)^{j-1}}{p} \sum_{i \in T_j}X_i \right. \\
  & \left.  \ \ \ \ \ \ \ \ \    \bigwedge
 \frac{(1 + \epsilon/4)^j}{p} \sum_{i \in T_j}X_i \leq \left(1 + \frac{\epsilon}{4}\right)^2  \sum_{i \in T_j}  \lis^{[\ell_i, r_i]}(\sa_i)\right]\\
 \geq & \Pr\left[
  \left(1 - \frac{\epsilon}{4}\right)^2  \left(1 + \frac{\epsilon}{4}\right)^j |T_j| \leq \frac{(1 + \epsilon/4)^{j-1}}{p} \sum_{i \in T_j}X_i  \right.\\
  & \left.  \ \ \ \ \ \ \ \ \  \bigwedge \ \ 
 \frac{(1 + \epsilon/4)^j}{p} \sum_{i \in T_j}X_i \leq \left(1 + \frac{\epsilon}{4}\right)^{j+1} |T_j|\right]\\
 \geq & \Pr\left[
  \left(1 - \frac{\epsilon}{4}\right) p |T_j| \leq  \sum_{i \in T_j}X_i  \ \ \bigwedge \ \ 
 \sum_{i \in T_j}X_i \leq \left(1 + \frac{\epsilon}{4}\right) p |T_j|\right]\\
 = & 1 - \exp(-\Omega(\log^3 n)),
\end{split}
\end{equation}
where the first inequality follows from Equation~\eqref{equ:lis_evaluation_4}, the second inequality follows from the definition of $T_j$ and the last inequality follows from Equation~\eqref{equ:lis_evaluation_3}.
By Equation~\eqref{equ:lis_evaluation_1}, \eqref{equ:lis_evaluation_2}, \eqref{equ:lis_evaluation_6} and union bound, we have 
\begin{equation}\label{equ:lis_evaluation_6}\begin{split}
& \Pr\left[ \left(1 - \frac{\epsilon}{4}\right)^2 \left(\quality(\ps) -\frac{ \epsilon\lambda n}{100t}\right) \leq \sum_{j : |T_j |\geq \Delta} \sum_{i \in T_j}  \frac{1}{p} \lis^{[\ell_i, r_i]}(\sa_i) X_i \leq  \left(1 + \frac{\epsilon}{4}\right)^2 \quality(\ps)\right] \\
\geq & \Pr\left[ \left(1 - \frac{\epsilon}{4}\right)^2 \sum_{j : |T_j |\geq \Delta} \sum_{i \in T_j}  \lis^{[\ell_i, r_i]}(\sa_i) \leq \sum_{j : |T_j |\geq \Delta} \sum_{i \in T_j}  \frac{1}{p} \lis^{[\ell_i, r_i]}(\sa_i) X_i \right. \\ 
& \left. \ \ \ \ \ \ \ \ \ \leq \sum_{j : |T_j |\geq \Delta} \left(1 + \frac{\epsilon}{4}\right)^2  \sum_{i \in T_j}  \lis^{[\ell_i, r_i]}(\sa_i)\right]\\
\geq & 1 - O\left(\frac{\log n}{\epsilon}\right)\exp(-\Omega(\log^3 n))\\
= & 1 - \exp(-\Omega(\log^3 n)).
\end{split}\end{equation}

Now we bound random variable $\sum_{j : |T_j| <  \Delta} \sum_{i \in T_j} \lis^{[\ell_i, r_i]}(\sa_i) X_i / p$.
If $|T_j| < \Delta$, then by Chernoff bound we have
\[\Pr\left[ \sum_{i \in T_j} X_i \leq \frac{2\log^3 n}{\epsilon^2} \right] \geq  1 - \exp(-\Omega(\log^3 n)).\]
By union bound we have
\begin{align}\label{equ:lis_evaluation_7} \Pr\left[  \sum_{j : |T_j| < \Delta }\sum_{i \in T_j} \frac{1}{p} \lis^{[\ell_i, r_i]}(\sa_i) X_i  \leq \frac{2\log^3 n}{\epsilon^2}\cdot \frac{\sqrt{n}}{p} \cdot  \left\lceil \log_{1+\epsilon} \sqrt{n} \right\rceil  < \frac{\epsilon \lambda n}{100 t}\right] &\geq\\ 1 - \exp(-\Omega(\log^3 n)))&.\end{align}

By Equation~\eqref{equ:lis_evaluation_6} and Equation~\eqref{equ:lis_evaluation_7}, we obtain Equation~\eqref{equ:estimation_lis}. 

\end{proof}

Putting all the previous lemmas together gives the following result:
\begin{corollary}[algorithm for \textsf{LIS} decision problem]\label{cor:estimate_lis_main}
Given a length-$n$ sequence $A$, let $\lambda \in [1/n,1]$. There is a randomized algorithm that runs in time $O (\lambda^{-7} \sqrt{n} \log^{O(1)} n)$ such that with probability $1-1/\poly(n)$ %and is able to confirm one of the two cases : \\
\begin{itemize}
\item The algorithm accepts  if $\lis(A) \geq n\lambda$. 
\item The algorithm rejects if  $\lis(A) = O(n\lambda^4)$. 
\end{itemize}
\end{corollary}

\begin{proof}
The correctness follows from Lemma \ref{lemma:estimate}, Lemma \ref{lem:construct_pseudo_solutions_quality}, and Lemma \ref{lem:construct_pseudo_solutions_time}.

{\bf Running time:}
The running time is 
\begin{align*}
\text{time} 
= & ~ \underbrace{ O(t k^2 \sqrt{n} \log^{O(1)} n ) }_{ \text{Lemma~\ref{lem:construct_pseudo_solutions_time}} } + \underbrace{ O( t ^2 \lambda^ {-1}\sqrt{n} \log^{O(1)} n  ) }_{ \text{Lemma~\ref{lemma:estimate}} } \\
= & ~ O( t^2 \lambda^{-1} \sqrt{n} \log^{O(1)} n ) & \text{~since $t \leq O(k^2/\lambda)$} \\
= & ~ O( k^4 \lambda^{-3} \sqrt{n} \log^{O(1)} n ) & \text{~since $k \leq O(1/\lambda)$} \\
= & ~ O( \lambda^{-7} \sqrt{n} \log^{O(1)} n )
\end{align*}
Thus, we complete the proof.
\end{proof}

Finally, by starting with $\lambda = 1$ and iteratively multiplying $\lambda$ by a $1/(1+\epsilon)$ factor until a solution is found, we can approximate $\lis(A)$ within an approximation factor of $O(\lambda^3)$.

\begin{theorem}[polynomial approximation for \textsf{LIS}]\label{theorem:lis}
Given a length-$n$ sequence $A$ such that $  \lis(A)  = n \lambda $ where $\lambda \in [1/n,1]$ is unknown to the algorithm. There is an algorithm that runs in time
$\tilde O( \lambda^{-7} \sqrt{n} )$
 and outputs a number $\mathsf{est}$ such that
\begin{align*}
 \Omega(  \lis(A)   \lambda^3 ) \leq \mathsf{est} \leq O(  \lis(A)  ).
\end{align*}
with probability at least\\ $1 - 1 / \poly(n)$.
\end{theorem}

We remark that one can turn Theorem \ref{theorem:lis} into an algorithm with running time $\tilde O(n^{17/20})$ by considering two cases separately. If $\lambda < n^{-1/20}$ we sample the array with a rate of $n^{-3/20}$ and compute the \textsf{LIS} for the sampled array. Otherwise, the running time of the algorithm is already bounded by $\tilde O(n^{17/20})$.

\newcommand{\factor}{\kappa}

\subsection{An $O(n^\factor)$ Time Algorithm via Bootstrapping}\label{sec:lis_extension}
In this section, we present a general framework to reduce the running time for LIS approximation at the expense of worse approx\-imation factor. 

Let us move a step backward and analyze the previous algorithm for obtaining an $O(\lambda^3)$ approximate solution. We first divide the input array into $\sqrt{n}$ subarrays of size $\sqrt{n}$ and after constructing the pseudo-solutions, we sample $O(\lambda^4)$ subarrays to estimate the size of the solution for pseudo-solutions. The reason we set the size of the subarrays to $\sqrt{n}$ is that there is a trade-off between the first and the last steps of the algorithm. More precisely, if we have more than $\sqrt{n}$ subarrays then the number of samples we draw in the beginning would exceed $O_{\lambda}(\sqrt{n})$. On the other hand, having fewer than $\sqrt{n}$ subarrays results in larger subarrays which would be costly in the last step.

If one favors the running time over the approximation factor, the following improvement can be applied to the algorithm: In the last step of the algorithm, instead of sampling the entire subarrays and computing $\lis$ for every pseudo-solution, we recursively call the same procedure to approximate the size of the solution for each subarray. This way, having large subarrays would no longer be an issue and therefore we can have fewer subarrays to improve the number of samples we draw in the first step of the algorithm.

More formally, in order to obtain a running time of $O(\poly(\lambda) n^{\factor})$ for any constant $0 < \factor < 1$,  we set the size of each subarray to $n^{1-\factor}$ and therefore after constructing the pseudo-solutions, the problem boils down to approximating the solution for $\poly(\lambda)$ many subarrays of length $n^{1 - \factor}$. By running the same algorithm, we would have $n^{\factor}$ subarrays of length $n^{1-2\factor}$ in the second iteration. After $1/\factor - 1$ iterations, the subarrays are small enough and we can access all their elements in time $O(\poly(\lambda) n^{\factor})$. Of course, this imposes a factor of $\poly(\lambda)$ to the approximation.

\begin{figure}[h!]
\begin{center}
\includegraphics[width=12cm]{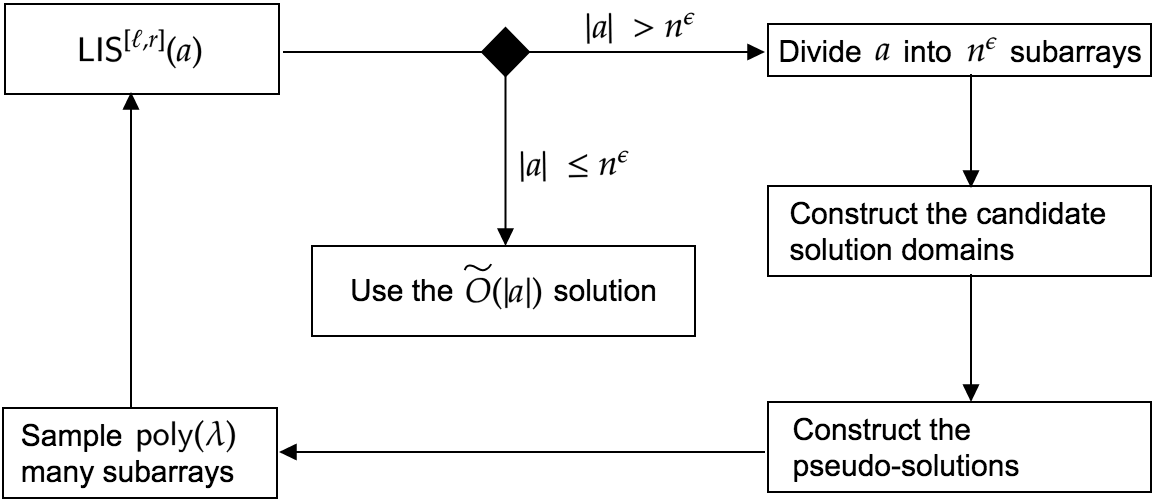}
\end{center}
\caption{The flowchart of the $O_{\lambda}(n^\epsilon)$ time algorithm is shown.}\label{fig:fig1}
\end{figure}

\begin{algorithm}[t!]
\begin{algorithmic}[1]\caption{Recursive Estimate \textsf{LIS} with Oracle}\label{alg:distinguish_with_oracle}
\Procedure{RecursiveEstimationWithOracle}{$\textsc{Oracle}, A, \lambda, \ell, r$} \Comment{Lemma~\ref{lem:approximation_with_oracle}}
	\State \Comment input: sequence $A$, parameter $\lambda$, domain interval $[\ell, r]$ 
	\State \Comment assume $\sa_1, \sa_2, \dots, \sa_{n^{\factor}}$ are subarrays of $A$
	\State \Comment subroutine $\textsf{Oracle}$  approximate \textsf{LIS} for subarrays with approximation factor $f(\lambda)$
	\For{$i \in [\zeta]$}
		\State $\cdi_i \leftarrow$ \textsc{ConstructCandidateDomains}($\sa_i$)
		\State discard all the intervals which are not in $[\ell, r]$ from $\cdi_i$
	\EndFor
	\State $\{\ps_1, \dots, \ps_t\} \leftarrow \textsc{ConstructPseudoSolutions}(\cdi_1,\dots, \cdi_{\zeta})$
	\State $\lambda_0 \leftarrow \left(\frac{\lambda}{2^8}\right)^4$ 	
	\State $p \leftarrow \frac{20 \log^4 n }{\lambda_0\zeta }$ 
	\State randomly sample each $i \in [\zeta]$ with probability $p$, and put all the samples in a set $Q$
	\For{$j \in [t]$} 
	\State $\mathsf{c} \leftarrow 0$
	\For{$ i \in W$ }
		 \If{$\exists [\ell_i, r_i] \in \ps_j$ and \textsc{Oracle}$(\sa_i, \lambda_0, \ell_i, r_i)$ accepts}
		 	 \State $\mathsf{c}  \leftarrow \mathsf{c} + 1$
		\EndIf
	\EndFor 
	\If{$\mathsf{c} \geq 3 \lambda_0 p \zeta / 4$ }
		\State \Return accept
	\EndIf
	\EndFor	
	\State \Return reject
\EndProcedure
\end{algorithmic}
\end{algorithm}

By generalizing the ideas from  previous subsections, 
we show that if there is an algorithm for \textsf{LIS} with approximation factor $f(\lambda)$, 
then we can get a $\left(f\left(\frac{\lambda^4}{2^{32}}\right) \cdot \frac{\lambda^4}{2^{33}}\right)$-approximate \textsf{LIS} algorithm with better running time using the $f(\lambda)$-approximate algorithm as a subroutine.

\begin{lemma}\label{lem:approximation_with_oracle}
Assume we partition the sequence into $\zeta$ subarrays, where $\zeta$ is polynomially related to the length of the input sequence. 
For parameter $\lambda \in (0, 1)$, let \textsc{Oracle} be a $f(\lambda)$-approximate \textsf{LIS} algorithm (with respect to a domain interval) with running time $g(n, \lambda)$ and success probability $1 - \exp(-\Omega(\log^2 n))$ where $n$ is the length of the input sequence.
Then  Algorithm~\ref{alg:distinguish_with_oracle} using \textsc{Oracle} as a subroutine 
is a $\left(f\left(\frac{\lambda^4}{2^{32}}\right) \cdot \frac{\lambda^4}{2^{33}}\right)$-approximate \textsf{LIS} algorithm with \[O\left(\lambda^{-4} g\left(\frac{n}{ \zeta}, \frac{\lambda^4}{2^{32}}\right) \log^{O(1)}n + \lambda^{-7}\zeta \log^{O(1)} n\right)\] running time and success probability $1 - \exp(-\Omega(\log^2 n))$, where $\zeta$ is the number of subarrays.
\end{lemma}
\begin{proof}
We first prove the correctness of the algorithm.
Let $A$ be a sequence of length $n$, and $\sa_1, \dots, \sa_{\zeta}$ be the subarrays.

Consider the case of $\lis^{[\ell, r]}(A) \geq \lambda n$. 
By Corollary~\ref{cor:existence_of_good_solution} and Lemma~\ref{lem:construct_pseudo_solutions_quality} with $\epsilon = \delta = 1/10$,
with probability $1 - \exp(-\Omega(\zeta \lambda))$, there exists a pseudo-solution $\ps_j$ within interval $[\ell, r]$ satisfying
\[\quality(\ps_j) \geq \frac{\lis^{[\ell, r]}(A)}{2t} \geq \frac{\lis^{[\ell, r]}(A)\lambda \eps}{2k^2} \geq \frac{\lis^{[\ell, r]}(A)\lambda \eps}{2\cdot 20^2 \log^2 (1/\delta) / ( \lambda^2 \epsilon^4 )}
\geq \frac{\eps^5 \lambda^4}{800 \log^2 (1 / \delta)} n \geq \frac{\lambda^4}{2^{31}} n.\]
Let $\alpha$ denote the number of subarrays $\sa_i$ such that $\lis^{[\ell_i, r_i]}(\sa_i) \geq \lambda_0 n / \zeta$ where $[\ell_i, r_i]$ is the interval for subarray $\sa_i$ in $\ps_j$.
We have \[\alpha \geq \frac{\quality(\ps_j) - \lambda^4 n / 2^{32}}{n/\zeta} \geq \frac{\lambda^4 \zeta}{2^{32}}.\]
By Chernoff bound, Step 14 to Step 22 of Algorithm~\ref{alg:distinguish_with_oracle} accepts on  $\ps_j$ with probability at least $1 -\exp(- \Omega(\log^2 n))$.

Consider the case of \[\lis^{[\ell, r]}(A) \leq f\left(\frac{\lambda^4} { 2^{32}}\right) \frac{\lambda^4}{2^{33}} n.\] 
Then for any pseudo-solution $\ps_j$, we have  \[\quality(\ps_j) \leq f\left(\frac{\lambda^4} { 2^{32}}\right) \frac{\lambda^4}{2^{33}} n.\] 
Let $\beta$ denote the number of subarrays $\sa_i$ such that $\lis^{[\ell_i, r_i]}(\sa_i) \geq f(\lambda^4 / 2^{32}) n / \zeta$ where $[\ell_i, r_i]$ is the interval for subarray $\sa_i$ in $\ps_j$.
We have 
 \[\beta \leq \frac{\quality(\ps_j)}{ f(\lambda^4 / 2^{32}) n / \zeta} \leq \frac{\lambda^4}{2^{33}}\zeta.\]
By Chernoff bound, Step 14 to Step 22 of Algorithm~\ref{alg:distinguish_with_oracle} do not accept on $\ps_j$ with probability at least $1 - \exp(-\Omega(\log^2 n))$.
By union bound, Algorithm~\ref{alg:distinguish_with_oracle} rejects with probability at least $1 - \exp(-\Omega(\log^2 n))$.

Hence, Algorithm~\ref{alg:distinguish_with_oracle} is a $\left(f\left(\frac{\lambda^4}{2^{32}}\right) \cdot \frac{\lambda^4}{2^{33}}\right)$-approximate algorithm for \textsf{LIS} of length $n$ with success probability at least $1 - \exp(-\Omega(\log^2 n))$.

Finally, we discuss the running time of Algorithm~\ref{alg:distinguish_with_oracle}.
By the definition of procedures \textsc{ConstructCandidateDomains} and \textsc{ConstructPseudoSolutions}, 
the running time of Step 5 to Step 9 of Algorithm~\ref{alg:distinguish_with_oracle} is $O(\lambda^{-9} \zeta \log^{O(1)}n)$.
By Lemma~\ref{lem:construct_pseudo_solutions_size}, $t = O(\lambda^{-3})$, and by Chernoff bound with probability $1 - \exp(-\Omega(\log^ 2 n))$ the size of $Q$ is at most $O(\lambda^{-4} \log^4 n)$. 
Hence, the running time of Step 12 to Step 24 of Algorithm~\ref{alg:distinguish_with_oracle} is $O(\lambda^{-7} g(n / \zeta, \lambda^4 / 2^{32}) \log^4 n)$.
\end{proof}

\begin{algorithm}[t!]
\begin{algorithmic}[1]\caption{Recursive Estimate  \textsf{LIS}}\label{alg:recursive_lis}
\Procedure{RecursiveLIS}{$A, \lambda, \ell, r$} \Comment{Lemma~\ref{lem:approximate_lis_recursive}}
	\State \Comment input: sequence $A$, parameter $\lambda$, domain interval $[\ell, r]$ 
	\State \Comment assume $\sa_1, \sa_2, \dots, \sa_{n^{\factor}}$ are subarrays of $A$
	\If {the length of $A$ is greater than $n^{2\factor}$}
	\State \Return \textsc{RecursiveLISWithOracle}($\textsc{RecursiveLIS}, A, \lambda, \ell, r$) with $\zeta = n^\kappa$
	\Else
	\For{$i \in [n^\factor]$}
		\State $\cdi_i \leftarrow$ \textsc{ConstructCandidateDomains}($\sa_i$)
		\State discard all the intervals which are not in $[\ell, r]$ from $\cdi_i$
	\EndFor
	\State $\{\ps_1, \dots, \ps_t\} \leftarrow \textsc{ConstructPseudoSolutions}(\cdi_1,\dots, \cdi_{n^\factor})$	
	\If{\textsc{EvaluatePseudoSolutions}($\ps_1, \ldots, \ps_{t}$) $\geq \lambda |A|$}
		\State \Return accept
	\Else
		\State \Return reject
	\EndIf 
	\EndIf
\EndProcedure
\end{algorithmic}
\end{algorithm}

By definition, we have the following basic fact about approximate ratio.
\begin{fact}\label{fact:approx_ratio}
Let $f$ and $f'$ be two functions mapping $(0, 1)$ to  $(0,1)$ such that $f(\lambda) \geq f'(\lambda)$ for any $\lambda \in (0, 1)$.
If there is a $f(\lambda)$-approximate \textsf{LIS} algorithm, then the algorithm is also $f'(\lambda)$-approximate.
\end{fact}

Now we present algorithm to approximate \textsf{LIS} using $\tilde O(n^{\kappa} \poly(\lambda^{-1}))$ space by applying the pseudo-solution construction-evaluation framework recursively. 
In particular, we use the same algorithm on subarrays as an oracle and apply Lemma~\ref{lem:approximation_with_oracle} recursively to approximate the entire sequence with slightly worse approximation ratio (compared with approximation ratio of the oracle). 
  
\begin{lemma}\label{lem:approximate_lis_recursive}
Let $\factor$ be a constant of $(0, 1)$ and $\lambda \in (0, 1)$. 
Algorithm~\ref{alg:recursive_lis} approximates \textsf{LIS} with approximation ratio 
\[\frac{\lambda^{2\cdot 4^{(\lceil1/\kappa\rceil - 1)}}}{256^{3\cdot 4^{(\lceil1/\kappa\rceil - 1)}}}\] 
and running time $O(n^\factor  \cdot \lambda^{-4^{O(1 / \kappa)}}\log^{O(1)} n)$ and success probability $1 - \exp(-\Omega(\log^3 n))$.
\end{lemma}
\begin{proof}
We first prove the correctness of the algorithm by induction. 
Without loss of generality, we assume $1/\factor$ is an integer.

For $i \in \{2, 3, \dots, 1/\kappa\}$, denote
\[ h_i(\lambda) = \frac{\lambda^{2\cdot 4^{(i - 1)} - 4}}{256^{ 2\cdot 4^{(i - 1)} + 3 \cdot 4^{(i-2)} - 7}}.\]
We show that Algorithm~\ref{alg:recursive_lis} is $h_i(\lambda)$-approximate if the length of the input sequence  is $n^{i \cdot \factor}$.

If the length of input sequence is $n^{2\factor}$, then $h_2(\lambda) = \frac{\lambda^4}{2^{32}}$.
By Corollary~\ref{cor:existence_of_good_solution}, Lemma~\ref{lem:construct_pseudo_solutions_quality}, and Lemma~\ref{lemma:estimate},
Algorithm~\ref{alg:recursive_lis} is $h_2(\lambda)$-approximate. 

In the induction step, for an integer $2 \leq i < 1 / \factor$,
we assume Algorithm~\ref{alg:recursive_lis} is $h_i(\lambda)$-approximate for input instance of length $n^{i \cdot \factor}$.
By Lemma~\ref{lem:approximation_with_oracle}, 
 Algorithm~\ref{alg:recursive_lis} is $\left( h_i\left(\frac{\lambda^4}{2^{32}}\right) \frac{\lambda^4}{2^{33}}\right)$-approximate for input instance of length $n^{(i+1) \cdot \factor}$.
 Since 
 \begin{align*} h_i\left(\frac{\lambda^4}{2^{32}}\right) \frac{\lambda^4}{2^{33}} =&  \frac{\lambda^{4 \cdot (2\cdot 4^{(i - 1)} - 4)}\cdot \lambda^4}{256^{4\cdot (2\cdot 4^{(i - 1)} - 4)}\cdot 256^{ 2\cdot 4^{(i - 1)} + 3 \cdot 4^{(i-2)} - 7} \cdot 2^{33}}  
\\
= &  \frac{\lambda^{2\cdot 4^{i}-12}}{256^{2\cdot 4^i + 2\cdot 4^{(i-1)} + 3\cdot 4^{(i-2)} - 18.875}} \\
> & \frac{\lambda^{2\cdot 4^{i} - 4}}{256^{ 2\cdot 4^{i} + 3 \cdot 4^{(i-1)} - 7}} \\
= & h_{i+1}(\lambda),
 \end{align*}
 by Fact~\ref{fact:approx_ratio}, 
  Algorithm~\ref{alg:recursive_lis} is $ h_{i+1}(\lambda)$-approximate for input instance of length $n^{(i+1) \cdot \factor}$.
  Since 
  \[h_{1/\kappa}(\lambda) > \frac{\lambda^{2\cdot 4^{((1/\kappa) - 1)}}}{256^{3\cdot 4^{((1/\kappa) - 1)}}},\]
 by Fact~\ref{fact:approx_ratio},  
  Algorithm~\ref{alg:recursive_lis} is $\left(\frac{\lambda^{2\cdot 4^{((1/\kappa) - 1)}}}{256^{3\cdot 4^{((1/\kappa) - 1)}}}\right)$-approximate for input instance of length $n$.
  
  By  
  Corollary~\ref{cor:existence_of_good_solution}, Lemma~\ref{lem:construct_pseudo_solutions_quality},  Lemma~\ref{lemma:estimate} and Lemma~\ref{lem:approximation_with_oracle},
  we have the desired running time.
  The success probability is obtained by same corollaries/lemmas and union bound.
\end{proof}

Finally, by starting with $\lambda = 1$ and iteratively multiplying $\lambda$ by a $1/(1+\epsilon)$ factor until a solution is found, we can approximate $\lis(A)$ within an approximation factor of $\lambda^{O(4^{1 / \kappa})}$.

\begin{theorem}
Let $\factor$ be a constant of $(0, 1)$ and $\lambda \in (0, 1)$. 
There exists a $\tilde O(n^\factor  \cdot \lambda^{-4^{O(1 / \kappa)}})$ time algorithm for $\lis$ with approximation factor $\lambda^{4^{O(1 / \kappa)}}$ and success probability $1 - \exp(-\Omega(\log^3 n))$  .
\end{theorem}

\section{Acknowledgement}
The authors would like to thank Barna Saha, Debarati Das, and anonymous reviewers for helpful feedback on earlier versions.

\bibliography{ref,lis}
\bibliographystyle{alpha}%{siamplain}

\appendix
% !TeX root = ../../main.tex

\newpage
\section{Probability, combinatorial, and Graph Tools}
In this section, we restate probability and graph tools that we use throughout this paper. All these theorems are proven in previous work.

\begin{theorem}[Chernoff Bounds \cite{c52}]\label{thm:chernoff}
Let $X = \sum_{i=1}^n X_i$, where $X_i=1$ with probability $p_i$ and $X_i = 0$ with probability $1-p_i$, and all $X_i$ are independent. Let $\mu = \E[X] = \sum_{i=1}^n p_i$. Then \\
1. $ \Pr[ X \geq (1+\delta) \mu ] \leq \exp ( - \delta^2 \mu / 3 ) $, $\forall \delta > 0$ ; \\
2. $ \Pr[ X \leq (1-\delta) \mu ] \leq \exp ( - \delta^2 \mu / 2 ) $, $\forall 0 < \delta < 1$. 
\end{theorem}

\begin{theorem}[Hoeffding bound \cite{h63}]\label{thm:hoeffding}
Let $X_1, \cdots, X_n$ denote $n$ independent bounded vari\-ables in $[a_i,b_i]$. Let $X= \sum_{i=1}^n X_i$, then we have
\begin{align*}
\Pr[ | X - \E[X] | \geq t ] \leq 2\exp \left( - \frac{2t^2}{ \sum_{i=1}^n (b_i - a_i)^2 } \right)
\end{align*}
\end{theorem}

\begin{theorem}[Blakley-Roy inequality, \cite{br65}, see also Proposition 3.1 in \cite{ksv13}]\label{lem:blakley_roy_inequality}
Let $G$ denote a graph that has $n$ vertices and average degree $d$. The number of walks of length $k$ in graph $G$ is at least $n d^k$.
\end{theorem}

\begin{theorem}[Tur{\'a}n theorem for bipartite graphs, \cite{kst54}, see also \cite{bbk13}]\label{lem:turan}
For a graph $G$ the Tur{\'a}n number $\mathrm{ex}(G,n)$ is the maximum number of edges that a graph on $n$ vertices can have without containing a copy of $G$. For any $s \leq t$, $\mathrm{ex}(K_{s,t}, n ) \leq \frac{1}{2} (t-1)^{1/s} n^{2-1/s} + o(n^{2-1/s})$
\end{theorem}
In particular, when $s = t$, Theorem~\ref{lem:turan41} implies that for large enough $n$ we have $\mathrm{ex}(K_{s,s}, n ) \leq n^{2-1/s}$.

\begin{theorem}[Tur{\'a}n theorem for cliques \cite{t41}]\label{lem:turan41}
Let $G$ be any graph with $n$ vertices, such that $G$ is $K_{r+1}$-free. Then the number of edges in $G$ is at most 
\begin{align*}
(1-\frac{1}{r}) \cdot \frac{n^2}{2}.
\end{align*}
\end{theorem}

\begin{corollary}[of Theorem \ref{lem:turan41}]
	Let $G$ be any graph with $n$ vertices, such that $G$ has no independent set of size $r+1$. Then the number of edges in $G$ is at least 
	\begin{align*}
		\binom{n}{2} - (1-\frac{1}{r}) \cdot \frac{n^2}{2} = \frac{nr+n^2}{2r}.
	\end{align*}
\end{corollary}

\begin{lemma}[application of Jensen's inequality]\label{lemma:math}%H\"{o}lder's
	Let $n_1, n_2, \ldots, n_k$ be a sequence of integer numbers of size $k$ and $\lambda_1, \lambda_2, \ldots, \lambda_k$ be $k$ real numbers in the interval $[0,1]$. Define $\lambda = (\sum n_i \lambda_i)/(\sum n_i)$. Then for any $y\geq 0$ we have 
	\begin{align*}
	\sum n_i \lambda_i^{1+y} \geq \sum n_i \lambda^{1+y}.
	\end{align*}
\end{lemma}

\begin{proof}
Let $\psi(x) = x^{1+y}$. Since $\psi$ is real convex function, by Jensen's inequality, we have
\begin{align*}
 \psi( \frac{ \sum_i n_i \lambda_i }{ \sum_i n_i } ) \leq \frac{ \sum n_i \psi(\lambda_i) }{  \sum n_i }
\end{align*}
Applying definition $\psi$, we have
\begin{align*}
	 ( \frac{ \sum_i n_i \lambda_i }{ \sum_i n_i } )^{1+y} \leq \frac{ \sum n_i \lambda_i^{1+y} }{  \sum n_i } .
\end{align*}
Using definition of $\lambda$, we have
\begin{align*}
\lambda^{1+y} \leq \frac{ \sum n_i \lambda_i^{1+y} }{  \sum n_i } .
\end{align*}
\end{proof}

\begin{theorem}[A well-known algorithm also used in~\cite{hsss19}]\label{theorem:smalllcs}
	Given two strings $A$ and $B$, one can with preprocessing time $O(|B| \log |B|)$  verify if the \textsf{LCS} of $A$ and $B$ is at least $q$ or not in time $O(|A|q \log |B|)$. In case the answer is positive, finding such a common subsequence can be done in time $O(|A|q \log |B|)$.
\end{theorem}
\begin{proof}
	In the preprocessing step, for each character $x$ in $B$, we construct a binary tree that keeps track of all the places that $x$ appears in $B$. This enables us to answer the following queries in $O(\log |B|)$ time: \textit{Given an index $i$ of $B$ and a character $x$, what is the smallest index $i' \geq i$ such that $B_{i'} = x$?}
	
	To find the \textsf{LCS} of $A$ and $B$, we slightly modify the conventional dynamic program for computing \textsf{LCS} and construct a two-dimensional array $T^*$ that stores the following information
	
	$$
	T^*[i][j]= \begin{cases}
	\text{ the smallest $k$ s.t.} & \hspace{-2mm}\text{ if }|\mathsf{lcs}(A[1,i],B)| \geq j \\
	|\mathsf{lcs}(A[1,i],B[1,k])| = j & \\
	\infty & \hspace{-2mm}\text{ otherwise } \end{cases}
	$$
	Using the above definition, we can construct table $T^*$ via the following recursive formula:
	
	\begin{equation}\label{dp2}
	T^*[i][j] := \min\Big\{\begin{array}{c} T^*[i-1][j],\\ f(T^*[i-1][j-1]+1, A_i)\end{array}\Big\}
	\end{equation} 
	where $f(T^*[i-1][j-1]+1, A_i)$ is the index of the first occurrence of $A_i$ in $B$ after position $T^*[i-1][j-1]$ (or $\infty$ if $A_i^*$ does not appear in $B$ after position $T^*[i-1][j-1]$). Since such queries can be answered in $O(\log |B|)$ time then the overall runtime of the algorithm is $O(|A|q\log |B|)$.
\end{proof}

\newpage

\end{document}